%% file: main.tex
\documentclass[11pt]{article}

\input{preamble}

\begin{document}

\thispagestyle{empty}
\begin{adjustwidth}{-1cm}{-1cm}
\begin{center}
    ~\vspace{9mm}
    
     {\LARGE \bf 
   3D near-de Sitter gravity and the soft mode of DSSYK 
   }
   \end{center}
\end{adjustwidth}
    
\begin{center}
   \vspace{0.4in}
    
    {\large \bf Tommaso Marini$^a$, Xiao-Liang Qi$^b$ and Herman Verlinde$^a$}\\
    
    \vspace{0.4in}
    
    {$^a$Department of Physics, Princeton University, Princeton, NJ 08544, USA} \\[3mm]
    {
    $^b$Leinweber Institute for Theoretical Physics, Stanford University, Stanford, CA 94305, USA}
\end{center}

\vspace{0.4in}

\begin{abstract}

\input{sections/abstract}
\end{abstract}

\def\nspc{\hspace{-1pt}}
\pagebreak
\setcounter{page}{1}
\tableofcontents

\newpage
\section{Introduction}\label{sec:introduction}

\input{sections/introduction}

\medskip

\section{Soft mode dynamics of DSSYK}\label{sec:soft_mode_of_DSSYK}
\vspace{-.5mm}

\input{sections/soft_mode_of_DSSYK}

\section{Boundary dynamics of 3D near-de-Sitter gravity}\label{sec:boundary_dynamics}
\vspace{-1mm}

\input{sections/boundary_dynamics}

\section{Thermodynamics of 3D near-de-Sitter gravity}\label{sec:boundary_thermodynamics}
\vspace{-1mm}

\input{sections/boundary_thermodynamics}

\section{Towards a holographic dictionary}\label{sec:towards_an_holographic_dictionary}
\vspace{-1mm}

\input{sections/towards_an_holographic_dictionary}

\section{Concluding remarks}\label{sec:discussion}
\vspace{-1mm}

\input{sections/discussion}

\appendix
\section{Spinor coordinates on 3D de Sitter}\label{app:spinor_coordinates}

\input{sections/spinor_coordinates}

\section{More on the effective action}\label{app:more_on_the_action}
\input{sections/more_on_the_action}

\printbibliography

\end{document}

%% file: preamble.tex
\pdfoutput = 1
\usepackage{xcolor}
\usepackage[utf8]{inputenc}
\usepackage{color,graphicx}
\usepackage{epsfig}
\usepackage{amsmath}
\usepackage{verbatim}
\usepackage{amssymb}
\usepackage{mathabx}
\usepackage{stmaryrd}
\usepackage{empheq}
\usepackage{url}
\usepackage{esint}
\usepackage{float}
\usepackage{physics} 
\usepackage{amsfonts}
\usepackage{array}
\usepackage{setspace}
\usepackage{braket}
\usepackage{float}
\usepackage{url}
\usepackage{mathtools}
\usepackage{tensor}
\usepackage{slashed}
\usepackage{tikz}
\usepackage{tikz,pgfplots}
\usepackage{tikz-3dplot}
\usepackage{epstopdf}
\usepackage{subcaption}
\usepackage[labelfont=bf,font={sf}]{caption}
\usepackage{pgfplots}
\pgfplotsset{compat=1.18}
\usepackage{mathrsfs}
\usepackage{fancybox}
\usepackage{eurosym}
\usepackage{tcolorbox}
\usepackage{tensor}
\allowdisplaybreaks
\usepackage{changepage}

\usepackage[mathcal]{eucal}

\usepackage[margin = 2.2cm]{geometry}
\usepackage[ragged]{footmisc}
\usepackage[maxbibnames=99,
    sorting=none,
    backend=biber,
    style=numeric,
]{biblatex}
\usepackage[bookmarks=true,bookmarksnumbered=false,hyperindex=true,bookmarksopen=true,hyperfigures=true,colorlinks=true,linkcolor=webblue,citecolor=webgreen,urlcolor=webblue,breaklinks]{hyperref}
\definecolor{webgreen}{rgb}{0, 0.5, 0}
\definecolor{webblue}{rgb}{0, 0, 0.5}
\definecolor{webred}{rgb}{0.5, 0, 0}
\definecolor{darkgreen}{rgb}{0,0.5,0}

%% file: sections/abstract.tex
We present a dual gravity interpretation of the complex reparametrization mode $\psi$ that governs the soft dynamics of double-scaled SYK in the presence of a time-dependent Maldacena-Qi coupling. We find that the dual gravity system takes the form of 2+1-dimensional Einstein-de Sitter gravity with an energy distribution localized on a dS$_2$ slice within dS$_3$. The effective SYK equations of motion take the form of the Israel junction conditions across the dS$_2$ slice. We study the 1D effective action of the SYK soft mode and show that it coincides with the effective action derived from 3D Einstein-de Sitter gravity with conformal boundary conditions  on $\mathscr{I}^\pm$. The boundary conditions split $\mathscr{I}^\pm$ into two hyperbolic $k=-1$ slices, and the holographic screen is placed at the intersection. 
We adapt the Gibbons-Hawking calculation of the Schwarzschild-de Sitter entropy to the case with $k=-1$ boundary conditions and find that it reproduces the semiclassical DSSYK entropy. The boundary-to-boundary Green functions in 3D de Sitter are equal to the square of DSSYK two-point functions. We give an alternative holographic interpretation of our results in terms of 3D AdS gravity with two time directions.

%% file: sections/introduction.tex
\def\sstrut{\mbox{\footnotesize ${{}^{\strut}_{\strut}}$}.}
\def\is{\! & \! =\!  &\!}
\def\spc{\hspace{1pt}}
\vspace{-2mm}

The SYK model \cite{ Sachdev_1993, kitaev2015simple} is one of few soluble models of low-dimensional holography. Its low energy dynamics reduces to that of a single one-dimensional reparametrization mode governed by 1D Schwarzian quantum mechanics \cite{Maldacena:2016hyu,Kitaev:2017awl}. The presence of this soft mode enables the identification of a holographic dual in terms of two-dimensional JT gravity theory on AdS$_2$ \cite{almheiri2015modelsads2backreactionholography, maldacena2016conformalsymmetrybreakingdimensional, Jensen:2016pah, Engels_y_2016,Kitaev:2017awl}. 
The aim of this paper is to extend this holographic duality to the double-scaled SYK model \cite{Berkooz:2018jqr, Berkooz:2018qkz, Berkooz:2024lgq} at {all energies}. 
 
We will focus on the non-linear dynamics in the semiclassical regime where the dimensionless DSSYK coupling $\lambda = 2q^2/N$, with $N$ the number of Majorana oscillators and $q$ the order of the SYK interaction, is taken to zero while energies are allowed to scale as $1/\lambda$. It was shown in \cite{Lensky_2021} that the DSSYK soft dynamics in this high energy regime is still governed by a reparametrization mode $\psi(t)$, but the mode is now {\it complex}. It is natural to look for a dual gravity interpretation of this complex soft mode. We will find that the relevant dual theory is 2+1 Einstein-de Sitter gravity 
 with a special type of conformal boundary conditions at past and future infinity $\mathscr{I}^\pm$. The boundary conditions are such that the gravitational soft dynamics is captured by a single complex mode $\psi(t)$ that parametrizes the shape of a one-dimensional curve ${\cal C}$ within $\mathscr{I}^\pm$ on which the SYK model lives. We will match the effective action and classical equations of motion on both sides of the duality and establish a semiclassical dictionary between the two-point functions of the DSSYK model with boundary-to-boundary propagators of local operators placed on the holographic screen.

\vspace{-.5mm}

\subsection{Soft mode of DSSYK} 
\vspace{-1mm}

 Let us briefly describe the 1D effective theory of the DSSYK model that will be our main object of study. More details are found in section \ref{sec:soft_mode_of_DSSYK} and in the original papers \cite{Lensky_2021,Lin_2022,blommaert2025qschwarzianliouvillegravity}. The 1D effective action describes the motion of two canonically conjugate variables $\phi$ and $p$ and takes the~form
\begin{eqnarray}
\label{action+hamiltonian}
S_{\rm SYK} \! \is \! \frac{2}{\lambda}\int\!dt\,(p\spc \dot\phi - H),
\qquad \quad
H \,=\,  -\sqrt{1\!-\!e^{-2\phi}}\spc \cos p\spc .
%-\spc \mu(t) \phi.    
\end{eqnarray}
%\footnote{Here we use the short-hand notation $\psi_1 = \psi(t_1)$, $\psi_2^* = \psi^*(t_2)$, etc.}
This Hamiltonian, introduced in \cite{Lensky_2021}, can be 
interpreted as an effective quantum mechanics description of the DSSYK chord rules via the identification \cite{Lin_2022}
\begin{eqnarray}
 a^\dagger = {e^{ip}}\sqrt{1-e^{-2\phi}}, \qquad a= \sqrt{1-e^{-2\phi}} e^{-ip}, \qquad e^{-2\phi} \, =\; e^{-\lambda n} 
\end{eqnarray} 
with $a,a^\dagger$ the $q$-deformed oscillators and $n$ the number operator that create, annihilate, and count the Hamiltonian chords \cite{Berkooz:2018jqr}. As explained in \cite{Lin_2022}, the Hamiltonian \eqref{action+hamiltonian} should be interpreted as acting on the Hilbert space of a two-sided SYK model. %The prefactor of the action \eqref{action+hamiltonian} is determined by matching with the chord analysis.  

On the gravity side, the Hamiltonian \eqref{action+hamiltonian} has been derived via a covariant phase space treatment of two-dimensional sine-dilaton gravity \cite{blommaert2025qschwarzianliouvillegravity, Blommaert:2023opb, Blommaert:2024whf_entropic_puzzle, Blommaert:2024ydx} 
(or complex Liouville gravity \cite{collier2025complexliouvillestring, Blommaert:2025eps}). The variable $\phi$ then represents the total boundary-to-boundary length of the dual 2D spacetime \cite{Lin_2022}. The chord rule interpretation of \eqref{action+hamiltonian} and the duality with sine-dilaton gravity both extend to the full quantum regime at finite $\lambda$. Our goal here is different: we will interpret the semiclassical 1D effective theory \eqref{action+hamiltonian} as a non-linear generalization of Schwarzian quantum mechanics and look for a dual interpretation in terms of a gravitational soft mode, generalizing \cite{Jensen:2016pah,maldacena2016conformalsymmetrybreakingdimensional,Engels_y_2016}. The first step was already done for us in \cite{Lensky_2021}. A holographic dual interpretation of this 1D effective theory in terms of gravitational soft dynamics, however, has not been given thus far. 

Following \cite{Lensky_2021}, we will generalize our setup to include a time-dependent perturbation to the effective Hamiltonian proportional to the total chord number variable $\phi$
\begin{eqnarray}
\label{eq:pertham}
H(t) = -\sqrt{1\!-\!e^{-2\phi}}\spc \cos p \, + \, \mu(t)\phi  . 
\end{eqnarray}
In terms of the microscopic SYK model, the extra interaction term represents a time-dependent Maldacena-Qi (MQ) bilinear deformation $H_{\rm int} \,\propto\, \mu(t) \sum_i \chi_i^L\chi_i^R$ of the two-sided SYK system \cite{maldacena2018eternaltraversablewormhole}. On the dual 2D gravity side, the coupling turns on a bilocal source for particles that travel from boundary to boundary. This leads to a non-trivial motion of the boundary trajectory. This motion was obtained on the SYK side in \cite{Lensky_2021} by solving for the two-point function of two same-side SYK operators ${\cal O}_\Delta(t_1)$ and ${\cal O}_\Delta(t_2)$  
with scale dimension $\Delta$ in the presence of the MQ coupling $\mu(t)$. This two-point function can be written as a functional integral over a complex reparametrization mode $\psi(t)$ 
\begin{eqnarray}
\label{twopoint}
\bigl\langle \mathcal{O}(t_1) \mathcal{O}(t_2) \bigr\rangle%_{\rm DSSYK} 
\is \int\! [d\psi]\, e^{\mbox{\footnotesize $i$}
S_{\rm SYK}[\psi]} \, G_\Delta(t_1,t_2) ,\\[2.25mm]
\label{greenone}
\  G_\Delta(t_1,t_2) \is
\left(\frac{\dot\psi(t_1)\dot\psi^{*}(t_2)}{\sin^2\bigl(\frac {\psi(t_1) - \psi^*(t_2)}2\bigr)}\right)^{\Delta}.
\end{eqnarray}
Here $S_{\rm SYK}[\psi]$ is given by the 1D effective action \eqref{action+hamiltonian} with time-dependent Hamiltonian \eqref{eq:pertham}, and $\psi(t)$ is expressed in term of the two real variables $p(t)$ and $\phi(t)$ via \cite{Lensky_2021}
\begin{equation}
\label{psiparam}
\dot\psi =  \frac{e^{i p}}{\sqrt{e^{2\phi}-1}}.
\end{equation}
The real part of the variable $\psi(t)$ can be viewed as the analog of the low energy Schwarzian mode. The imaginary part of $\psi(t)$ represents a dynamical UV regulator that allows $G_\Delta$ to look like a thermal CFT$_1$ two-point function while satisfying the boundary condition that $G_\Delta =1$ at coincident points.\footnote{This boundary condition is responsible for the `fake temperature' phenomenon of large $q$ SYK. Im $\psi(t)$ can be interpreted as a time-dependent demarcation point between the fake and real thermal circle.}
The above formula was derived in \cite{Lensky_2021} in the leading order semiclassical limit $\lambda\to 0$.\footnote{It may be possible to lift \eqref{twopoint}-\eqref{greenone} to a geometric path-integral representation of the exact SYK chord rules \cite{Berkooz:2018jqr,Berkooz:2024lgq,Lin_2022}. We will not attempt to develop this line of reasoning here, as our interest is in the semiclassical theory.}
Our goal is to find the exact gravity dual that reproduces the above non-linear 1D SYK effective theory.

In the parametrization \eqref{psiparam}, the low temperature limit of the SYK model corresponds to the regime $p\to 0$, $\phi \to \infty$. The $\psi(t)$ variable then becomes real and the 1D effective theory \eqref{action+hamiltonian} reduces to the familiar Schwarzian quantum mechanics that describes the dynamical AdS$_2$ boundary \cite{Jensen:2016pah,maldacena2016conformalsymmetrybreakingdimensional,Engels_y_2016} while the bi-local observable $G_\Delta(t_1,t_2)$ in \eqref{greenone} represents the boundary-to-boundary propagator of a massive bulk particle in AdS$_2$. As we will see, this interpretation extends to all energies, provided one replaces the dual near-AdS$_2$ JT gravity theory by 2+1-D near-de Sitter gravity.

\def\darkblue{blue}
\def\darkred{red}

\subsection{2+1-D near-de Sitter gravity} \vspace{-1mm}

There are several indications that the gravitational theory that captures DSSYK correlation functions at high temperature must involve 3D de Sitter dynamics \cite{narovlansky2023doublescaledsyksitterholography, lin2022infinitetemperatureshot, susskind2022sitterspacedoublescaledsyk, rahman2024dsjtgravitydoublescaled,  verlinde2024doublescaledsykchordssitter, Gaiotto_verlinde:2024kze, Mengyang_Verlinde:2024zrh, tietto2025microscopicmodelsitterspacetime, Narovlansky:2025tpb}. %First, there exists an exact correspondence between the DSSYK partition and  correlation functions and those of 3D SL(2,$\mathbb{C}$) Chern-Simons theory on a suitably chosen geometry.\footnote{ A key element in this map is the isomorphism between the chord rules and the skein algebra of Wilson lines. Both are encoded via the 1D effective action \eqref{action+hamiltonian}. } Secondly, the bi-local DSSYK effective theory and the edge theory of 3D de Sitter gravity both take the form of complex Liouville/sine-dilaton gravity. 
In spite of these clear hints, formulating a comprehensive holographic dS$_3$/SYK dictionary remains an outstanding challenge. 
Here we will add some key new elements to the correspondence. 

There are two popular notions of de Sitter holography. The first is the dS/CFT dictionary \cite{Strominger:2001pn}, that places a CFT at past or future infinity $\mathscr{I}^\pm$. For 2+1 de Sitter space, this would require starting with a 2D boundary theory. %\footnote{For a relevant recent proposal, see \cite{Collier}} 
Another approach is worldline holography \cite{Anninos_2012, Chandrasekaran:2022cip,narovlansky2023doublescaledsyksitterholography}, which places the holographic screen along the timelike worldline of a model observer. This viewpoint has some advantages, especially since our candidate dual theory is one-dimensional.  Our proposal combines elements of both approaches, but our philosophy is different: we are not guided by any preferred holographic viewpoint but by evidence provided by the DSSYK model itself. The idea is as follows.

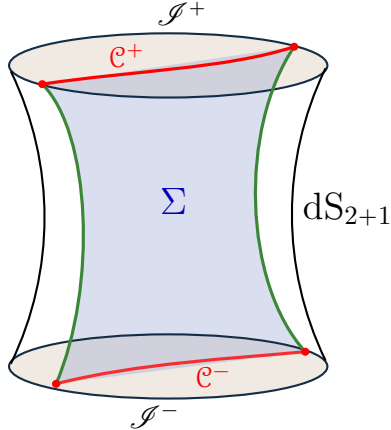
\begin{figure}[t]
\centering
\input{images/near_dS2}
\caption{The near-dS$_2$ geometry $\Sigma$ is embedded as a gluing surface inside 3D de Sitter spacetime with conformal boundary conditions at $\mathscr{I}^\pm$. 
%The boundary conditions restrict the energy flux at $\mathscr{I}^\pm$ to be located on the 1D holographic screen $\mathcal{C}^\pm$. 
The shape of the gluing surface $\Sigma$ and the holographic screen $\mathcal{C}^\pm$ are determined by the 3D Einstein equations, which reduce to the Israel junction conditions on $\Sigma$. The resulting equations of motion match the Lensky-Qi equations that govern the SYK boundary dynamics.}
\vspace{-3mm}
\end{figure}

The 1D SYK effective action \eqref{action+hamiltonian} governs the dynamics of a 1D trajectory $\psi(t) = x(t)+ iy(t)$ moving on a complex plane parametrized by two coordinates $x$ and $y$ with the same signature. 
Holography then adds a third direction with the opposite signature. So the total emergent spacetime is three-dimensional and has either (2,1) or (1,2) signature. For most of this paper, we will choose the $x$ and $y$ coordinates to be both spacelike. We then find that the dual theory describes 3D de Sitter gravity. We will comment on the other choice of signature momentarily and in section \ref{subsec:twotime}.  

From now on, we will denote the SYK (real) time by $u$. The shape of the $\psi(u)$ trajectory is influenced by the coupling $\mu(u)$. It is natural to think of $\psi(u)$ as a gravitational degree of freedom and of $\mu(u)$ as a matter source. Suppose the source is localized on a 1D trajectory inside a 2D boundary. Adding the third holographic direction,  the matter source then creates a 2D energy sheet $\Sigma$ that extends into the 3D bulk. Away from $\Sigma$, the bulk is empty and looks like pure 3D de Sitter space. All non-trivial gravitational dynamics takes place right at the surface $\Sigma$, where the Einstein equations reduce to the Israel junction conditions (here $[K_{ab}]$ denotes the discontinuity of $K_{ab}$ at $\Sigma$)
\begin{eqnarray}
\label{eq:israelj}
  [K_{ab}]- [K] h_{ab} = 8\pi G_N T_{ab}.
\end{eqnarray}
Our hypothesis, which we confirm in section \ref{sec:boundary_dynamics}, is that the  equations of motion derived from the 1D effective action \eqref{action+hamiltonian} describe the Israel junction conditions in 3D Einstein-de Sitter gravity across a tensionless energy sheet sourced by the coupling $\mu(u)$. This correspondence between the dynamical equations of SYK and dS$_3$ gravity remains non-trivial even when we turn off the coupling $\mu$.

The match between the equations of motion of the  soft mode $\psi(t)$ and the Israel junction conditions \eqref{eq:israelj} motivates us to introduce a 3D generalization of near-AdS$_2$ gravity, which we call 3D near-de Sitter gravity. It is defined by pure Einstein-de Sitter gravity  with conformal boundary conditions on $\mathscr{I}^\pm$.
%\footnote{Extrinsic curvatures are computed with the inward pointing normal; this conventions implies the non standard signs of the boundary terms in the action.}
 As indicated in figure \ref{fig:near-desitter}, the boundary conditions specify that future and past infinity split up into two hyperbolic $k=-1$  slices. We place our 1D holographic screen at the equator $\mathcal{C}$ where the two hyperbolic slices meet. The 1D screen $\mathcal{C}$ can emit and absorb energy by means of local operator insertions. In the basic setting, with only the MQ coupling turned on, the energy-momentum in the bulk forms a stationary mass density localized on a 2D  surface $\Sigma$. Gravitational backreaction leads to a discontinuity that splits the embedding equation of the surface into two surfaces $
    \Sigma \, =\,
    (\tau, x(u),y(u))$ and $\Sigma^*=(\tau, x(u),-y(u))$
that stretch between ${\cal C}^{\pm}$. The physical spacetime is obtained by gluing $\Sigma$ and $\Sigma^*$.

\begin{figure}[t]
\centering
\raisebox{-2.9cm}{\input{images/action}}~~~~~~~~~~~~~~
\fbox{\hspace{-5mm}
\parbox{7.5cm}{
${}$
\smallskip
\begin{eqnarray}\notag
   \qquad\ S_{EH} =\frac{1}{16\pi G_N}
\int_{\mbox{\footnotesize\color{purple}$\mathcal{V}$}}\!\! \sqrt{-g}\spc(R-2)\qquad\quad \\[4mm]\notag
  ~~~~~+\, \frac{1}{16\pi G_N}\left[2
\int_{\mbox{\scriptsize\color{blue}$\Sigma$}}\!\! \sqrt{-h}\spc K_h
    -\int_{\mbox{\scriptsize\color{darkgray}$\mathscr{I}^{\pm}$}}\!\!\sqrt{\sigma}\spc K_\sigma \right]~~~~
    \\[4mm]\notag
   \qquad   +\,\;  \frac{1}{8\pi G_N}
\int_{\mbox{\scriptsize\color{red}$\mathcal{C}^\pm$}}
   \!\!\sqrt{\gamma}\spc \eta_H\qquad \qquad\qquad
\end{eqnarray}}}
\caption{The Einstein-Hilbert action of 2+1D near-de Sitter gravity splits up into the usual bulk term, two Gibbons-Hawking-York boundary terms on $\Sigma^\pm$ and $\mathscr{I}^\pm$, and Hayward corner terms on $\mathcal{C}^\pm$. The normalization of the boundary terms at $\mathscr{I}^\pm$ is fixed by our choice of conformal boundary conditions.}
\label{fig:near-desitter}
\vspace{-2mm}
\end{figure}

The Einstein-Hilbert action of near-dS$_3$ gravity consists of the usual bulk action, two Gibbons-Hawking-York boundary terms \cite{Gibbons:1976ue, PhysRevLett.28.1082}, and a Hayward corner term \cite{Hayward:1993my}.  
%The normalization of the GHY term on $\mathscr{I}^\pm$ \cite{BlauGRNotes, Krishnan_2017} is prescribed by the conformal boundary conditions with $K<0$ fixed.  
As our main result, we show 
that the reduced 1D action that governs the shape $\psi(u)=x(u)+iy(u)$ of the gluing surface $\Sigma$ for a given energy-momentum distribution $\mu(u)$, matches the 1D effective action of the DSSYK soft mode 
\begin{eqnarray}
   \boxed{\ \ S_{EH}[\psi] = %\frac{1}{\lambda}\,
   S_{\rm SYK}[\psi],\qquad \lambda = 4\pi G_N \sstrut\ }
\end{eqnarray}

%\begin{figure}[t]
 % \centering
  % \includegraphics[width=0.285\textwidth]{images/ads-flat}~~~~~~~~
  % \includegraphics[width=0.28\textwidth]{images/ads-middle2}~~~~~~~~
  % \includegraphics[width=0.30\textwidth]{images/ads-full2}\\[2mm]
%$\mbox{${}$~A.~Low~temperature~~~~~~~~~~~~~~~~~~~~~B.~High temperature~~~~~~~~~~~~~~~~~C.~Infinite~temperature }
%$\caption{The euclidean geometry for three different temperatures. At low energies, the boundary equations reduce to those of Schwarzian quantum mechanics.}
% \label{fig:threetees} \end{figure}

\vspace{-2mm}

\def\darkgreen{green!30!black}
\subsection{dS$_{2+1}$ or AdS$_{1+2}$\,?}
\vspace{-1mm}

\noindent
In sections \ref{sec:boundary_dynamics} through \ref{sec:towards_an_holographic_dictionary}, we will analyze the 1D effective action of $2+1$-dimensional de Sitter gravity. In the absence of matter, the $2+1$-D de Sitter spacetime is specified by the embedding equation %
\begin{eqnarray}
{\rm dS}_{d,1}\, : \  \qquad \ \ -X_0^2 + \sum_{i=1}^{d+1} X_i^2 = 1
\label{sitter-embed}
\end{eqnarray}
in $\mathbb{R}^{1,d+1}$ with $d=2$. A potentially worrying feature of our proposed holographic dictionary is that the time-direction of the SYK model gets mapped to a spacelike direction along $\mathscr{I}^\pm$.
Indeed, one could convincingly argue that the dual gravity theory of DSSYK should have two time and one space directions, rather than two space and one time direction. The $\psi=x+iy$ plane is Euclidean, but the boundary time $u$ is real time, so it looks like $x$ and $y$ should both be timelike.  
Since holography then adds an extra spatial dimension, one would conclude that the dual gravity theory should be 1+2-dimensional AdS gravity with two time directions. Its vacuum spacetime manifold is then given by
\begin{eqnarray}
{\rm AdS}_{d-1,2}\ :\  \qquad   -X_0^2 - X_{d+1}^2\! - X_{d}^2 + \sum_{i=1}^{d-1} X_i^2 = -1
\label{asitter-embed}
\end{eqnarray}
in $\mathbb{R}^{3,d-1}$  with $d=2$. Equation \eqref{asitter-embed} for $d=2$ coincides with the de Sitter embedding equation \eqref{sitter-embed}. Indeed the gravitational dynamics of  dS$_{2+1}$ and AdS$_{1+2}$ gravity are equivalent and trivially mapped to each other by putting an overall minus sign in front of the metric.  

\begin{figure}[t]
\centering
    \includegraphics[width=3.3in]{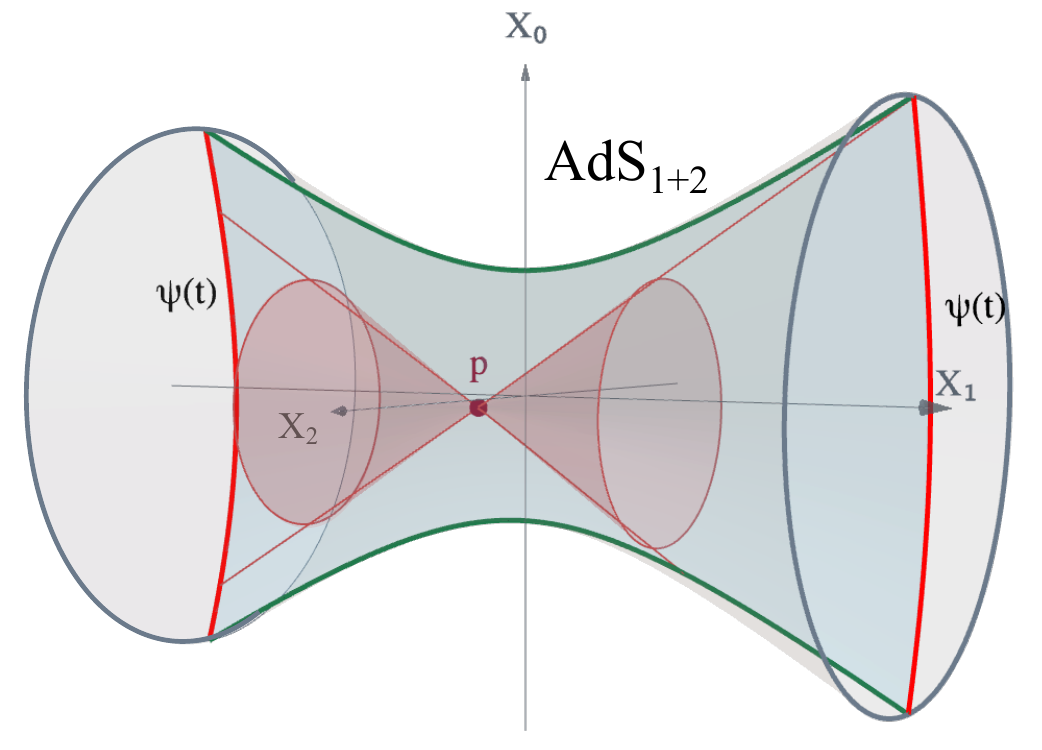}
    \caption{Schematic depiction of the AdS$_{1+2}$ holographic perspective with two time directions. The soft mode $\psi(t)$ traces a 1D path (\textcolor{red}{red}) on the 2D boundary  of the emergent 3D geometry. The 2D gluing surface $\Sigma$ spans between the 1D paths. The bulk light cone centered at $p$ is oriented towards the 2D boundary. The 1D curves lie at spacelike future and past infinity relative to the bulk lightcone.}
      \label{fig:ads2plus1}
\end{figure}

The two-time perspective has the advantage that time evolution on both sides of the duality is identified. However, the two-time bulk theory has a rotated lightcone, as shown in figure 
\ref{fig:ads2plus1}. Since future and past are now related by a continuous isometry, this rotates the notion of bulk causality. Assuming one can construct semiclassical local bulk fields $\Phi(x)$ in 3D, the only robust notion of bulk causality would be to ensure that bulk commutators vanish outside the lightcone 
\begin{eqnarray}
 \qquad\qquad   [\Phi(x), \Phi(y)] = 0\qquad \ {\rm for}\ \ \left\{\begin{array}{c} {(x-y)^2>0 \ \  {\rm in\ \, dS}_{2+1}} \\[1.5mm]{\   (x-y)^2<0 \  \; {\rm in \ AdS}_{1+2}}\end{array}\right..
\end{eqnarray}
Hence, from the perspective of 3D bulk causality, the de Sitter perspective where future and past are defined as the regions inside the lightcone looks most naturally self-consistent. 
   
We believe that this tension between two choices of 3D spacetime signature is not a weakness of our proposed 1D/3D holographic dictionary but an intrinsic physical feature. With either choice of overall spacetime signature, the main result obtained in the following sections, that the 1D effective theory of SYK matches the 1D effective theory derived from pure 3D Einstein gravity, stands firm. From here on, we will adopt the 3D de Sitter gravity perspective and return to the discussion on overall spacetime signature in Sec. \ref{subsec:twotime}.

This paper is organized as follows. In Sec.~\ref{sec:soft_mode_of_DSSYK}, we review some aspects of the large $q$ SYK model. In Sec.~\ref{sec:boundary_dynamics}, we derive the DSSYK effective action from the Einstein-Hilbert action of 2+1D near-de Sitter gravity and match the Israel junction conditions with the DSSYK 1D equations of motion; we show that finite temperature states correspond to a Schwarzschild-de Sitter spacetime with a conical deficit. In Sec. \ref{sec:boundary_thermodynamics}, we compute the gravitational partition function and spectral density and show that they match those of DSSYK. In Sec. \ref{sec:towards_an_holographic_dictionary}, we show that the gravitational two-point functions are given by the square of the DSSYK two-point functions and discuss aspects of the resulting holographic dictionary.
In Appendix~\ref{app:spinor_coordinates}, we review some aspects of three-dimensional de Sitter spacetime and compute geodesic lengths between points at $\mathscr{I}^\pm$. In Appendix~\ref{app:more_on_the_action}, we detail some steps in our evaluation of the Einstein-Hilbert action.

%% file: images/near_dS2.tex
\begin{tikzpicture}[xscale=.92,yscale=.89,line cap=round,line join=round]
  \definecolor{outline}{RGB}{20,45,70}
  \definecolor{capfill}{RGB}{236,228,219}
  \definecolor{slicefill}{RGB}{150,165,210}
  \definecolor{greenline}{RGB}{60,135,55}
  \def\yt{2.25}
  \def\yb{-2.25}
  \def\rx{2.3}
  \def\ry{0.48}
  \def\hx{1.20}
  \def\hy{2.75}
  \def\ang{55}
  \coordinate (A) at (-1.82,\yt-0.28);   % top-left
  \coordinate (B) at ( 1.82,\yt+0.28);   % top-right
  \coordinate (C) at (-1.62,\yb-0.26);   % bottom-left
  \coordinate (D) at ( 1.98,\yb+0.22);   % bottom-right
  \fill[capfill,opacity=.78] (0,\yt) ellipse[x radius=\rx,y radius=\ry];
  \fill[capfill,opacity=.78] (0,\yb) ellipse[x radius=\rx,y radius=\ry];
  \fill[slicefill,opacity=.35]
    (A)
      .. controls (-1.12,1.25) and (-1.02,-1.05) ..
    (C)
      -- (D)
      .. controls (1.08,-1.00) and (1.02,1.30) ..
    (B)
      -- cycle;
  \draw[outline,thick] (0,\yt) ellipse[x radius=\rx,y radius=\ry];
  \draw[outline,thick] (0,\yb) ellipse[x radius=\rx,y radius=\ry];

  \draw[greenline,very thick]
    (A) .. controls (-1.12,1.25) and (-1.02,-1.05) .. (C);
  \draw[greenline,very thick]
    (B) .. controls (1.02,1.30) and (1.08,-1.00) .. (D);
  \draw[red,very thick]
    (A) .. controls (-0.55,\yt-0.08) and (0.85,\yt-0.05) .. (B);
  \draw[red,very thick,opacity=.78]
    (C) .. controls (-0.45,\yb+0.05) and (0.95,\yb+0.12) .. (D);
    
  \fill[red] (A) circle (1.7pt);
  \fill[red] (B) circle (1.7pt);
  \fill[red] (C) circle (1.7pt);
  \fill[red] (D) circle (1.7pt);

\path[draw=black,  thick] (-2.27,-2.2) arc (-25:25:5.15cm);
\path[draw=black,  thick] (2.27,2.2) arc (155:205:5.15cm);

  \node[text=blue!70!black] at (0.08,0.22) {\Large $\Sigma$};
  \node[red] at (-0.6,\yt+0.15) { $\mathcal{C}^{+}$};
  \node[red] at (0.65,\yb-0.14) {$\mathcal{C}^{-}$};
  \node[right] at (1.8,0.10) {\Large $\mathrm{dS}_{2+1}$};
    \node[above] at (0.2,2.7) {\large $\mathscr{I}^+$};
    \node[below] at (-0.2,-2.65) {\large $\mathscr{I}^-$};
\end{tikzpicture}

%% file: images/action.tex
\begin{tikzpicture}[xscale=1.3, yscale=1.2];
%\path[draw= cyan, line width=0.01mm, snake it] (2,1.5) arc (40:-40:2.2cm);
%\path[draw=blue,very thick,<-] (5.32,.06) arc (179:181:1cm);
\draw[thick,dashed,gray] (2.25,-2) -- (4,-0.25);
\draw[thick,dashed,gray] (5.75,-2) -- (4,-0.25);
\draw[thick,dashed,gray] (2.25,2) -- (4,0.25);
\draw[thick,dashed,gray] (5.75,2) -- (4,0.25);
\draw[lightgray,fill=lightgray,opacity=.75] (3.99,-2) -- (3.99,2) -- (4.01,2) -- (4.01,-2)
-- cycle;
\draw[blue] (3.99,-2) -- (3.99,2);
\draw[blue] (4.01,-2) -- (4.01,2);
\draw[lightgray,fill=lightgray,opacity=.5] (2.25,1.65) -- (3.8,1.95) --(4,2) -- (2.25,2) -- cycle;
\draw[lightgray,fill=lightgray,opacity=.5] (2.25,-1.65)-- (3.8,-1.95) -- (4,-2) -- (2.25,-2) -- cycle;
\draw[lightgray,fill=lightgray,opacity=.5] (5.75,1.65) -- (4.2,1.95) -- (4,2) -- (5.75,2) -- cycle;
\draw[lightgray,fill=lightgray,opacity=.5] (5.75,-1.65) -- (4.2,-1.95) -- (4,-2) -- (5.75,-2) -- cycle;
%\filldraw[\darkblue] (6,-1.55) circle (2pt);
%\draw[thick, \darkblue] (6,1.5) circle (2pt);
\draw (3.82,0) node {\rotatebox{0}{\textcolor{blue}{
$\Sigma$}}};
%\draw (4.34,0) node {\rotatebox{0}{\textcolor{blue}{
%$\Sigma^+$}}};
\draw (2.75,0) node {\rotatebox{0}{\textcolor{purple}{\large 
${\cal V}$}}};
\draw (5.25,0) node {\rotatebox{0}{\textcolor{purple}{\large 
${\cal V}^*$}}};
\draw[red] (4,-2.3) node {\rotatebox{0}{{\large
$\mathcal{C}^-$}}};
\draw[red] (4,2.3) node {\rotatebox{0}{{\large
$\mathcal{C}^+$}}};
\draw[color=gray]  (3,-2.3) node {\rotatebox{0}{{
$\mathscr{I}^-$}}};
\draw[color=gray]  (3,2.3) node {\rotatebox{0}{{
$\mathscr{I}^+$}}};
\draw[color=gray]  (5,-2.3) node {\rotatebox{0}{{
$\mathscr{I}^-$}}};
\draw[color=gray] (5,2.3) node {\rotatebox{0}{{
$\mathscr{I}^+$}}};
\draw[gray, thick] (2.25,-2) -- (5.75,-2);
\draw[gray, thick] (2.25,2) -- (5.75,2);
\draw[thick,gray] (5.75,-2) -- (5.75,2);
\draw[thick,gray] (2.25,-2) -- (2.25,2);
\path[draw=black,thick, fill=lightgray,opacity=.5] (2.25,1.65) arc (270:292:4.65cm);
\path[draw=black, fill=lightgray,opacity=.5,thick] (5.75,1.65) arc (270:248:4.65cm);
\path[draw=black, fill=lightgray,opacity=.5, thick] (2.25,-1.65) arc (90:68:4.65cm);
\path[draw=black, fill=lightgray,opacity=.5, thick] (5.75,-1.65) arc (90:112:4.65cm);
\draw[thick, fill=red] (4,2) circle (1.5pt);
\draw[thick, fill=red] (4,-2) circle (1.5pt);
%\draw (2.77,1.86) node {\rotatebox{0}{{\scriptsize $k=-1$}}};
%\draw (2.77,-1.86) node {\rotatebox{0}{{\scriptsize $k=-1$}}};
%\draw (5.23,1.86) node {\rotatebox{0}{{\scriptsize $k=-1$}}};
%\draw (5.23,-1.86) node {\rotatebox{0}{{\scriptsize $k=-1$}}};
\draw (1.7,0) node {\rotatebox{0}{{\Large dS$_3$}}};
\end{tikzpicture}

%% file: sections/soft_mode_of_DSSYK.tex
In this section, we summarize the derivation of the effective action of the soft mode $\psi$ of the large $q$ limit of SYK. 
A more detailed discussion can be found in \cite{Lensky_2021}.

The SYK system \cite{Sachdev_1993,kitaev2015simple,Maldacena:2016hyu} is the solvable model of quantum chaotic dynamics of  $N$ Majorana oscillators $\{\chi_i,\chi_j\} = 2\delta_{ij}$ interacting via the $q$-th order Hamiltonian 
%\tm{Here and below, I followed the normalization of \cite{Lensky_2021}, up to a sign.}
\begin{eqnarray}
    H=\frac{i^{q/2}}{q!} \sum_{i_1,\dots, i_q} J_{i_1\ldots i_q} \;\chi_{i_1}\dots \chi_{i_q}
    %\qquad \{\chi_{i}, \chi_j\} = \delta_{ij};
\end{eqnarray}
with gaussian random couplings $J_{i_1\ldots i_q}$.
%\begin{equation}
%    \overline{J_{i_1\dots i_p}}=0,\qquad \overline{J_{i_1\dots i_p}J_{i'_1\dots i'_p}} = \mathcal{J}^2\frac{2^{p-1}}{p}\frac{(p-1)!}{N^{p-1}}\delta_{i_1i'_1}\dots\delta_{i_pi'_p}.
%\label{eq:DSSYK_Hamiltonian_Majorana}
%\end{equation}
We henceforth choose units such that $\mathcal{J}=1/2$\footnote{Here $\mathcal{J}$ determines the mean of the random couplings, $\overline{J_{i_1\dots i_p}J_{i'_1\dots i'_p}} = \mathcal{J}^2\frac{2^{p-1}}{p}\frac{(p-1)!}{N^{p-1}}\delta_{i_1i'_1}\dots\delta_{i_pi'_p}$.}.  The double-scaled model (DSSYK) is obtained by taking the limit $N\to \infty$, $q\to\infty$ with $\lambda\equiv 2q^2/N$ fixed.
In this limit, the model can be exactly solved. The parameter $\lambda$ controls the quantum fluctuations, or on the holographic dual side, the gravitational coupling. In this paper, we will be interested in the non-linear semiclassical regime where $\lambda\to 0$ but in which energy and temperature can be of order $1/\lambda$. This semiclassical high energy theory \cite{Lensky_2021, Lin_2022,goel2023semiclassicalgeometrydoublescaledsyk} is a non-linear modification of the more familiar Schwarzian low energy theory \cite{kitaev2015simple,Maldacena:2016hyu}.

\begin{figure}[t]
  \centering

  % ---------- Row 1 ----------
  ~~~~~~~~~~~~~~~\subfloat{\includegraphics[width=0.32\textwidth]{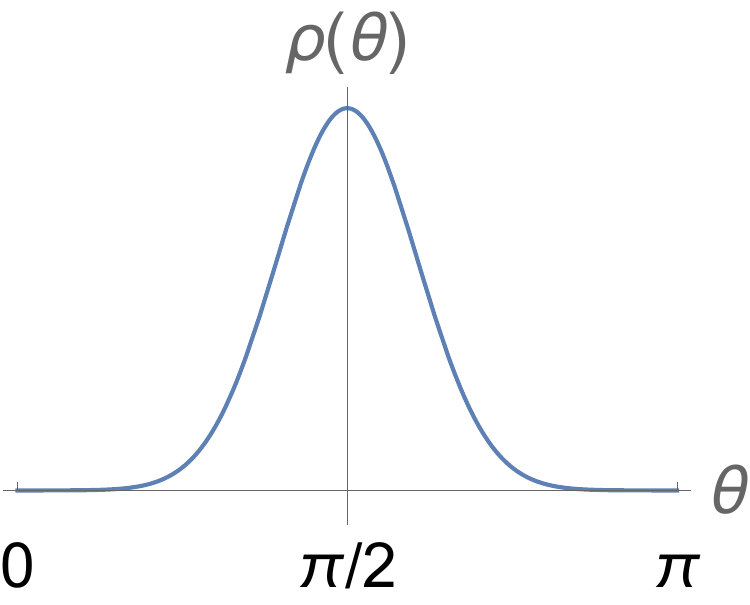}}\hfill
    \subfloat{\includegraphics[width=0.32\textwidth]{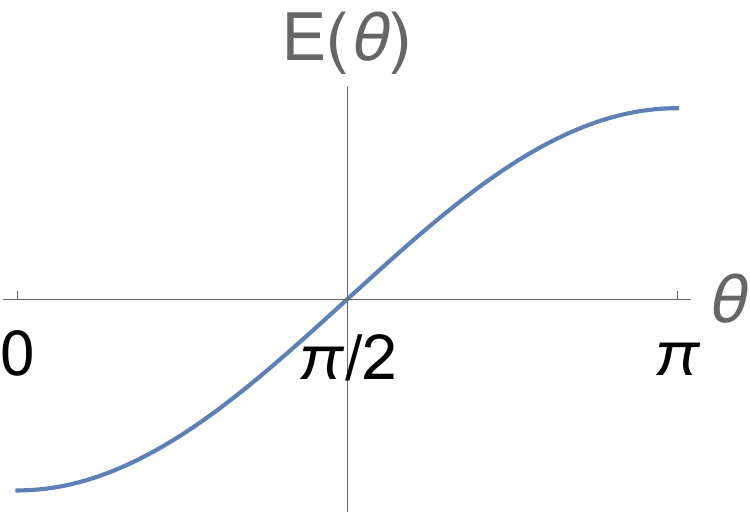}}~~~~~~~~~~~~~

  \caption{The DSSYK spectrum and energy as a function of the variable $\theta$. We see that $\theta\in[0, \pi/2]$, $v\in[0, 1]$, corresponds to positive temperature ($\rho$ increases as $E$ increases), while $\theta\in[\pi/2, \pi]$, $v\in[-1, 0]$, corresponds to positive temperature ($\rho$ decreases as $E$ decreases).}
  \label{fig:spectrum}
  \vspace{-.5mm}
\end{figure}

In the semiclassical limit, fixed energy and fixed temperature expectation values are equivalent. The energy and temperature of DSSYK are parametrized by a single dimensionless parameter $v$ via \cite{Maldacena:2016hyu}
\begin{equation}
\label{eq:energy-temperature}
    E(\theta)=- \frac 2 \lambda \cos \theta , \qquad \beta = \frac{\pi v}{\cos\frac{\pi v}{2}}, \qquad \theta = \frac{\pi}{2}(1-v). 
    %, \qquad  
\end{equation}
%\rho(\theta)d\theta = (e^{\pm2i\theta}, q)_{\infty}d\theta.
The DSSYK spectral density is exactly known \cite{Berkooz:2018jqr} and plotted in figure~\ref{fig:spectrum}. 
%The relation between temperature and energy is given by \cite{goel2023semiclassicalgeometrydoublescaledsyk} \begin{equation}    \qquad \theta = \frac{\pi}{2}(1+v). \label{eq:temperature_energy} \end{equation} 
The parameter $v$ runs over the range $v\in[-1, 1]$.
The temperature is positive for $v\in[0, 1]$, which corresponds to $\beta\in[0, \infty]$, and negative over the range $v\in[-1, 0]$, where the entropy decreases with energy.

We are interested in fixed temperature expectation values of the operator $g(u_1,u_2)$ defined by
\begin{equation}
    \frac{1}{2}\text{sgn}(u_{12})e^{g(u_1, u_2)/q}=\frac{1}{N}\sum_{i=1}^N\chi_i(u_1)\chi_i(u_2).
\end{equation}
Here $u$ denotes real time. From the Majorana algebra we deduce that $ g(u_1, u_2)=g(u_2, u_1)$, and $ g(u, u) = 0$. %The DSSYK action can be written as a Liouville action in terms of the field $g$ \cite{Maldacena:2016hyu,Lin:2023trc} 
 %\begin{equation}
   %  S= 
   %  \frac{1}{8\lambda}\int du_1 du_2 \left( \partial_1 g(u_1, u_2)\partial_2 g(u_1, u_2) - e^{g(u_1, u_2)}\right). 
% \end{equation}
The equation of motion for $g$ reduces to the Liouville equation in the double-scaled limit \cite{Maldacena:2016hyu, goel2023semiclassicalgeometrydoublescaledsyk, Lin_2023}; we will review its solution in the next subsection, after introducing the MQ coupling.

\vspace{-2mm}

\subsection{Derivation of the 1D effective action}\label{sec:MQ_coupling}

\vspace{-2mm}

In this subsection, we review some results of \cite{Lensky_2021}. Following \cite{maldacena2018eternaltraversablewormhole}, we consider two coupled copies of the SYK system, which we call the left and right models, with Hamiltonian
\begin{equation}
    H_{\rm tot} =H_R + (-1)^{q/2}H_L+ H_{\rm MQ}, \qquad \ \  H_{\rm MQ} =  \mbox{\Large $\frac i q$} \; \mu(u) \sum_{i=1}^N\chi^L_i(u)\chi^R_i(u).
\end{equation}
Here, $H_{L(R)}$ and  $\chi^{L(R)}_i$ are the Hamiltonian and Majorana oscillators for the left (right) system. We refer to the  $\mu(u)$ term as the time-dependent MQ coupling. 

\begin{figure}[t]
  \begin{center}
    \resizebox{0.9\textwidth}{!}{\input{images/contour}}
        \caption{On the left, in blue, the contour used to compute expectation values in the TFD state. The horizontal and vertical directions are Euclidean and Lorentzian time, respectively. The red points denote the operator locations in $g_{RL}(u_1, u_2)$ and the wiggly red line represents the MQ  coupling $\mu(u)$. On the right, the domain of $u_1, u_2$. We can use reflection symmetry to restrict to the shaded region $u_1\geq u_2$. The $\mu(u)$ coupling is inserted at the boundary of the domain. When we extend beyond this domain, the normal derivative to the boundary changes sign.}
    \label{fig:contour_liouville}
    \vspace{-2mm}
  \end{center}
\end{figure}
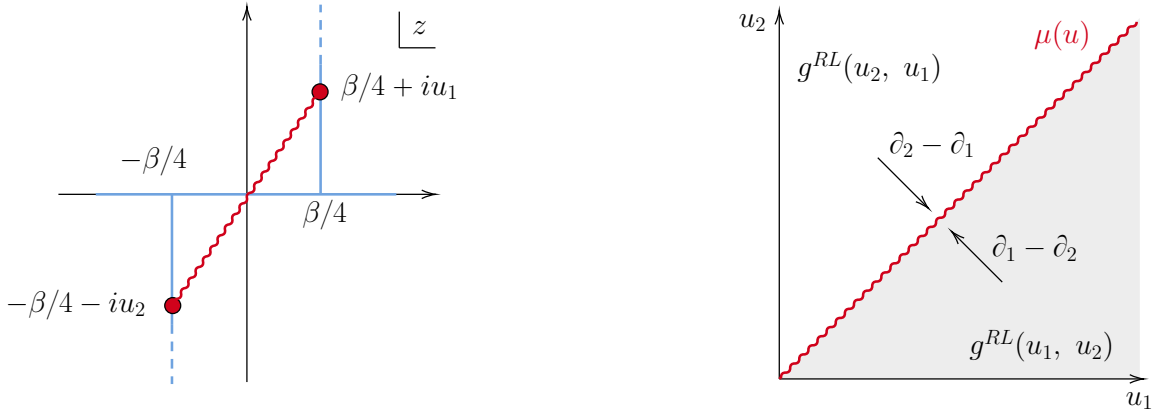

We can now consider two types of operators,
\begin{equation}
\begin{split}
    \mbox{\Large $\frac 1 2$}\spc \text{sgn}(u_{12})\, e^{g_{RR}(u_1, u_2)/q}&=\frac{1}{N}\sum_{i=1}^N\chi^R_i(u_1)\chi^R_i(u_2),\\
    \mbox{\Large $\frac 1 2$}\spc \text{sgn}(u_{12})\, e^{ g_{RL}(u_1, u_2)/q}&=\frac{1}{N}\sum_{i=1}^N\chi^R_i(u_1)\chi^L_i(u_2).
\end{split}
\end{equation}
To study the coupled system in the TFD state, it is convenient to introduce a field $g(z_1, z_2)$ such that
\begin{equation}
    g_{RR}(u_1, u_2) = g(\beta/4 + iu_1, \beta/4 + iu_2),\qquad g_{RL}(u_1, u_2) = g(\beta/4 + iu_1, -\beta/4 - iu_2),
\end{equation}
and consider $z_{1,2}$ on the contour depicted in figure~\ref{fig:contour_liouville}. The horizontal part of the contour corresponds to the Euclidean time evolution that prepares the state, while the vertical part corresponds to Lorentzian time evolution. We also extend $\mu$ to this contour, and take it to be non-zero only on the Lorentzian part, with $\mu(z)=\mu(-z)$. 

  The MQ coupling acts as a source term in the Liouville equation of motion, and %\tm{normalization}
 %\begin{eqnarray}
  %   S_{\rm tot}\is
  %   \frac{1}{8\lambda}\int dz_1dz_2 \left( \partial_1 g\partial_2 g- e^{g(z_1,z_2)}\right) \; +\; S_{\rm MQ}
   %  \\[2mm] 
   %  S_{\rm MQ} \is \frac{i}{\lambda}\int\! du \,\mu (u)g\bigl(\beta/4+iu , -\beta/4 -iu).
% \end{eqnarray}
the semiclassical expectation value of $g$ now satisfies the equation
\begin{eqnarray}
    \partial_1\partial_2g(z_1, z_2)+\frac{1}{2}e^{g(z_1, z_2)}=-\mu(u)\delta(z_1+z_2).
    \label{eq:liouville_delta}
\end{eqnarray}
Recall that $\mu(z)$ only has support for $z = \pm(\beta/4+iu)$. We want to solve this in the Lorentzian part of the domain,
\begin{equation}
    z_1=\beta/4+iu_1\qquad z_2 = \pm(\beta/4+iu_2),\qquad u_1, u_2\geq 0.
\end{equation}
Given the symmetries of $g$, we can further restrict to $u_1\geq u_2$\footnote{There is a small subtlety when $z_1=\beta/4+iu_1$, $z_2=-\beta/4-iu_2$, and $g(z_1, z_2) = g_{RL}(u_1, u_2)$. Indeed, we need to use $g_{RL}(u_1, u_2)= g_{LR}(u_1, u_2)$ to extend to $u_2>u_1$ by equating
    $g_{RL}(u_1, u_2) = g_{LR}(u_2, u_1)=g_{RL}(u_2, u_1)$.}. We see that the delta function term is at the boundary of this domain; away from this boundary, $g$ is a solution to  the Liouville  equation.

A general solution to the Liouville equation\footnote{In general, we can have different functions for $u_1$ and $u_2$. %, an for $g_{RR}$ and $g_{RL}$, i.e.
%\begin{equation}
 %   g_{RR}(u_1, u_2) = {\rm log} \frac{\psi_1'(u_1)\psi_2'(u_2)}{\sin^2\frac{\psi_1(u_1)-\psi_2(u_2)}{2}}-i \pi, \qquad  g_{LR}(u_1, u_2) = {\rm log} \frac{\tilde\psi_1'(u_1)\tilde\psi_2'(u_2)}{\cos^2\frac{\tilde\psi_1(u_1)-\tilde\psi_2(u_2)}{2}}.
%\end{equation}
The authors of \cite{Lensky_2021} show that we can take $\psi_1=\tilde\psi_1=\psi_2^*=\tilde\psi_2^*$.} 
can be written in terms of a single function $\psi(u)\in\mathbb{C}$,
\begin{equation}
\begin{split}
    g_{RR}(u_1, u_2) =& \log \biggl(\frac{\psi'(u_1)\psi'^{*}(u_2)}{\sin^2((\psi(u_1)-\psi^*(u_2))/2)}\biggr) -i \pi, \\[2mm]  
    g_{RL}(u_1, u_2) =& \log \biggl(\frac{\psi'(u_1)\psi'^{*}(u_2)}{\cos^2((\psi(u_1)-\psi^*(u_2))/2)}\biggr),
 \end{split}
    \label{eq:g_psi}
\end{equation}
where $\psi^*$ is the complex conjugate of $\psi$, and a prime denotes differentiation with respect to $u$. The explicit form of the complex function $\psi$ is obtained by imposing the correct boundary conditions.  First, we have $g_{RR}(u, u)=0$, which gives
\begin{eqnarray}
\label{eq:lensky-qi-zero}
    \boxed{\ |\psi'| = \sinh |\text{Im}\psi|.\scriptsize{{}^{\strut}_{\strut}}\ }
\end{eqnarray}

Second, we have to impose the delta function boundary condition. To this end, we can study the equation near $z_1=-z_2$, with $z_1=\beta/4+iu_1$, $z_2=-\beta/4-iu_2$, $u_1\simeq u_2$; here $g(z_1, z_2)= g_{RL}(u_1, u_2)$, and we must satisfy equation \eqref{eq:liouville_delta}.
%\begin{eqnarray}
 %   \partial_1\partial_2g_{RL}(u_1, u_2)+\frac{1}{2}e^{g_{RL}(u_1, u_2)}=-i\mu(u)\, \delta(u_1\!+\!u_2).
%\label{eq:liouville_delta}
%\end{eqnarray}
The field $g$ is continuous in the domain $u_1>u_2$, and the delta function arises from extending beyond this domain, see figure~\ref{fig:contour_liouville}. In particular, the discontinuity comes from the derivative normal to the boundary. Writing $
    \partial_1\partial_2 =  \partial_+^2 - \partial_-^2$ with $\partial_\pm  \equiv \frac 1 2 
  ({\partial_1\pm\partial_2})$ and using that we have extended the  domain by 
\begin{equation}
    g_{RL}(u_1, u_2)\to \Theta(u_{12})g_{RL}(u_1, u_2)+\Theta(u_{21})g_{RL}(u_2, u_1), 
\end{equation}
we can replace $
   \partial_-% \left( \frac{\partial_1-\partial_2}{2} \right)
   \to \text{sgn}(u_{12})\partial_-$
   %\left( \frac{\partial_1-\partial_2}{2} \right)
 and $\partial_-^2 %\left( \frac{\partial_1-\partial_2}{2} \right)^2
   \to\delta(u_{12}) \partial_- + \ldots $
   %\left( \frac{\partial_1-\partial_2}{2} \right)
We then compute near $u_1\simeq u_2$
\begin{equation}
   \partial_1\partial_2g_{RL}(u_1, u_2) = i\bigl( p'+ %|\psi'| 
   {\rm Re} \,{\psi'}\, \tanh(\text{Im}\psi)
   %\cos p
   \bigr)\, \delta(u_1\!+\!u_2) + \ldots,  
\end{equation}
where we introduced the variable $p$ via $\psi'=|\psi'|e^{ip}$. Then, from equation \eqref{eq:liouville_delta}, we find that the boundary condition takes the form\footnote{Equation (2.31) of \cite{Lensky_2021} reads
$p'-|\psi'|\tanh |\text{Im}\,\psi| \cos p = -\mu.$ This matches our equation \eqref{eq:lensky-qi-one} when $\text{Im}\psi<0$. For all the solutions we consider below, we have $\text{Im}\psi<0$.}
\begin{eqnarray}
\label{eq:lensky-qi-one}
   \boxed{ \ 
   p'+|\psi'|\tanh (\text{Im}\,\psi) \cos p = -\mu\small{{}^{\strut}_{\strut}}.\ }
\end{eqnarray}
Combined, the equations \eqref{eq:lensky-qi-zero} and \eqref{eq:lensky-qi-one} specify the equations of motion of a complex boundary trajectory $\psi(u)$. Via equations \eqref{eq:g_psi}, this trajectory determines the semiclassical two-point functions of the SYK model for arbitrary time-dependent MQ coupling $\mu(u)$.

%Let us sum up what we have reviewed so far. The semiclassical expectation value of the field $g$ in the presence of a Maldacena-Qi coupling can be expressed in terms of a function $\psi$ as in Eq.~\ref{eq:g_psi}. This function is determined by the following equations ($\psi'=|\psi'|e^{ip}$)
%\begin{equation}
 %   |\psi'|=\sinh |\text{Im}\psi|\qquad  p'+|\psi'|\tanh (\text{Im}\,\psi) \cos p = -\mu. \label{eq:LQ_equation_of_motion}
%\end{equation}
%In the thermofield double state, the solution to these equations must satisfy the initial condition
%\begin{equation}
 %   \psi'(0)= \left|\cot \frac{\pi v}{2} \right|.
%\end{equation}
%In the next section, we will review the solution to these equations for a constant $\mu$ at low temperature, and for $\mu =0$ at generic temperature.

We conclude this section by summarizing the derivation of equation \eqref{action+hamiltonian}. Let us introduce two new variables $p$ and $\phi$ as follows \cite{Lensky_2021}
\begin{eqnarray}
\label{eq:phi_definition}  
\boxed{\  \psi' = \frac{e^{ip}}{\sqrt{e^{2\phi}-1}}\sstrut\ }
\end{eqnarray}
We will interpret $\phi$ and $p$ as canonical conjugate variables with classical Poisson brackets $\{\phi, p\}=1$. Combining the above definition with equation~\eqref{eq:lensky-qi-zero}, we have $\phi=\log\coth |\text{Im}\psi|$. The equations \eqref{eq:lensky-qi-zero} and \eqref{eq:lensky-qi-one} for $\psi(u)$ then turn into the classical Hamilton equations derived from the Hamiltonian
\begin{eqnarray}
    \boxed{\ H=-\sqrt{1-e^{-2\phi}}\cos p-\mu\phi\sstrut\ }
    \label{eq:DSSYK_hamiltonian_coupling}
\end{eqnarray} 
This remarkably compact formula takes a familiar form \cite{Lin_2022}: the first term in 
\eqref{eq:DSSYK_hamiltonian_coupling} can be recognized as the sum of the chord creation and chord annihilation operators $a_\pm = \sqrt{1-e^{-2\phi}}e^{\pm ip}$. 
%\begin{equation}
%    H=\frac{1}{\lambda}\sqrt{1-e^{-2\phi}}\cos p,
%%\end{equation}
In the chord context, the variable $\phi$ is identified with the chord number $n$ via $\phi=\lambda n$ and $p$ is its canonically conjugate momentum. The correspondence with the chord rules suggests that when $\lambda = 2q^2/N$ is finite, we should replace the Poisson bracket between $p$ and $\phi$ by the commutator relation
\begin{equation}
\boxed{\ [\phi,p] = \frac{i \lambda}{2} \raisebox{-.5mm}{\Large$\strut$}. \;}
\end{equation}

In the following section, we will give a dual gravity interpretation to the classical trajectories $\psi(u)$ that follow from the Hamiltonian \eqref{eq:DSSYK_hamiltonian_coupling}. For illustration, let us write three special solutions for the complex boundary trajectory $\psi(u)$. 
\vspace{-1 mm}

\paragraph{Thermal solution.}
The solution in the thermofield double state at inverse temperature $\beta$ is %\footnote{This is the expression for $u_1>u_2$; we can extend in the full domain using the symmetry $g(u_1, u_2)=g(u_2, u_1)$.} 
\begin{equation}
    g(u_1, u_2) = 2\log\;\frac{\mathsf{c}}{\cosh\left(\mathsf{c}\, \frac{u_{1}-u_2}{2}+\frac{i \pi v}{2}\right)}, \qquad \quad  \mathsf{c} \equiv \cos \frac {\pi v} 2 .
    \label{eq:thermal_g} 
\end{equation}
The definition of the parameter $v$, and the relation with fixed energy expectation values are given in equation~\eqref{eq:energy-temperature}.
In terms of the canonical variables $\phi$ and $p$, the TFD initial conditions are $\phi(0)=-\log\cos\frac{\pi v}{2},$ and $p(0)=0$.
Using the parameterization \eqref{eq:g_psi}, we have
\begin{equation}
    \psi(u) = 2\arctan\tanh\left(\mathsf{c}\, \frac{u}{2} -\frac{i\pi}{4}(1-v) \right).
    \label{eq:thermal_solution_psi}
\end{equation}
Note that the decay of the thermal correlator is set by the fake temperature
\begin{equation}\label{eq:fake_temperature}
    \beta_{\rm fake} = \frac{2\pi}{\mathsf{c}} = \frac{2\pi}{\cos\frac{\pi v}{2}}.
\end{equation}

\paragraph{Delta function source.}
We can obtain the solution for a delta function source $\mu(u) = \hat{\mu}\,\delta(u)$ by patching together two TFD solutions 
\begin{eqnarray}
\label{eq:delta_function_mu}
   & & \quad  \psi(u) = \begin{cases}
        & 2\arctan\tanh\left(\mathsf{c}\, \frac{(u-u_0)}{2} -i\frac{\pi}{4}(1-v) \right) + a\quad \text{for } u<0,\\
        &2\arctan\tanh\left(\mathsf{c}\, \frac{(u+u_0)}{2} -i\frac{\pi}{4}(1-v) \right)-a\quad \text{for } u>0,\\
    \end{cases}\\[3mm]
    a\is \text{Re}\Bigl\{\arctan\tanh\left(\mathsf{c}\, \frac{u_0}{2} -i\frac{\pi}{4}(1-v) \right)\Bigr\}-\text{Re}\Bigl\{\arctan\tanh\left(-\,\mathsf{c}\,\frac{u_0}{2} -i\frac{\pi}{4}(1-v) \right)\Bigr\}.
\end{eqnarray}
Note that, to preserve the condition \eqref{eq:lensky-qi-zero}, we can only add a real constant to the TFD solution. Then, it would seem that the imaginary part of the equation above is discontinuous. However, the TFD solution $\psi$ has the property $\text{Im}\,\psi(u)=\text{Im}\,\psi(-u)$, and equation~\eqref{eq:delta_function_mu} is continuous.
The value of $\hat{\mu}$ is given by the jump in $p$ at $u=0$. This gives a relation between $\hat{\mu}$, $v$, and $u_0$, which can be expressed as
\begin{equation}
    e^{-i\hat \mu} = \frac{\sin(\frac{\pi v}{2} + i\mathsf{c}u_0)}{\sin(\frac{\pi v}{2}-i\mathsf{c} u_0)}.
\end{equation}

\vspace{1mm}

\paragraph{Constant $\mu$ solution.} Finally, we consider a (fined tuned) solution for constant $\mu$. In this case, the dynamics of $\phi$ and $p$ preserves the energy \eqref{eq:DSSYK_hamiltonian_coupling}. Furthermore, the Hamiltonian admits a minimum for $\mu>0$; on this minimum energy solution, $\phi$ is constant and $\psi(u)$ in linear in $u$ \cite{Lensky_2021}
\begin{eqnarray}
    \phi(u)\is \phi_{\mu} \equiv -\frac{1}{2}\ln\left(\frac\mu 2 \Bigl(\sqrt{{\mu^2}+4}-\mu \Bigr)\right),\\[2mm]
    \psi(u) \is \frac{1}{\sqrt{e^{2\phi_\mu}-1}}(u-u_0) -i\tanh^{-1}e^{-\phi_\mu}.
\label{eq:constant_mu_solution_psi}
\end{eqnarray}

\noindent
We will return to the above special solutions in the next section, where we will interpret them in terms of the holographic dual gravity theory.

\bigskip

%% file: images/contour.tex
\tikzset{every picture/.style={line width=0.75pt}} %set default line width to 0.75pt        

\begin{tikzpicture}[x=0.75pt,y=0.75pt,yscale=-1,xscale=1]
%uncomment if require: \path (0,344); %set diagram left start at 0, and has height of 344

%Shape: Right Triangle [id:dp6320403355417771] 
\draw  [draw opacity=0][fill={rgb, 255:red, 237; green, 237; blue, 237 }  ,fill opacity=1 ] (907.38,13.98) -- (627.78,294.1) -- (907.38,294.1) -- cycle ;
%Straight Lines [id:da2016102189890162] 
\draw    (68.74,150.83) -- (359.43,150.83) ;
\draw [shift={(361.43,150.83)}, rotate = 180] [color={rgb, 255:red, 0; green, 0; blue, 0 }  ][line width=0.75]    (10.93,-3.29) .. controls (6.95,-1.4) and (3.31,-0.3) .. (0,0) .. controls (3.31,0.3) and (6.95,1.4) .. (10.93,3.29)   ;
%Straight Lines [id:da19618250770002288] 
\draw    (214.91,296.53) -- (214.91,6.97) ;
\draw [shift={(214.91,4.97)}, rotate = 90] [color={rgb, 255:red, 0; green, 0; blue, 0 }  ][line width=0.75]    (10.93,-3.29) .. controls (6.95,-1.4) and (3.31,-0.3) .. (0,0) .. controls (3.31,0.3) and (6.95,1.4) .. (10.93,3.29)   ;
%Straight Lines [id:da1386904723711747] 
\draw [color={rgb, 255:red, 111; green, 163; blue, 224 }  ,draw opacity=1 ][line width=1.5]    (156.79,150.83) -- (156.79,251.57) ;
%Straight Lines [id:da2707646944950295] 
\draw [color={rgb, 255:red, 111; green, 163; blue, 224 }  ,draw opacity=1 ][line width=1.5]  [dash pattern={on 5.63pt off 4.5pt}]  (156.79,251.57) -- (156.79,298.07) ;
%Straight Lines [id:da1023697159736443] 
\draw [color={rgb, 255:red, 111; green, 163; blue, 224 }  ,draw opacity=1 ][line width=1.5]    (272.06,150.83) -- (272.06,50.1) ;
%Straight Lines [id:da12606013162854846] 
\draw [color={rgb, 255:red, 111; green, 163; blue, 224 }  ,draw opacity=1 ][line width=1.5]  [dash pattern={on 5.63pt off 4.5pt}]  (272.06,50.1) -- (272.06,3.6) ;

%Straight Lines [id:da4369542526220642] 
\draw [color={rgb, 255:red, 111; green, 163; blue, 224 }  ,draw opacity=1 ][line width=1.5]    (97.7,150.83) -- (330.82,150.83) ;
%Shape: Ellipse [id:dp585552881640491] 
\draw  [fill={rgb, 255:red, 208; green, 2; blue, 27 }  ,fill opacity=1 ] (277.84,71.29) .. controls (277.84,67.88) and (275.07,65.11) .. (271.65,65.11) .. controls (268.23,65.11) and (265.47,67.88) .. (265.47,71.29) .. controls (265.47,74.71) and (268.23,77.48) .. (271.65,77.48) .. controls (275.07,77.48) and (277.84,74.71) .. (277.84,71.29) -- cycle ;
%Straight Lines [id:da21344498392332723] 
\draw    (333.08,9.01) -- (333.08,37.1) ;
%Straight Lines [id:da1055208017878333] 
\draw    (627.78,294.1) -- (908.55,294.1) ;
\draw [shift={(910.55,294.1)}, rotate = 180] [color={rgb, 255:red, 0; green, 0; blue, 0 }  ][line width=0.75]    (10.93,-3.29) .. controls (6.95,-1.4) and (3.31,-0.3) .. (0,0) .. controls (3.31,0.3) and (6.95,1.4) .. (10.93,3.29)   ;
%Straight Lines [id:da33652765554325237] 
\draw    (627.78,294.1) -- (627.78,9.99) ;
\draw [shift={(627.78,7.99)}, rotate = 90] [color={rgb, 255:red, 0; green, 0; blue, 0 }  ][line width=0.75]    (10.93,-3.29) .. controls (6.95,-1.4) and (3.31,-0.3) .. (0,0) .. controls (3.31,0.3) and (6.95,1.4) .. (10.93,3.29)   ;
%Straight Lines [id:da12294049323555889] 
\draw [color={rgb, 255:red, 208; green, 2; blue, 27 }  ,draw opacity=1 ][line width=1.5]    (904.93,16.96) .. controls (904.93,19.31) and (903.75,20.49) .. (901.39,20.49) .. controls (899.03,20.49) and (897.85,21.67) .. (897.86,24.03) .. controls (897.86,26.39) and (896.68,27.57) .. (894.32,27.56) .. controls (891.96,27.56) and (890.78,28.74) .. (890.78,31.1) .. controls (890.78,33.45) and (889.6,34.63) .. (887.25,34.63) .. controls (884.89,34.63) and (883.71,35.81) .. (883.71,38.17) .. controls (883.71,40.52) and (882.53,41.7) .. (880.18,41.7) .. controls (877.82,41.7) and (876.64,42.88) .. (876.64,45.24) .. controls (876.65,47.6) and (875.47,48.78) .. (873.11,48.78) .. controls (870.75,48.77) and (869.57,49.95) .. (869.57,52.31) .. controls (869.58,54.67) and (868.4,55.85) .. (866.04,55.85) .. controls (863.68,55.84) and (862.5,57.02) .. (862.5,59.38) .. controls (862.5,61.74) and (861.32,62.92) .. (858.96,62.92) .. controls (856.61,62.92) and (855.43,64.1) .. (855.43,66.45) .. controls (855.43,68.81) and (854.25,69.99) .. (851.89,69.99) .. controls (849.54,69.99) and (848.36,71.17) .. (848.36,73.52) .. controls (848.36,75.88) and (847.18,77.06) .. (844.82,77.06) .. controls (842.47,77.06) and (841.29,78.24) .. (841.29,80.59) .. controls (841.29,82.95) and (840.11,84.13) .. (837.75,84.13) .. controls (835.39,84.13) and (834.21,85.31) .. (834.22,87.67) .. controls (834.22,90.03) and (833.04,91.21) .. (830.68,91.2) .. controls (828.32,91.2) and (827.14,92.38) .. (827.15,94.74) .. controls (827.15,97.1) and (825.97,98.28) .. (823.61,98.27) .. controls (821.25,98.27) and (820.07,99.45) .. (820.07,101.81) .. controls (820.07,104.16) and (818.89,105.34) .. (816.54,105.34) .. controls (814.18,105.34) and (813,106.52) .. (813,108.88) .. controls (813,111.23) and (811.82,112.41) .. (809.47,112.41) .. controls (807.11,112.41) and (805.93,113.59) .. (805.93,115.95) .. controls (805.94,118.31) and (804.76,119.49) .. (802.4,119.49) .. controls (800.04,119.48) and (798.86,120.66) .. (798.86,123.02) .. controls (798.87,125.38) and (797.69,126.56) .. (795.33,126.56) .. controls (792.97,126.55) and (791.79,127.73) .. (791.79,130.09) .. controls (791.79,132.45) and (790.61,133.63) .. (788.25,133.63) .. controls (785.9,133.63) and (784.72,134.81) .. (784.72,137.16) .. controls (784.72,139.52) and (783.54,140.7) .. (781.18,140.7) .. controls (778.83,140.7) and (777.65,141.88) .. (777.65,144.23) .. controls (777.65,146.59) and (776.47,147.77) .. (774.11,147.77) .. controls (771.75,147.77) and (770.57,148.95) .. (770.58,151.31) .. controls (770.58,153.67) and (769.4,154.85) .. (767.04,154.84) .. controls (764.68,154.84) and (763.5,156.02) .. (763.51,158.38) .. controls (763.51,160.74) and (762.33,161.92) .. (759.97,161.91) .. controls (757.61,161.91) and (756.43,163.09) .. (756.43,165.45) .. controls (756.43,167.8) and (755.25,168.98) .. (752.9,168.98) .. controls (750.54,168.98) and (749.36,170.16) .. (749.36,172.52) .. controls (749.36,174.87) and (748.18,176.05) .. (745.83,176.05) .. controls (743.47,176.05) and (742.29,177.23) .. (742.29,179.59) .. controls (742.3,181.95) and (741.12,183.13) .. (738.76,183.13) .. controls (736.4,183.12) and (735.22,184.3) .. (735.22,186.66) .. controls (735.23,189.02) and (734.05,190.2) .. (731.69,190.2) .. controls (729.33,190.19) and (728.15,191.37) .. (728.15,193.73) .. controls (728.15,196.09) and (726.97,197.27) .. (724.61,197.27) .. controls (722.26,197.27) and (721.08,198.45) .. (721.08,200.8) .. controls (721.08,203.16) and (719.9,204.34) .. (717.54,204.34) .. controls (715.19,204.34) and (714.01,205.52) .. (714.01,207.87) .. controls (714.01,210.23) and (712.83,211.41) .. (710.47,211.41) .. controls (708.11,211.41) and (706.93,212.59) .. (706.94,214.95) .. controls (706.94,217.31) and (705.76,218.49) .. (703.4,218.48) .. controls (701.04,218.48) and (699.86,219.66) .. (699.87,222.02) .. controls (699.87,224.38) and (698.69,225.56) .. (696.33,225.55) .. controls (693.97,225.55) and (692.79,226.73) .. (692.79,229.09) .. controls (692.79,231.44) and (691.61,232.62) .. (689.26,232.62) .. controls (686.9,232.62) and (685.72,233.8) .. (685.72,236.16) .. controls (685.72,238.51) and (684.54,239.69) .. (682.19,239.69) .. controls (679.83,239.69) and (678.65,240.87) .. (678.65,243.23) .. controls (678.66,245.59) and (677.48,246.77) .. (675.12,246.77) .. controls (672.76,246.76) and (671.58,247.94) .. (671.58,250.3) .. controls (671.59,252.66) and (670.41,253.84) .. (668.05,253.84) .. controls (665.69,253.83) and (664.51,255.01) .. (664.51,257.37) .. controls (664.51,259.73) and (663.33,260.91) .. (660.97,260.91) .. controls (658.62,260.91) and (657.44,262.09) .. (657.44,264.44) .. controls (657.44,266.8) and (656.26,267.98) .. (653.9,267.98) .. controls (651.55,267.98) and (650.37,269.16) .. (650.37,271.51) .. controls (650.37,273.87) and (649.19,275.05) .. (646.83,275.05) .. controls (644.48,275.05) and (643.3,276.23) .. (643.3,278.58) .. controls (643.3,280.94) and (642.12,282.12) .. (639.76,282.12) .. controls (637.4,282.12) and (636.22,283.3) .. (636.23,285.66) .. controls (636.23,288.02) and (635.05,289.2) .. (632.69,289.19) .. controls (630.33,289.19) and (629.15,290.37) .. (629.16,292.73) -- (627.78,294.1) -- (627.78,294.1) ;
%Shape: Boxed Line [id:dp08820223412341976] 
\draw [color={rgb, 255:red, 0; green, 0; blue, 0 }  ,draw opacity=1 ][line width=0.75]    (800.7,218.93) -- (762.52,180.75) ;
\draw [shift={(761.1,179.33)}, rotate = 45] [color={rgb, 255:red, 0; green, 0; blue, 0 }  ,draw opacity=1 ][line width=0.75]    (10.93,-3.29) .. controls (6.95,-1.4) and (3.31,-0.3) .. (0,0) .. controls (3.31,0.3) and (6.95,1.4) .. (10.93,3.29)   ;
%Shape: Ellipse [id:dp145622122225874] 
\draw  [fill={rgb, 255:red, 208; green, 2; blue, 27 }  ,fill opacity=1 ] (163.57,237.17) .. controls (163.57,233.75) and (160.8,230.98) .. (157.38,230.98) .. controls (153.97,230.98) and (151.2,233.75) .. (151.2,237.17) .. controls (151.2,240.58) and (153.97,243.35) .. (157.38,243.35) .. controls (160.8,243.35) and (163.57,240.58) .. (163.57,237.17) -- cycle ;
%Straight Lines [id:da9675293503217881] 
\draw [color={rgb, 255:red, 208; green, 2; blue, 27 }  ,draw opacity=1 ][line width=1.5]    (271.65,71.29) .. controls (272.08,73.61) and (271.13,74.98) .. (268.81,75.41) .. controls (266.49,75.84) and (265.55,77.21) .. (265.98,79.53) .. controls (266.41,81.85) and (265.46,83.22) .. (263.14,83.65) .. controls (260.82,84.07) and (259.87,85.44) .. (260.3,87.76) .. controls (260.73,90.08) and (259.79,91.45) .. (257.47,91.88) .. controls (255.15,92.31) and (254.2,93.68) .. (254.63,96) .. controls (255.06,98.32) and (254.12,99.69) .. (251.8,100.12) .. controls (249.48,100.54) and (248.53,101.91) .. (248.96,104.23) .. controls (249.39,106.55) and (248.44,107.92) .. (246.12,108.35) .. controls (243.8,108.78) and (242.86,110.15) .. (243.29,112.47) .. controls (243.72,114.79) and (242.77,116.16) .. (240.45,116.59) .. controls (238.13,117.01) and (237.18,118.38) .. (237.61,120.7) .. controls (238.04,123.02) and (237.1,124.39) .. (234.78,124.82) .. controls (232.46,125.25) and (231.51,126.62) .. (231.94,128.94) .. controls (232.37,131.26) and (231.42,132.63) .. (229.1,133.06) .. controls (226.79,133.49) and (225.84,134.86) .. (226.27,137.17) .. controls (226.7,139.49) and (225.75,140.86) .. (223.43,141.29) .. controls (221.11,141.72) and (220.16,143.09) .. (220.59,145.41) .. controls (221.02,147.73) and (220.08,149.1) .. (217.76,149.53) .. controls (215.44,149.95) and (214.49,151.32) .. (214.92,153.64) .. controls (215.35,155.96) and (214.4,157.33) .. (212.08,157.76) .. controls (209.76,158.19) and (208.82,159.56) .. (209.25,161.88) .. controls (209.68,164.2) and (208.73,165.57) .. (206.41,166) .. controls (204.09,166.42) and (203.14,167.79) .. (203.57,170.11) .. controls (204,172.43) and (203.06,173.8) .. (200.74,174.23) .. controls (198.42,174.66) and (197.47,176.03) .. (197.9,178.35) .. controls (198.33,180.67) and (197.38,182.04) .. (195.06,182.47) .. controls (192.75,182.9) and (191.8,184.27) .. (192.23,186.58) .. controls (192.66,188.9) and (191.71,190.27) .. (189.39,190.7) .. controls (187.07,191.13) and (186.12,192.5) .. (186.55,194.82) .. controls (186.98,197.14) and (186.04,198.51) .. (183.72,198.94) .. controls (181.4,199.36) and (180.45,200.73) .. (180.88,203.05) .. controls (181.31,205.37) and (180.37,206.74) .. (178.05,207.17) .. controls (175.73,207.6) and (174.78,208.97) .. (175.21,211.29) .. controls (175.64,213.61) and (174.69,214.98) .. (172.37,215.41) .. controls (170.06,215.84) and (169.11,217.21) .. (169.54,219.52) .. controls (169.97,221.84) and (169.02,223.21) .. (166.7,223.64) .. controls (164.38,224.07) and (163.43,225.44) .. (163.86,227.76) .. controls (164.29,230.08) and (163.35,231.45) .. (161.03,231.88) .. controls (158.71,232.3) and (157.76,233.67) .. (158.19,235.99) -- (157.38,237.17) -- (157.38,237.17) ;
%Shape: Boxed Line [id:dp9946413562038099] 
\draw    (361.17,37.1) -- (333.08,37.1) ;
%Shape: Boxed Line [id:dp5399572529992147] 
\draw [color={rgb, 255:red, 0; green, 0; blue, 0 }  ,draw opacity=1 ][line width=0.75]    (704.17,122.4) -- (742.36,160.59) ;
\draw [shift={(743.77,162)}, rotate = 225] [color={rgb, 255:red, 0; green, 0; blue, 0 }  ,draw opacity=1 ][line width=0.75]    (10.93,-3.29) .. controls (6.95,-1.4) and (3.31,-0.3) .. (0,0) .. controls (3.31,0.3) and (6.95,1.4) .. (10.93,3.29)   ;

% Text Node
\draw (341.66,15) node [anchor=north west][inner sep=0.75pt]  [font=\LARGE]  {$z$};
% Text Node
\draw (114.84,111.95) node [anchor=north west][inner sep=0.75pt]  [font=\LARGE]  {$-\beta /4$};
% Text Node
\draw (256.59,153.31) node [anchor=north west][inner sep=0.75pt]  [font=\LARGE]  {$\beta /4$};
% Text Node
\draw (286.29,54.82) node [anchor=north west][inner sep=0.75pt]  [font=\LARGE]  {$\beta /4+iu_{1}$};
% Text Node
\draw (26.56,223) node [anchor=north west][inner sep=0.75pt]  [font=\LARGE]  {$-\beta /4-i u_{2}$};
% Text Node
\draw (895.02,302.81) node [anchor=north west][inner sep=0.75pt]  [font=\LARGE]  {$u_{1}$};
% Text Node
\draw (595,9.18) node [anchor=north west][inner sep=0.75pt]  [font=\LARGE]  {$u_{2}$};
% Text Node
\draw (773.7,254.18) node [anchor=north west][inner sep=0.75pt]  [font=\LARGE]  {$g^{RL}( u_{1} ,\ u_{2})$};
% Text Node
\draw (640.37,38.54) node [anchor=north west][inner sep=0.75pt]  [font=\LARGE]  {$g^{RL}( u_{2} ,\ u_{1})$};
% Text Node
\draw (791,181.47) node [anchor=north west][inner sep=0.75pt]  [font=\LARGE,color={rgb, 255:red, 0; green, 0; blue, 0 }  ,opacity=1 ]  {$\partial _{1} -\partial _{2}$};
% Text Node
\draw (711.35,101.13) node [anchor=north west][inner sep=0.75pt]  [font=\LARGE,color={rgb, 255:red, 0; green, 0; blue, 0 }  ,opacity=1 ]  {$\partial _{2} -\partial _{1}$};
% Text Node
\draw (823.99,14.59) node [anchor=north west][inner sep=0.75pt]  [font=\LARGE,color={rgb, 255:red, 208; green, 2; blue, 27 }  ,opacity=1 ]  {$\mu ( u)$};

\end{tikzpicture}

%% file: sections/boundary_dynamics.tex
In this section, we develop a holographic dual interpretation of the DSSYK soft mode in terms of three-dimensional de Sitter gravity (or, upon changing the overall sign of the metric, a three-dimensional anti-de Sitter gravity with two time directions). Our dictionary generalizes the familiar near-AdS$_2$ holography map between the low energy soft mode dynamics of SYK and  JT gravity \cite{Kitaev:2017awl,maldacena2016conformalsymmetrybreakingdimensional,Jensen:2016pah,Engels_y_2016} to all energies. The 3D spacetime dynamics of interest is obtained by placing an energy-momentum distribution on a two-dimensional dS$_2$ slice $\Sigma$ within global 3D de Sitter and including the backreaction via the Israel junction conditions.  In the following, we show that the embedding of the slice $\Sigma$ is determined by a complex function $\psi(u)$ that solves the same equation of motion \eqref{eq:lensky-qi-one} as the DSSYK soft mode. We then derive a 1D effective action that determines the embedding of this surface in the 3D spacetime by explicit reduction of the Einstein-Hilbert action, and match it to the 1D effective action \eqref{action+hamiltonian}.
% %At low temperatures, our equations reduce to those of Schwarzian quantum mechanics. 
%The $\mu$ coupling sources an energy flux on $\Sigma$ and determines its embedding in the 3D spacetime via the Israel junction condition derived from the 3D Einstein-de Sitter equations of motion. 

\subsection{Israel junction 
condition}\label{subsec:cutting_ds}

\vspace{-1mm}
  In this subsection, we solve the Einstein-de Sitter equations in the presence of a 1D matter source localized on the equator of past and future asymptotic infinity. The matter source is arranged such that the resulting energy distribution in the bulk is confined to a 2D surface, with a static energy density $T_{00}$ proportional to the MQ coupling $\mu(u)>0$.  
  
Three-dimensional gravity has no local degrees of freedom and the spacetime outside of the surface is locally 3D de Sitter. Thus, we can embed our setup in global 3D de Sitter spacetime, with metric
\begin{eqnarray}
    ds^2=-d\tau^2+\cosh^2\tau(d\theta^2+\sin^2\theta d\varphi^2).
\end{eqnarray}
Introduce the coordinates $x=\varphi$ and $ y = \log\cot\theta/2$.
   % \label{eq:global_to_complex}\end{equation}
The metric then becomes
\begin{eqnarray}
    ds^2=-d\tau^2+\frac{\cosh^2\tau}{\cosh^2y}(dx^2+dy^2).
    \label{eq:metricXY}
\end{eqnarray}
Next, we place an energy distribution localized on  a 2D  surface $\Sigma$ at the equator slice of the spatial sphere, thus dividing the de Sitter space into two halves. 

\begin{figure}[t]
 \begin{center}
\resizebox{0.35\textwidth}{!}{\input{images/cutds}}~~~~~~~~~~~~~~~~~~~~~~~~\raisebox{-.9cm}{\input{images/dS3_cut}}~~~~~~~~
        \caption{The future region of the spacetime obtained by cutting and gluing global 3D de Sitter along hypersurfaces $\Sigma$, $\Sigma^*$ given by the function $\psi(u)= x(u) \pm i y(u)$. We remove the dark gray region and glue together its boundaries. %On the left, the spacetime in global coordinates;  on the right, the spacetime in $x, y$ coordinates. 
        The vertical direction is the global time $\tau$.}
    \label{fig:maldacenaqi}
  \end{center}
  \vspace{-1.5mm}
\end{figure}
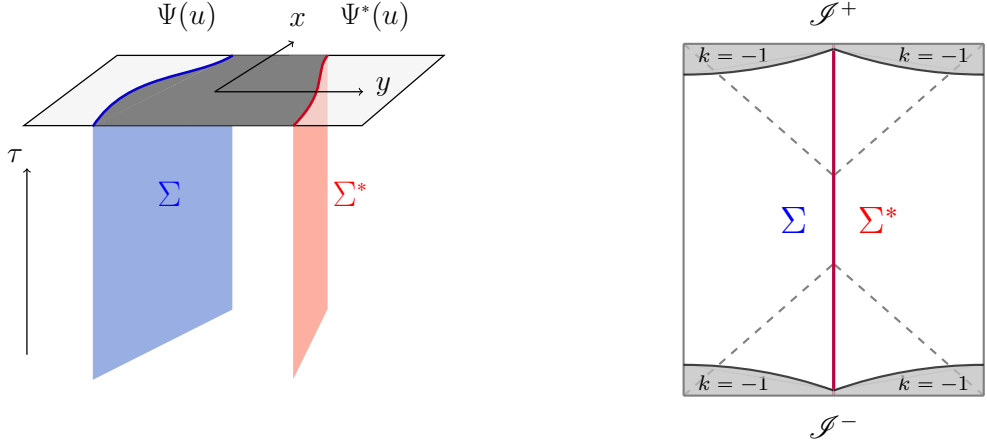

The gravitational backreaction due to an energy source can be implemented by means of a cut-and-paste procedure. First, we partition the spacetime into two half-de Sitter spacetimes $\mathcal{V}$, $\mathcal{V}^*$, and choose the metric to have the form \eqref{eq:metricXY} on both sides. The coordinate systems on ${\cal V}$, $\mathcal{V}^*$ are related via a non-trivial diffeomorphism and the embedding equation of the slice $\Sigma$ as seen from the two sides will thus be different. We represent the two embedding equations as two different surfaces $\Sigma, \Sigma^*$ placed inside a global de Sitter spacetime with one single coordinate patch $(\tau,x,y)$. We can take the two surfaces to be symmetric with respect to the equator, and parameterize them as 
\begin{eqnarray}
\label{eq:twosigmas}
   \Sigma = (\tau, x(u), y(u)), \qquad \ \  \Sigma^* = (\tau, x(u), -y(u)).
   %\bigr| (\tau,u) \in \mathbb{R}^2\bigr\} 
\end{eqnarray}
 The physical spacetime is then obtained by cutting global de Sitter spacetime along $\Sigma$ and $\Sigma^*$, removing the interior region (or the exterior region in case $\mu(u)$ is negative) and then gluing the two half-spacetimes $\mathcal{V}$, $\mathcal{V}^*$ back together, as shown in figure~\ref{fig:maldacenaqi}. The shape of $\Sigma$ and $\Sigma^*$ is taken to be static: their intersection with constant $\tau$ slices  does not depend on $\tau$.

%When $\mu=0$, we still have some non trivial global dynamics. We will argue  that constant finite temperature states in SYK with $\mu=0$ correspond to a Schwarzschild-de Sitter spacetime with a conical deficit %obtained by cutting and gluing every spatial slice along two great semicircles with relative opening angle $\pi(1-v)$% of $\pi(1+v)$. In the low temperature limit $v\to 1$, the opening angle goes to zero, which means that the two surfaces $\Sigma_\pm$ given in \eqref{eq:twosigmas} approach the $x$-axis~$\{y=0\}$ from opposite sides. For general $\mu(u)>0$ the surfaces follow an arbitrarily shaped convex trajectory.

%\begin{eqnarray}
%S_{EH} = \frac{1}{16\pi G_N}\int\! du \Bigl(p\spc  \phi' + \sqrt{1-e^{-2\phi}} \cos p - \mu \phi\Bigr)
%\end{eqnarray}

 The Einstein equations reduce to the Israel junction conditions \cite{Israel:1966rt, Poisson:2009pwt}
\begin{eqnarray}
    [K_{ab}] -[K]h_{ab} = -8\pi G_NT_{ab},
    \label{eq:einsdisc}
\end{eqnarray}
where $h_{ab}$ is the induced metric on the surface, $K_{ab}$ is the second fundamental form, and $K=K_{ab}h^{ab}$ is the mean extrinsic curvature. A square bracket denotes a discontinuity, i.e. $[K] = K-K^*$
where $K$ ($K^*$) is the extrinsic curvature as seen from the right (left)  of the gluing surface. As we will see below,
the induced metric on the slice will always be that of two-dimensional de Sitter space. We wish to write the junction conditions \eqref{eq:einsdisc} as an explicit 1D equation of motion that determines the shape \eqref{eq:twosigmas} of the gluing surface. 

Turning on a non-zero MQ coupling $\mu(u)$ on the SYK side corresponds to placing a static energy-momentum tensor on the gluing surface $\Sigma$  
\begin{eqnarray}
    T_{ab}=\begin{pmatrix}
         \, T_{00}\! &\! 0\, \\
       \, 0\! &\! 0\,
    \end{pmatrix}.
\end{eqnarray}
Here $T_{00}$ is delta-function supported on $\Sigma$ and may depend on the spatial coordinate $u$. The $\tau$ dependence is then uniquely determined by the energy conservation equation $\nabla^aT_{ab} = 0$. Physically, we can think of this static energy distribution as a collection of non-interacting dust particles that follow geodesics at constant $x, y$, or equivalently, at constant polar coordinates $\theta, \varphi$ on the spatial sphere\footnote{The geodesic equations for the metric \eqref{eq:metricXY} are 
 \begin{eqnarray}
    \ddot\tau=-\frac{\cosh\tau\sinh\tau}{\cosh^2 y}(\dot x^2\!+\dot y^2),\quad\quad
         \frac{d}{d\lambda}\left(\frac{\cosh^2\tau}{\cosh^2y}\dot x\right)=0,\quad\quad
         \frac{d}{d\lambda}\left(\frac{\cosh^2\tau}{\cosh^2y}\dot y\right)=-\frac{\sinh y\cosh^2\tau}{\cosh^3y}(\dot x^2\! +\dot y^2).
 \end{eqnarray} 
We are considering the solution $\dot x=\dot y =0$.}. We are interested in computing the backreaction on the geometry due to this energy-momentum distribution.

First, we compute the extrinsic curvature of the timelike surface $\Sigma$. Recalling that $\psi(u)= x(u)+iy(u)$ parametrizes $\Sigma$ via \eqref{eq:twosigmas} with $\psi'=|\psi'|e^{ip}$, we find that the normalized tangent and normal vectors to  $\Sigma$ are
\begin{eqnarray}
\label{eq:tangent_normal}
    e_0^\mu = (1, 0, 0),\qquad e_1^\mu = \frac{\cosh{y}}{\cosh\tau}(0, \cos p, \sin p),\qquad n^\mu =\frac{\cosh{y}}{\cosh\tau}(0, \sin p, -\cos p).
\end{eqnarray}
Here, $n^\mu$ is the normal to $\Sigma$, which, by convention, points from $\mathcal{V}^*$ to $\mathcal{V}$. As we will see later, the sign of $n^\mu$ will determine which portion of the spacetime bounded by $\Sigma$ is physical and which portion is removed by the Israel junction conditions.

The induced metric and the extrinsic curvature of the surface $\Sigma$ are ($a, b=0,1$)\footnote{Here we use a locally flat coordinate system. However, the induced metric on $\Sigma$ is dS$_2$, as shown below in \eqref{eq:metric_membrane}.} 
\begin{eqnarray}
    h_{ab}\is e^\mu_ae^{\nu}_bg_{\mu\nu}= \begin{pmatrix}
        {-1} \!&\! 0\\[-.5mm]
        0\! &\! 1
    \end{pmatrix},\qquad K_{ab} = e^{\mu}_ae^{\nu}_b\nabla_\mu n_\nu = \begin{pmatrix}
       \, 0 \!&\! 0 \\[-.25mm]
        0\! & \! K
    \end{pmatrix},\\[2.5mm]
 \label{eq:kexp}   K\is\frac{1}{\cosh\tau}\left(\cosh y\Bigl(\cos p \,\frac{\partial p}{\partial x}+\sin p\,\frac{\partial p}{\partial y}\Bigr)+ \sinh y\cos p\right).
\end{eqnarray}
Since $h_{ab}$ is the unit metric, $K$ above is the trace of the extrinsic curvature $K=K_{ab}h^{ab}$. Substituting the relation
  %  \psi'\is x'+iy'\equiv |\psi'|e^{ip} \qquad \ 
 $p' = %\partial_x p\, x' + \partial_y p{\partial y}\, y' = 
  |\psi'|\bigl(\frac{\partial p}{\partial x}\, \cos p + \frac{\partial p}{\partial y}\, \sin p\bigr)$
into the expression \eqref{eq:kexp} for $K$, we obtain
\begin{eqnarray}
    \boxed{\  
   K=\frac{\cosh y}{\cosh \tau |\psi'|}\Bigl(p' +|\psi'|\tanh y\cos p\Bigr) \sstrut\ }
\label{eq:extrinsic_curvature}
\end{eqnarray}
The surface $\Sigma^*$ is similarly parameterized by $\psi^*$; since $\psi^*$ is the complex conjugate of $\psi$, the extrinsic curvature of $\Sigma^*$ is $-K$. Then, the left hand side of equation~\eqref{eq:einsdisc} reads
\begin{equation}
     [K_{ab}] -[K]h_{ab} =\begin{pmatrix}
         2K&0\\
         0&0
     \end{pmatrix}.
\end{equation}

The nonzero extrinsic curvature of $\Sigma$ is sourced by a non-zero energy density of the stationary matter distribution. We take the energy density on $\Sigma$ to be of the form 
\begin{eqnarray}
    4\pi G_NT_{00} = \frac{\mu(u)}{\ell'}\qquad \ell'= \frac{\cosh \tau |\psi'|}{\cosh y },
    \label{eq:stress_tensor}
\end{eqnarray}
where $\ell$ is the line element of the curve $(x(u), y(u))$ in the $x, y$ plane. This definition implies that $\mu$ is an energy density in the $u$ coordinate:
\begin{equation}
    \mu = \frac{dE}{du}\qquad \frac{dE}{d\ell} = \frac{\mu}{\ell'}.
\end{equation}
Putting everything together, we find that the Israel junction conditions \eqref{eq:einsdisc} read
\begin{equation}
    \boxed{\ p'+|\psi'|\tanh (\text{Im}\psi) \cos p = -\mu \sstrut\, }
    \label{eq:LQ_equation_of_motion}
\end{equation}
This exactly matches equation \eqref{eq:lensky-qi-one} for the trajectory of the soft mode $\psi(u)$ of the SYK model.

\subsection{1D effective action of 3D near-de Sitter gravity}\label{subsec:effective_action}
\vspace{-1mm}

In this section, we compute the effective action that governs the embedding of the cut-and-glued dS$_2$ slice $\Sigma$ in the 3D spacetime. We start from 3D de Sitter gravity and impose boundary conditions that reduce the dynamics to that of a two-dimensional surface located at the equator of the global dS$_3$ spacetime. The bulk dynamics of de Sitter gravity is topological, so any non-trivial dynamics will arise from the explicit breaking of diffeomorphism invariance due to boundary conditions.
%or coupling to matter source terms localized on the equator of past and future infinity. %The two equators are connected via a two-dimensional gluing surface with a static energy density $T_{00}$ proportional to the Maldacena-Qi coupling $\mu$. The matter source $T_{00}$ can be thought of as a collection of static dust particles that stay at a fixed location on the spatial sphere. In the classical background, the time dependence of $T_{00}$ is fixed by energy conservation.  

As explained, the gluing surface splits de Sitter spacetime into two half spacetimes bounded by the timelike surfaces $\Sigma$ and $\Sigma^*$. We parametrize their shape as in equation~\eqref{eq:twosigmas}
%\begin{eqnarray}
 %   \Sigma^\pm(\tau, u)\is (\tau, x(u), \pm y(u))
    %\nonumber \\[-3mm]\\[-3mm]\nonumber\Sigma^-(\tau, u) \is (\tau, x(u), -y(u)),\end{eqnarray}
and use the definition $\psi(u)=x(u)+iy(u)$, $\psi'=|\psi'|e^{ip}$.  The spatial section of $\Sigma$ and $\Sigma^*$ specify 1D curves 
\begin{equation}
 %\Sigma^\pm=\bigl\{ (\tau, x(u), \pm y(u))\bigr| (\tau,u) \in \mathbb{R}^2\bigr\} \qquad \quad 
 {\cal C} = 
 %\bigl\{ 
 ( x(u),  y(u)), \qquad \quad {\cal C}^* = 
 %\bigl\{ 
 ( x(u),  -y(u)) %\bigr| u \in \mathbb{R}\bigr\} 
 \label{eq:curves}
\end{equation} 
in the $(x,y)$ plane.
  We wish to determine the $\psi$ dependence of the on-shell gravitational action of the total spacetime. 
To write the gravitational action, we will allow for general ``off shell" configurations for which the metric on $\Sigma$ is continuous but its normal derivative may include a discontinuity that reflects the presence of the localized matter source.

%\pagebreak

\subsubsection{Boundary conditions} \vspace{-2mm}
To make the gravitational dynamics uniquely defined, we need to fix boundary conditions  at past and future infinity $\mathscr{I}^\pm$, and on the gluing surface $\Sigma$. 

At $\mathscr{I}^\pm$, we choose conformal boundary conditions
that fix the conformal structure of the metric and the trace of the extrinsic curvature.     
These boundary conditions are standard in (A)dS/CFT (see, e.g. \cite{Anninos:2023epi, Anninos:2024wpy, allameh2025timelikeliouvilletheoryads3}). Define $\sigma_{ij} = e^{\omega(x^i)}\hat\sigma_{ij}$, and introduce a Fefferman-Graham-like expansion for the boundary metric and the trace of the extrinsic curvature
\begin{eqnarray}
\label{eq:FG}
  \hat\sigma_{ij} \is e^{2T}\hat\sigma_{ij}^{(0)}+\hat\sigma_{ij}^{(2)}
 %+e^{-2T}\hat\sigma_{ij}^{(4)}
 +\dots 
 \notag \\[-2mm]  & & \qquad\qquad \qquad \qquad     \qquad \quad \quad T\to\infty.
 %\Pi^{ij} \is %\Pi^{ij\,(2)}+e^{-2T}\Pi^{ij\,(4)}+\dots  
\\[-2mm] \notag  \qquad
 \qquad 
 \ \ K_\sigma \is K_\sigma^{(2)}\,+e^{-2T}K_\sigma^{(4)}+\dots \qquad
\end{eqnarray}
 In near-de Sitter gravity,  we choose
 \begin{eqnarray}
%\qquad\qquad \qquad
\boxed{
\begin{aligned}
\ \ \hat\sigma_{ij}^{(0)}\;  = \; \delta_{ij},
%\hat\sigma_{xx}^{(2)} = 2\tanh^2 y,\;\;\;\hat\sigma_{yy}^{(2)}=\frac{2\text{ch}^2y-6}{\text{ch}^2y},\;\;\;\hat\sigma_{xy}^{(2)}=0;
%\\ & \qquad 
\quad %K^{(2)}_{\sigma} = 2, 
\quad K^{(4)}_\sigma = 4. \ \ 
%\qquad K^{(4)}_{\sigma} = 4.
\small{}_{\strut}^{\strut}
\end{aligned}
}
%\qquad \quad %\mbox{(conformal b.c.)}\qquad
\label{eq:conformalbc}
\end{eqnarray}
There are good reasons to believe that fixing the leading term of the metric $\hat{\sigma}^{(0)}_{ij}$ and the subleading term of the extrinsic curvature $K_{\sigma}^{(4)}$ turns solving for the bulk geometry into a well-posed problem, but we will not attempt to prove this here. Note that the leading term of the extrinsic curvature $K_{\sigma}^{(2)}=2$ is not independently specified, but rather fixed by the Fefferman-Graham expansion \eqref{eq:FG}. As we will see below, the above boundary conditions imply that $\mathscr{I}^\pm$ are covered by two $k=-1$ slices, one for the left and one for the right patch.
%The third condition prescribes the form of the subleading FG coefficients of the metric.\footnote{In our coordinate system, the extra boundary condition on the subleading metric components reads
%\begin{equation}
%  \hat\sigma_{xx}^{(2)} = 2\tanh^2 y,\;\;\;\hat\sigma_{yy}^{(2)}=\frac{2\text{ch}^2y-6}{\text{ch}^2y},\;\;\;\hat\sigma_{xy}^{(2)}=0.  
%\end{equation} }
%Note that choosing boundary conditions also fixes the coordinate system on $\mathscr{I}^\pm$.

As a side remark, we note that an alternative set of boundary conditions, which also appear to be compatible with our setup, is Neumann boundary conditions that fix all components of the canonical momentum\footnote{Here the canonical momentum is defined as $
 16\pi G_N\Pi^{ij}\equiv\sqrt{\sigma}(K^{ij}_{\sigma}-K_\sigma\sigma^{ij}) ,$
where $\sigma_{ij}$ is the induced metric and $K^{ij}_{\sigma}$ is the extrinsic curvature of $\mathscr{I}^\pm$.} $\Pi^{ij}$ dual to the 2D metric $\sigma_{ij}$. %Defining $\Pi^{ij}\equiv\pi^{ij}/ 16\pi G_{N}$, we choose %Let us decompose the canonical momentum $\Pi^{ab}$ into a trace part and a traceless part
%\begin{eqnarray}
%    \Pi^{ab} \is \Pi\, \sigma^{ab} + \hat{\Pi}^{ab}
%\end{eqnarray}
%We choose the following Neumann boundary condition at past and future infinity
%\begin{eqnarray}
% \qquad \qquad   \boxed{\ \  \Pi = -\frac{1}{16\pi G_N}, \quad    %\hat{\Pi}^{ab} = 0 \qquad\quad {\rm at} \ \ \mathscr{I}^\pm \sstrut\ }
%\end{eqnarray} 
%\begin{eqnarray}
%\quad\qquad \ \ \   \boxed{\ \ 
%   \pi^{ij \,(2)}=-\delta^{ij} 
  %, \quad \pi^{xx\,(4)}=-2\tanh^2y,\quad \pi^{yy\,(4)}=-\frac{2\text{sh}^2y+4}{\text{ch}^2y}, \quad\pi^{xy\,(4)}=0.
%\ \ , \qquad R[\sigma^{(0)}]=-2\quad (\mbox{on-shell}) \small %{}^{\strut}_{\strut}}\quad\qquad \mbox{(Neumann b.c.)}
%   \label{eq:neumannbc}
%\end{eqnarray}
%This condition also fixes our coordinate system on $\mathscr{I}^\pm$.
%Note that the Brown-York stress tensor \cite{Brown_1993} is proportional to $\Pi^{ij}/\sqrt{\sigma}$. Neumann boundary conditions thus fix the holographic stress-energy flux \cite{Balasubramanian_1999, Balasubramanian:2001nb} at infinity everywhere except at the location of the curve $\mathcal{C}$ where we place our holographic screen. Hence the bulk dynamics is restricted to be compatible with this boundary condition. Note that in the absence of matter sources on $\mathscr{I}^\pm$ (as is the case here away from the 1D holographic screen $\mathcal{C}$) equation \eqref{eq:neumannbc} automatically solves the 2D diffeomorphism constraint $\nabla_i\pi^{ij} = 0$.
Neumann boundary conditions can be obtained from conformal boundary conditions by promoting
 the conformal class $\hat{\sigma}_{ij}$ of $\sigma_{ij}$ to a dynamical degree of freedom. Integrating over all $\hat{\sigma}_{ij}$ amounts to setting the variation with respect to the off-diagonal components of $\sigma_{ij}$ equal to zero. Fixing $K_\sigma$ then also fixes the trace $\tr\Pi$ of the canonical momentum. Combined, this adds up to a full set of Neumann boundary conditions. 

For both types of boundary conditions, one needs to supplement the Einstein-Hilbert action with a surface term \cite{allameh2025timelikeliouvilletheoryads3,BlauGRNotes, Krishnan_2017}
\begin{eqnarray}
S_{\rm bd} =  -\frac{1}{16\pi G_N} \int_{\mathscr{I}^\pm} \!\sqrt{\sigma}\, K_\sigma .
\end{eqnarray}
This is the familiar Gibbons-Hawking-York term \cite{Gibbons:1976ue}, except that its normalization is 1/2 times smaller than for Dirichlet boundary conditions. We review the derivation of this surface term in Appendix \ref{app:more_on_the_action}.

Semiclassically, the boundary conditions \eqref{eq:conformalbc} imply that future and past infinity split up into two hyperbolic $k=-1$ slices that meet at the intersection curve $\mathcal{C}^\pm$. For this reason, we refer to \eqref{eq:conformalbc} as $k=-1$ boundary conditions. Parameterizing the 3D global de Sitter geometry by means of the coordinates $(\tau, x,y)$ as before, one finds that the boundary conditions amount to placing the cutoff surface at the location (c.f. equation \eqref{eq:phi_definition})
\begin{equation}
\label{eq:scridef}
   \mathscr{I}^\pm:\;  \tau= \pm \bigl(\spc  T+\phi(y)\spc \bigr),\qquad\phi(y) \equiv \log \coth|y|, \qquad T\to \infty
\end{equation}
as shown in figure~\ref{fig:maldacenaqi}.
%On this cutoff surface, and away from the gluing surface, we impose the Neumann boundary conditions that fix the normal derivative of the metric on $\mathscr{I}^\pm$ to a negative constant. Since the bulk geometry is standard 3D de Sitter, this condition implies that the intrinsic metric also has constant negative curvature. %
This choice of cutoff \eqref{eq:scridef} fixes the 2D metric on $\mathscr{I}^\pm$ to be of the form
\begin{eqnarray}
\label{eq:metric_scri}
   %\boxed{\ \ 
   \sigma_{ij}dx^idx^j = \frac{e^{2T}}{4}\,  {\frac{dx^2 + dy^2}{\sinh^2 y}} \qquad {\rm at} \ \mathscr{I}^\pm .%\sstrut\ \ }
\end{eqnarray}
The $k=-1$ slicing of de Sitter naturally splits up $\mathscr{I}^\pm$ into two halves, $y>0$ and $y<0$ \cite{Maldacena:2012xp}.

Finally,  we choose identical Dirichlet boundary conditions on the gluing surfaces $\Sigma$ and $\Sigma^*$, and fix the induced metric $h_{ab}$. With our choice of boundary conditions on metric and matter field, the classical saddle point metric on $\Sigma$ is that of a dS$_2$ slice,
\begin{eqnarray}
\label{eq:metric_membrane}
    h_{ab}dx^adx^b = -d\tau^2 + \frac{ \cosh^2\tau}{\cosh^2 y(u)}\; {|\psi'(u)|^2} du^2.
\end{eqnarray}
The matter source $\mu(u)$ is located at the intersection curve ${\cal C}^\pm = \mathscr{I}^\pm \cap \Sigma$ between the cutoff surface $\mathscr{I}^\pm$ and the gluing surface $\Sigma$, and  determines the embedding of the dS$_2$ slice inside dS$_3$.

\subsubsection{Gravitational action} 
\label{subsec:gravact}
\vspace{-1mm}
The gravitational action of 3D near-de Sitter gravity reads\footnote{The integrals are performed over both sides, i.e. on $\cal V$, $\cal V^*$, $\Sigma$, $\Sigma^*$, $\dots$; we omit the star to simplify the notation.}
\begin{eqnarray}
\label{eq:action}
S_{EH} =\frac{1}{16\pi G_N} \left[\int_{\mathcal{V}}\!\! \sqrt{-g}\spc(R-2)+2\int_{\Sigma}\!\! \sqrt{-h}\spc K_h - \int_{\mathscr{I}^\pm}\!\!\sqrt{\sigma}\spc K_\sigma+2\int_{\mathcal{C}^{\pm} }\!\!\!\sqrt{\gamma}\spc \eta_H\right].
\end{eqnarray}
 The first term is the bulk Einstein-Hilbert action. The second term and third term are accompanied by Dirichlet boundary conditions\footnote{This term can also be obtained by integrating the $\delta$ function discontinuity of the Ricci scalar across the gluing surface.} on $\Sigma$ and conformal (or Neumann) boundary conditions \cite{allameh2025timelikeliouvilletheoryads3, BlauGRNotes, Krishnan_2017}
 at~$\mathscr{I}^\pm$. The last term is a Hayward corner term \cite{Hayward:1993my} that lives on the intersections $\mathcal{C}^\pm = \Sigma \cap\mathscr{I}^\pm$,  and is necessary to have a well defined variational principle in the presence of corners. The boost parameter $\eta_H$ is defined as $\eta_H=\sinh^{-1}n^\mu m_\mu$ with $n^\mu$ the normal to $\Sigma$ and $m_\mu$ the normal to $\mathscr{I^\pm}$. 
 
 In Appendix \ref{app:more_on_the_action}, we show that the  action \eqref{eq:action} is the appropriate one for the type of boundary conditions we outlined above. To derive an effective action for $\psi$, we want to compute each term of the action as a function of the  shape of $\Sigma$.
%\footnote{A similar calculation, with a coordinate dependent cutoff, was performed in \cite{Chandra_2022}.}.  The resulting effective action is a functional of the complex path $\psi(u)$. 
 
\paragraph{Outline of the computation.}\vspace{-2mm} In Appendix \ref{app:more_on_the_action}, we explicitly evaluate each of the individual terms of the Einstein-Hilbert action and boundary term. The procedure is straightforward. Since the bulk geometry away from the gluing surface is pure de Sitter, and the gluing surface is stationary, we can evaluate each term in the action by plugging in the classical metric on each component and performing the integrals over the respective subregions. Here we collect the results
\begin{eqnarray}
\label{eq:action1}
    \int_{\mathcal{V}}\!\!\!\sqrt{-g}\,(R-2) \is%4\int_{-T-\phi(y) }^{T+\phi(y)}  \frac{d\tau dx  dy}{\cosh^2y}\cosh ^2\tau =% 
    {\color{red}\boxed{\color{black}{e^{2T}\int \frac{dxdy}{\sinh^2y}}}}\, +\, {\color{blue}\boxed{\color{black}{4T\int \frac{dxdy}{\cosh^2y}}}}\, +\, {\color{darkgreen}\boxed{\color{black}{4\int \frac{dxdy}{\cosh^2y}\phi(y)}}} \;, \\[3.5mm]
\label{eq:action2}
    \int_{\Sigma}\!\!\sqrt{-h}\, K_h %&=2\int_{-T-\phi(y(u))}^{T+\phi(y(u))} d\tau \int du(p'\!+|\psi'|\tanh |\text{Im}\psi|\cos p)\\&% 
    \is {\color{blue}\boxed{\color{black}-4T\int du (p'+|\psi'|\tanh \text{Im}\psi\cos p)}}
    -\! {\color{white}\boxed{\color{black} 4\!\int \!du\,\phi(u)(p'}}\! +{\color{darkgreen}\boxed{\vphantom{\int}\color{black}|\psi'|\tanh \text{Im}\psi\cos p)}}\;,\qquad \\[3.5mm]
\label{eq:action3}
    \int_{\mathscr{I}^\pm}\!\!\sqrt{\sigma}\, K_\sigma\is {\color{red}\boxed{\color{black} e^{2T}\int\frac{dxdy}{\sinh^2 y}}} \, +\, {\color{darkgreen}\boxed{\color{black}{4\int\frac{dxdy}{\cosh^2y}}}}\;,\\[3mm]
\label{eq:action4}
\int_{\mathcal{C}^\pm}\!\!\!\sqrt{\gamma}\, \eta_H\is%\int_{\Sigma^\pm\cap\mathscr{I}^\pm}\sqrt{\sigma}\sinh^{-1}m_\mu n^\mu=%
    4\int\! du \, \frac{|\psi'|}{\sinh |\text{Im}\psi|}\frac{\cos p}{\cosh \text{Im}\psi}=\color{white}\boxed{\color{black}4\int\! du \, \frac{|\psi'|}{\sinh |\text{Im}\psi|} \sqrt{1-e^{-2\phi}}\cos p.}
\end{eqnarray}
Here, all the $(x,y)$ integrals run over the two half planes bounded by the curves  \eqref{eq:curves}. In the last line we used the relation $ \cosh(y) = 1/\sqrt{1-e^{-2\phi}}$ to write the result in terms of $\phi\equiv\phi(y(u)) = \log\coth|y(u)|$. We also used the definition $\psi'=|\psi'|e^{ip}$.

The terms inside boxes of the same color cancel against each other (up to constants). The terms proportional to $e^{2T}$ in red explicitly cancel. To simplify the terms linear in $T$, we note that the term in blue in \eqref{eq:action1} is the area of the cut-and-glued spatial sphere. This is (twice) the area of the hemisphere with boundary given by ${\cal C}$. Since a sphere has constant curvature, we can use the Gauss-Bonnet theorem to relate this area to the integral of the extrinsic curvature of the boundary, 
\begin{eqnarray}
    \int \frac{dxdy}{\cosh^2y} = 4\pi + 2\int du\bigl(p'+|\psi'|\tanh \text{Im}\psi\cos p\bigr). 
\end{eqnarray}
This cancels the blue term in \eqref{eq:action2}, up to a constant proportional to $T$, which we drop.

Finally, we combine the terms in green in \eqref{eq:action1} and \eqref{eq:action3} and recognize a total derivative,
\begin{eqnarray}
    \int \frac{dx dy}{\cosh^2y}(\phi(y)-1) 
    \is \int \frac{dxdy }{\cosh^2y}(\log\coth |y|-1) = \int dxdy\frac{d}{dy}\bigl(\log(\coth|y|)\tanh y\bigr)\notag \\[-1.5mm]\\[-2mm]\notag
    \is 2\int dx\, \phi(\text{Im}\psi)\tanh\text{Im}\psi = 2\int du\,\phi(u) |\psi'|\tanh\text{Im}\psi\cos p,
\end{eqnarray}
which cancels the term in green in \eqref{eq:action2}. 

Collecting the remaining terms, we have
\begin{eqnarray}
\label{eq:onedseff}
\boxed{\ S_{EH}=\frac{1}{2\pi G_N}\int du \Bigl(p\phi'+\frac{|\psi'|}{\sinh| \text {Im}\psi|}\sqrt{1-e^{-2\phi}}\cos p\Bigr). {}^{\large \strut}_{\large \strut} \ } %+ \text{constant} 
\end{eqnarray}
Note that the term $p\phi'$ in the Lagrangian arises as a contribution from the extrinsic curvature of the gluing surface $\Sigma$, while the Hamiltonian term corresponds to the corner term at the intersection of the gluing surface and $\mathscr{I}^\pm$.
This is our main result announced in the introduction. 

\medskip

A few short comments are in order.

\vspace{-3mm}

\paragraph{Proper time gauge.} 

The action \eqref{eq:onedseff} is invariant under reparametrization of the $u$ coordinate. Indeed, up to now, we have not yet fixed the coordinate choice on the 1D curves ${\cal C}^\pm$.  To make contact with the DSSYK effective action, we need to choose the gauge
\begin{equation}
\label{eq:gauge_choice}
    |\psi'| = \sinh |\text{Im} \psi|.
\end{equation}
Comparing with equations \eqref{eq:metric_scri} and \eqref{eq:metric_membrane}, we see that the DSSYK time coordinate has a natural meaning on the gravity side as $e^{-T}$ times the proper distance along the curve ${\cal C}^\pm$. 

\vspace{-2mm}

\paragraph{Angle deficit.}  

Treating $u$ as the time coordinate and transitioning to the quantum theory, the first term in the effective action implies the commutation relation 
\begin{equation}
\label{eq:pphicom}
    [\phi, p] =  2\pi i  \spc  G_N.
\end{equation}
This equation has a natural gravity interpretation. The variable $\phi(u) = \log\coth| y(u)|$ represents half the proper time of the particle trajectory that connects the point $u$ on ${\cal C}^-$ to the same point on ${\cal C}^+$.  Adding a particle of mass $m$ at $u=u_0$ amounts to inserting its worldline action $e^{ 2 i m \phi(u_0)}$ into the gravity path integral. This adds a delta-function source $\mu_m(u) = m \spc \delta(u-u_0)$ to the mass density. The backreaction creates a small conical deficit  
%\alpha = 2\pi \sqrt{1-8 G_N m} \simeq 2\pi - \delta \alpha, \qquad 
\begin{eqnarray}
\delta \alpha = 8\pi G_N m.
\end{eqnarray}
This angle deficit looks like a step function discontinuity in $p(u)$. Since the two surfaces $\Sigma$ and $\Sigma^*$ both have a kink, the size of the discontinuity in $p(u)$ is $4\pi G_N m$, which matches the result derived from the canonical commutation relation \eqref{eq:pphicom}.
%, including the factor of $i$. 

\subsection{Examples}\label{subsec:examples_gravity}

In this subsection, we see how the dictionary we proposed applies to the three explicit solutions of the SYK equation of motions we reviewed in Sec.~\ref{sec:soft_mode_of_DSSYK}.

\vspace{-2mm}

\begin{figure}[hbtp]
  \centering
  % ---------- Row 1 ----------
\subfloat{\includegraphics[width=0.465\textwidth,height=0.24\textheight]{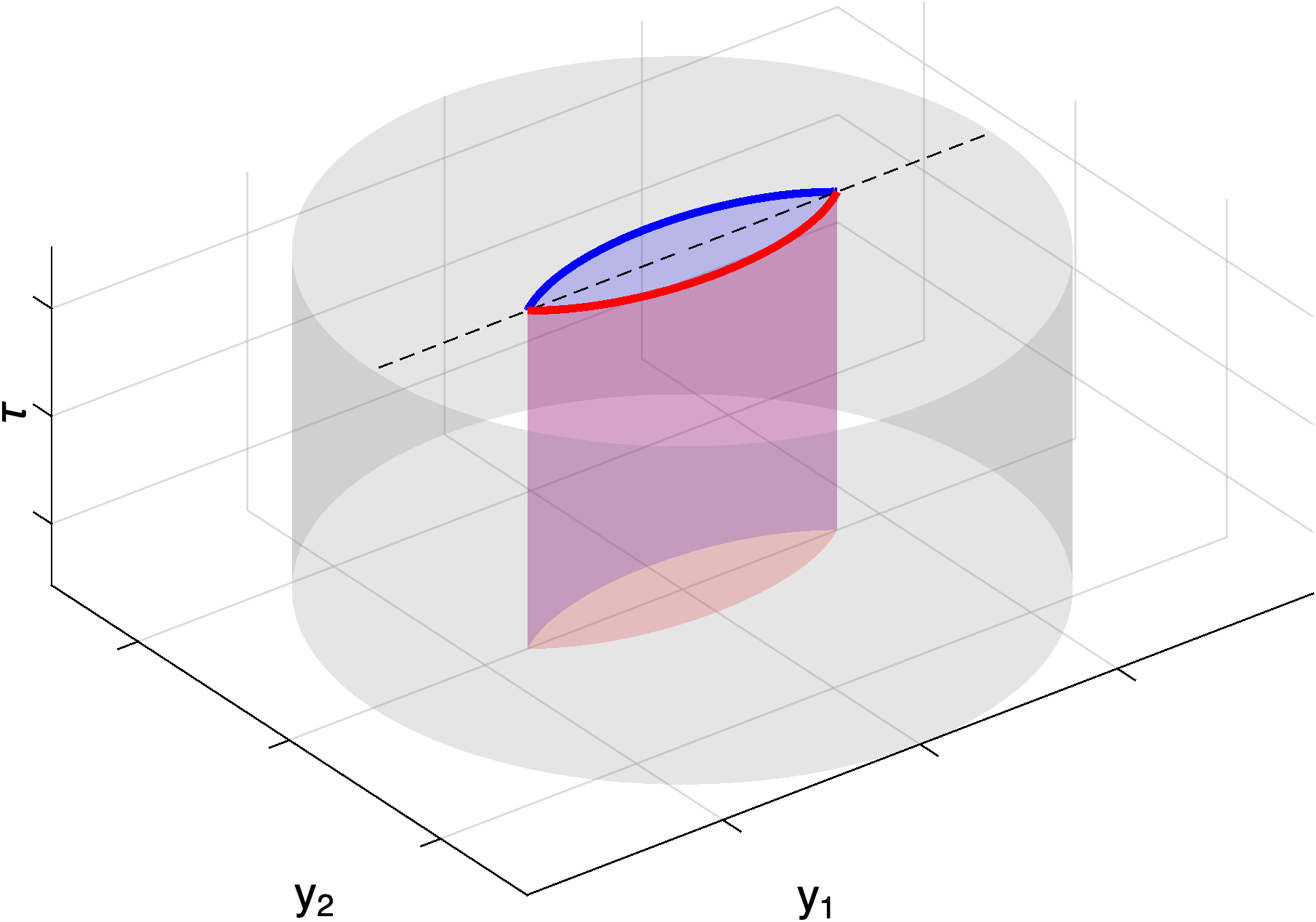}~~~~~~~~~~~~~~~~~~~~\raisebox{8mm}{\includegraphics[width=0.24\textwidth]{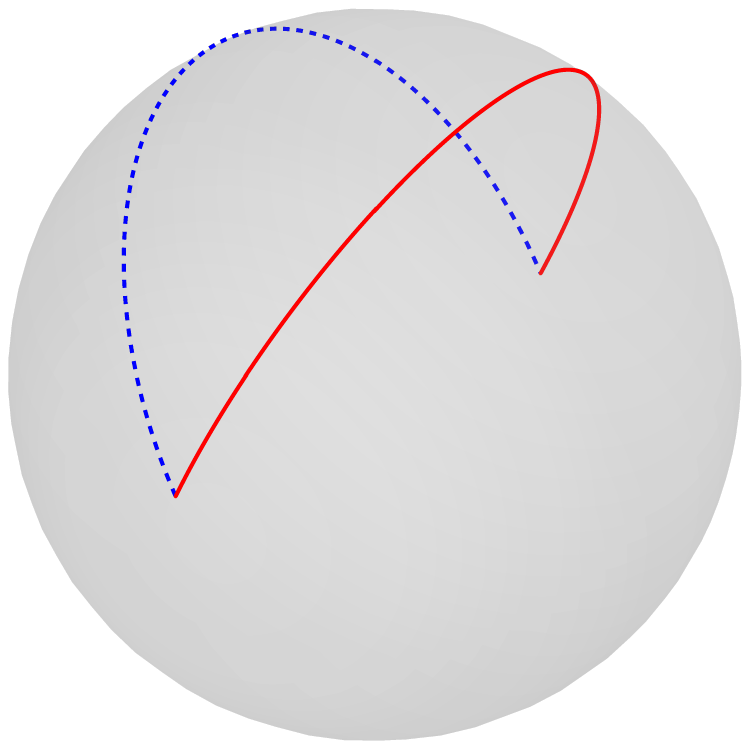}}}\
  \caption{The gluing surfaces $\Sigma$ and $\Sigma^*$ for the TFD solution (left) and their intersection with a constant $\tau$ slice 
  (right). The conical angle at the cusps is equal to $2\theta$. Depending on the sign of the temperature, i.e. on whether $\theta <\frac \pi 2$ or $\theta>\frac \pi 2$, either the interior/exterior of the wedge is physical/removed or vice versa.   }
  \label{fig:membrane1}
 \vspace{-1mm}
 \end{figure}

\noindent
\paragraph{Thermal solution.} 
First, we set $\mu=0$ and turn on a finite temperature $\beta$. We plot\footnote{To plot the gluing membrane, we projected the spatial sphere to the coordinates $$y_{1,2} \equiv \cosh^{-1}(\vec{n}\cdot \vec{e_3}) \frac{\vec{n}\cdot \vec{e}_{1,2}}{\sqrt{1-\vec{n}\cdot \vec{e}_3}},$$ with $\vec{n}$ the vector on the sphere, and $\vec{e}_{i}$ unit reference vectors. The dashed line is the intersection of the two $k=-1$ slices.} this TFD solution \eqref{eq:thermal_solution_psi} below
%\begin{equation}
%\psi(u) = 2\arctan\tanh\left(\cos\frac{\pi v}{2}\frac{u}{2} -i\frac{\pi}{4}(1-v) \right)
 %   \label{eq:thermal_solution_psi_dual}
%\end{equation}
 in figure~\ref{fig:membrane1}. Previous works have proposed that finite temperature states in DSSYK are dual to a 2+1-D Schwarzschild-de Sitter spacetime with a conical deficit \cite{verlinde2024doublescaledsykchordssitter, tietto2025microscopicmodelsitterspacetime}. Such a spacetime is obtained by cutting and gluing the spatial sphere along two great semicircles that end at the poles.  We can now verify this proposed holographic interpretation based on figure \ref{fig:membrane1}.
 We can read off the deficit angle from the imaginary part of $\psi$ at the equator, using our dictionary and the relation $x=\varphi$ and $ y = \log\cot\theta/2$ between the global and flat coordinates. We see that the dual spacetime to a state of energy $E(\theta) = -2\cos\theta/\lambda$ has deficit angle 
 \begin{equation}
2\pi-2\theta=\pi(1+ v).
 \end{equation}
 The thermal solution is supported by a localized energy source at the poles. We see that, as the parameter $v$ traverses its full range from 1 to -1, the deficit angle $2\theta$ goes from $2\pi$ to $0$. Note that the location of the gluing surface exhibits a symmetry under the $\mathbb{Z}_2$ operation of sending $\theta \to 2\pi-\theta$ and exchanging the interior and exterior geometry. In the SYK model, this $\mathbb{Z}_2$ operation flips the sign of the energy and temperature. As we will see, the same energy sign flip occurs in the gravity dual.

 \medskip
 
\medskip

\begin{figure}[hbtp]
  \centering
  % ---------- Row 1 ----------
  \subfloat{
    \resizebox{0.47\textwidth}{0.24\textheight}{\input{images/shockwave_with_line_2}}
    ~~~~~~~~~~~~~~~~~%
    \raisebox{8mm}{\includegraphics[width=0.24\textwidth]{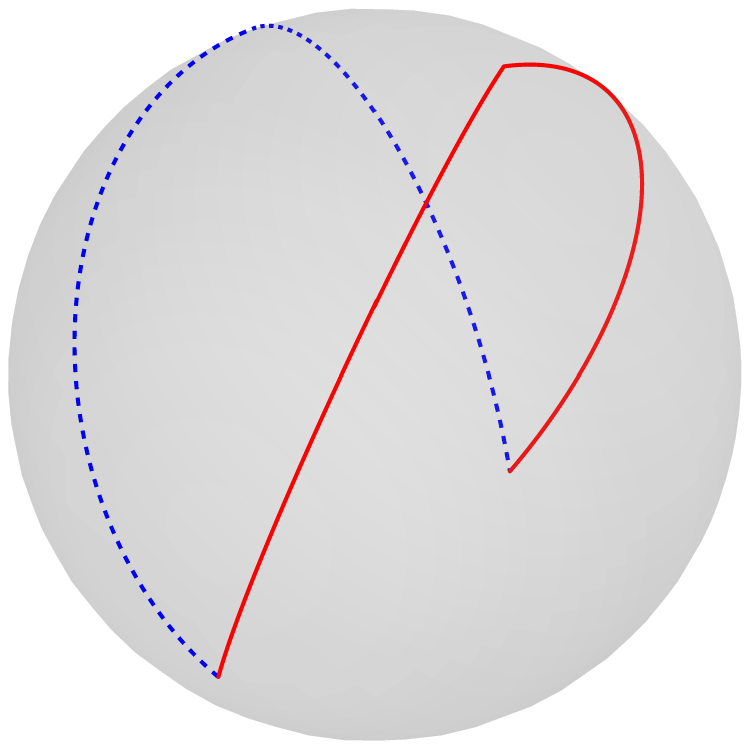}}
  }%

  \caption{The gluing surfaces $\Sigma$ and $\Sigma^*$ for the shockwave solution with $\hat\mu <0$.}
  \label{fig:membrane2b}
  \vspace{-2mm}
\end{figure}

\paragraph{Delta function source with $\hat{\mu}>0$.} Next, we consider the solution \eqref{eq:delta_function_mu} for a delta function source $\mu=-\hat\mu\,\delta(u)$ and finite temperature.
%\begin{equation}
%\label{eq:delta_function_mu}
 %   \psi(u)=\begin{cases}      &2\arctan\tanh\left(\cos\frac{\pi v}{2}\frac{(u-u_0)}{2} -i\frac{\pi}{4}(1-v) \right) + c\quad \text{for } u<0,\\       &2\arctan\tanh\left(\cos\frac{\pi v}{2}\frac{(u+u_0)}{2} -i\frac{\pi}{4}(1-v) \right)-c\quad \text{for } u>0,\\   \end{cases}
%\end{equation}
%with
%\begin{equation}
%    c=\text{Re}\{\arctan\tanh\left(\cos\frac{\pi v}{2}\frac{u_0}{2} -i\frac{\pi}{4}(1-v) \right)\}-\text{Re}\{\arctan\tanh\left(-\cos\frac{\pi v}{2}\frac{u_0}{2} -i\frac{\pi}{4}(1-v) \right)\}.
%\end{equation}
We expect this solution to describe the backreaction of a particle moving on a static trajectory, in the Schwarzschild-de Sitter spacetime with a conical deficit with some fixed temperature. We verify this expectation by plotting this solution in figure~\ref{fig:membrane2b}. Here, we chose the positive sign of the stress-energy source $\hat{\mu}$ to correspond to a worldline of a massive particle with positive mass as seen from the de Sitter region inside the wedge. Hence, the interior of the wedge should be interpreted as the physical region of the 3D spacetime, while the exterior region is removed by the gluing procedure that identifies the two sides of the wedge. On the SYK side, this regime corresponds to the positive temperature regime $0<v<1$. The green trajectory creates a conical defect on the inside and thus shortens the wedge. This makes it possible to send signals from the  bottom left corner to the  top right corner of the wedge.

\medskip

\begin{figure}[hbtp]
  \centering
  % ---------- Row 1 ----------
  \subfloat{
    \resizebox{0.5\textwidth}{0.255\textheight}{\input{images/shockwave_with_line}}
    ~~~~~~~~~~~~~~~~~%
    \raisebox{8mm}{\includegraphics[width=0.24\textwidth]{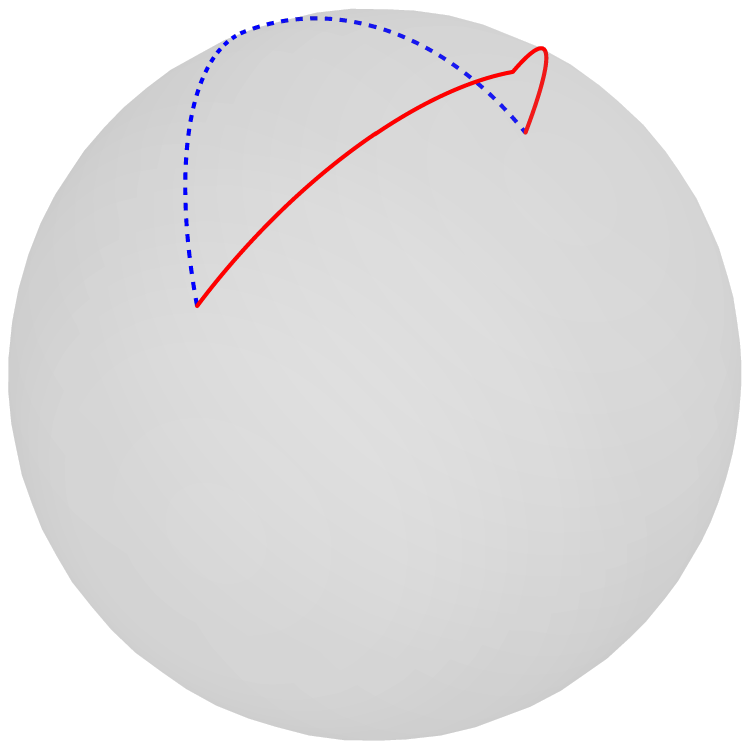}}
  }%

  \caption{The gluing surfaces $\Sigma$ and $\Sigma^*$ for the shockwave solution with $\hat\mu>0$.}
  \label{fig:membrane2}
\end{figure}

\paragraph{Delta function source with $\hat{\mu}<0$.} We can also choose the sign of $\hat{\mu}$ such that it corresponds to a particle with positive mass as seen from the de Sitter region outside of the wedge. On the SYK side, this regime corresponds to the negative temperature regime $-1<v<0$. The green trajectory creates a conical deficit on the outside and a conical excess on the inside and thus elongates the interior wedge. From the outside, the lower left corner and top right corner are timelike separated, so one can send a signal via the exterior geometry. Hence, for both signs of $\hat{\mu}$, the backreaction makes communication possible between the two poles.
\smallskip

\begin{figure}[h]
  \centering
  % ---------- Row 1 ----------
\subfloat{\includegraphics[width=0.47\textwidth,height=0.235\textheight]{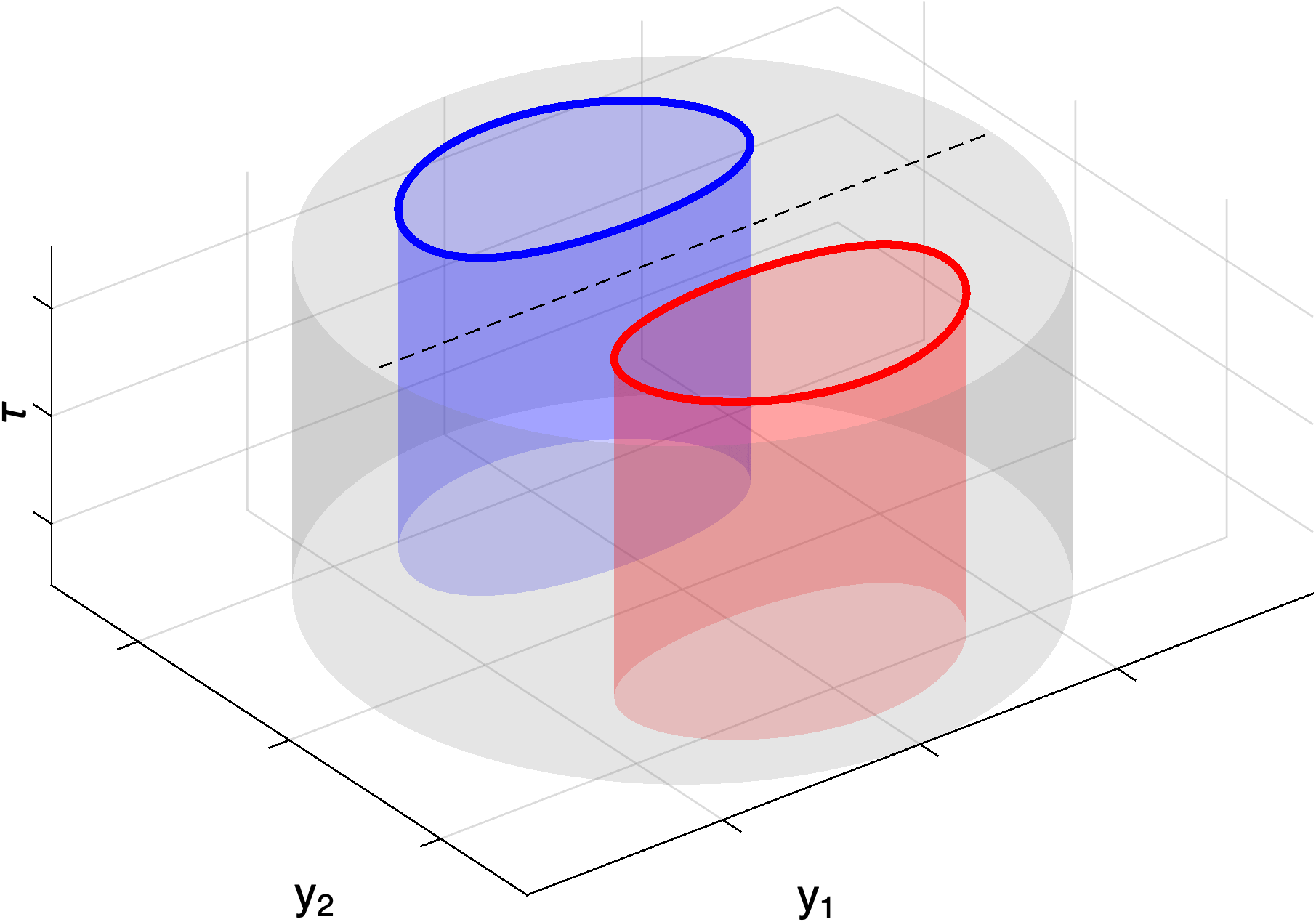}~~~~~~~~~~~~~~~~~~\raisebox{5mm}{\includegraphics[width=0.24\textwidth]{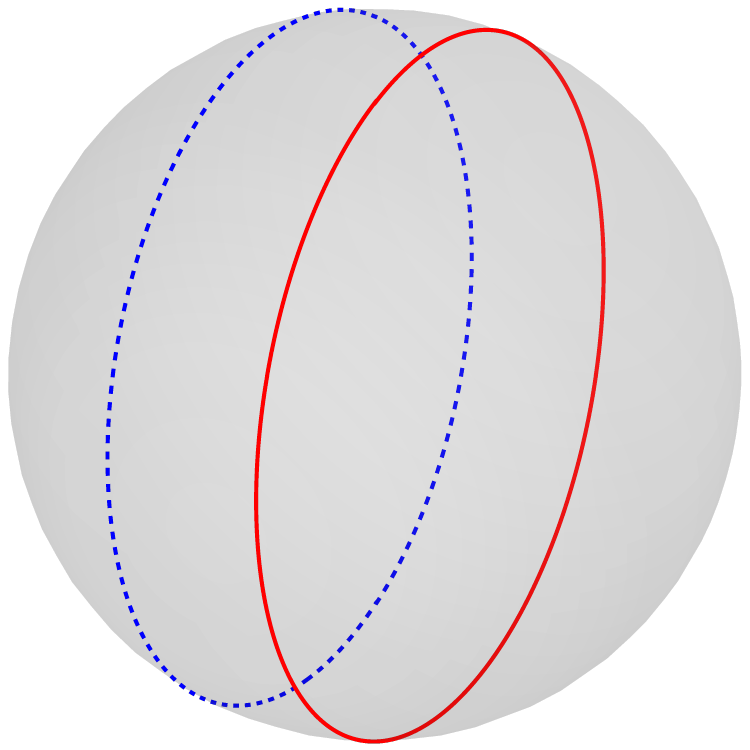}}}

  \caption{The gluing membranes $\Sigma$  and $\Sigma^*$ for the solution with constant $\hat{\mu}$.  }
  \label{fig:membrane3}
 \end{figure}

\paragraph{Constant $\mu$ solution.} Finally, we consider the constant $\mu$ solution. For a constant energy density, we expect to obtain the gravitational solution by cutting the spatial sphere on two circles parallel to the equator and gluing the two spherical caps together. In the $x, y$ coordinates, this corresponds to cutting on a constant $y$ surface. According to our dictionary, the corresponding solution for $\psi$ should have constant imaginary part. Indeed, we see that the constant $\mu$ solution in equation~\eqref{eq:constant_mu_solution_psi} 
%\begin{equation}
%    \psi(u) = \frac{1}{\sqrt{e^{2\phi_\mu}-1}}(u-u_0) -i\tanh^{-1}e^{-\phi_\mu}
%\end{equation}
has this property. We plot these curves and the resulting spacetime in figure~\ref{fig:membrane3}. Note that this special solution has no conical defects and consists of two disconnected trajectories. 
%We believe this is not a generic feature, but is related to the fine tuning of the SYK solution. 
%Upon Wick rotation to the two time signature, this solution describes an eternal AdS$_2$ spacetime.

\smallskip

\subsection{Schwarzian limit}

\vspace{-1.5mm}

The low energy limit of the SYK model is known to be dual to JT gravity on an AdS$_2$ spacetime. 
%In this section, we explain how to recover this limit in our setup.   The JT gravity action on a two dimensional manifold $\mathcal{M}$ with boundary $\partial\mathcal{M}$ reads \cite{JACKIW1985343, TEITELBOIM198341, almheiri2015modelsads2backreactionholography} 
%\begin{equation}
 %   S_{JT}= S_0 -\frac{1}{16\pi G_N}\int_{\mathcal{M}} d^2x\sqrt{g}\,\phi(R+2)  - \frac{1}{8\pi G_N}\int_{\partial\mathcal{M}} dx\sqrt{h}\,\phi_bK.  \label{eq:JT_action}
%\end{equation}
% Here $h$ and $K$ are respectively the induced metric and extrinsic curvature of the boundary and $\phi_b$ denotes the boundary value of the dilaton field $\phi$. The equation of motion of $\phi$ fix the metric to have constant curvature $R=-2$, so that the spacetime is a portion of $AdS_2$. 
Here we show how to recover this limit from our setup. We set $\mu=0$ for simplicity. Consider the $v\to 1$ ($\beta\to \infty$) limit of the thermal solution in equation~\eqref{eq:thermal_solution_psi}. In this limit, the $y=\text{Im}\psi$ location of the surface $\Sigma$ goes to zero, and the two sides $\Sigma=\Sigma^*$ coincide. To make contact with AdS JT gravity\footnote{If we do not swap the timelike and spacelike directions, we obtain dS JT gravity \cite{maldacena2025dimensionalnearlysittergravity}.} we momentarily swap the timelike and spacelike directions of our spacetime. The metric on the surface $\Sigma$, given in equation \eqref{eq:metric_membrane} thus becomes
\begin{equation}
    ds^2\to-ds^2 = d\tau^2 - \cosh^2\tau\tanh^2 y\;du^2,
\end{equation}
and can be recognized as the metric on AdS$_2$. We introduce global coordinates
\begin{equation}
    \sigma= \cot^{-1}\sinh\tau,\quad \nu(u)=\int^udu'\tanh y(u'), \qquad ds^2=\frac{-d\nu^2+d\sigma^2}{\sin^2\sigma}.
    \label{eq:metric_membrane_JT}
\end{equation}
The intersection between the AdS$_2$ slice and the cutoff \eqref{eq:scridef} determines the boundary curve of the AdS$_2$ spacetime\footnote{Note that the metric on the boundary of the surface $\Sigma$ is constant in $u$ once we fix the gauge \eqref{eq:gauge_choice}.}. Fluctuations of the $\Sigma$ embedding translate into fluctuations of the boundary curve. In terms of the canonical variable $\phi$, the boundary curve is
\begin{equation}
    \nu'(u)=\tanh y(u)=e^{-\phi(u)}\qquad\sigma(u)=\epsilon\nu'(u),
    \label{eq:bdy_trajectory}
\end{equation}
where we defined $\epsilon = 2 e^{-T}$. To recover the Schwarzian action, we first take the low energy limit, $y\sim0$, $p\sim 0$, of  the Hamiltonian~\eqref{eq:DSSYK_hamiltonian_coupling} \cite{Lin_2022} 
\begin{equation}
\label{eq:lham}
%16\pi G_NH=
-\sqrt{1-e^{-2\phi}}\cos p\sim\frac{p^2}{2} + \frac{e^{-2\phi}}{2}+\text{constant}.
\end{equation}
We recognize the  Liouville Hamiltonian of JT gravity \cite{Engels_y_2016,harlow2019factorizationproblemjackiwteitelboimgravity}. From equation \eqref{eq:bdy_trajectory} we derive
$e^{-2\phi}=\nu'(u)^2$ and $\phi' = -\frac{\nu''}{\nu'}$ and
a simple calculation shows that the Liouville action associated to \eqref{eq:lham} reduces to the Schwarzian action %\eqref{eq:schwartzian_action_JT}
\begin{equation}   
     S=\frac{1}{16\pi G_N}\int du\left(\frac{\phi'^2}{2}-\frac{e^{-2\phi}}{2}\right)
%=\frac{1}{16\pi G_N}\int du \left(\frac{1}{2}\left(\frac{\nu''}{\nu'}\right)^2-\frac{1}{2}\nu'^2\right) 
= -\frac{1}{16\pi G_N}\int du\left(\{\nu(u), u\}+\frac{1} {2}\nu'^2\right) .%+\text{bdy terms}.
\end{equation}
A geometric intuition of why the gravitational dynamics becomes two-dimensional in this limit is that the gluing surfaces $\Sigma$ and $\Sigma^*$ coincide and all boundary-to-boundary geodesics stay on the same 2D (anti-)de Sitter slice.
\bigskip

\begin{figure}[t]
\begin{center}

\resizebox{0.2\textwidth}{!}{\input{images/ds3_with_slice}}
\hspace{20mm} % push right
\raisebox{3mm}{%
\resizebox{0.55\textwidth}{!}{%
  \input{images/JT_limit_wick}
}}
        \caption{In the low energy limit, the two sides of the gluing surface $\Sigma = \Sigma^*$ coincide. We identify the surface $\Sigma$ with the AdS$_2$ geometry of JT gravity, after changing the overall sign of the metric. 
        %The cutoffs at $\mathscr{I}^\pm$ in the three dimensional spacetime translates into boundary curves (black lines on the right) in the $AdS_2$ geometry.
        }
   \label{fig:JT_limit}
  \end{center}
  \vspace{-2mm}
\end{figure}
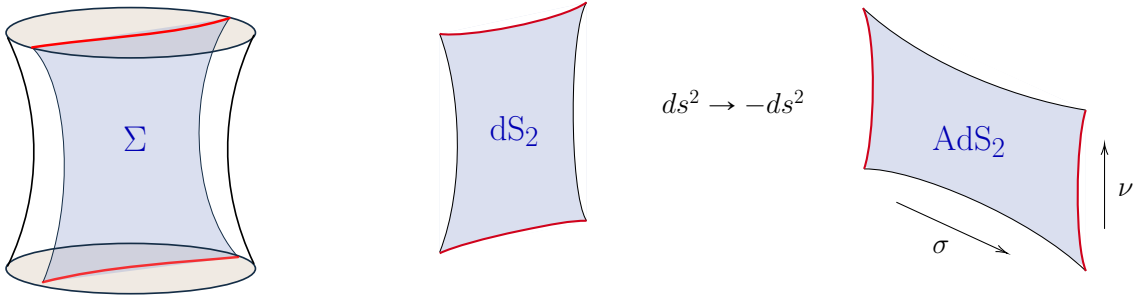 

%% file: images/cutds.tex
\tikzset{every picture/.style={line width=0.75pt}} %set default line width to 0.75pt        

\begin{tikzpicture}[x=0.75pt,y=0.75pt,yscale=-.65,xscale=.8]
\draw  [draw opacity=0][fill={rgb, 255:red, 251; green, 95; blue, 70 }  ,fill opacity=.5 ] (721.47,175.16) -- (721.47,478.54) -- (754.95,394.38) -- (754.95,91) -- cycle ;
%Shape: Parallelogram [id:dp41224028563120174] 
\draw  [draw opacity=0][fill={rgb, 255:red, 51; green, 95; blue, 200 }  ,fill opacity=.5 ] (527,176) -- (527,479) -- (662.47,395) -- (662.47,92) -- cycle ;
%Shape: Arc [id:dp12124068758301676] 
%\draw  [draw opacity=0][fill={rgb, 255:red, 155; green, 155; blue, 155 }  ,fill opacity=1 ] (145.98,205.39) .. controls (136.39,208.63) and (126.12,210.39) .. (115.43,210.39) .. controls (102.25,210.39) and (89.69,207.71) .. (78.27,202.87) -- (115.43,115.15) -- cycle ; \draw  [draw opacity=0] (145.98,205.39) .. controls (136.39,208.63) and (126.12,210.39) .. (115.43,210.39) .. controls (102.25,210.39) and (89.69,207.71) .. (78.27,202.87) ;  
%Shape: Rectangle [id:dp23294931672242736] 
%\draw  [draw opacity=0][fill={rgb, 255:red, 155; green, 155; blue, 155 }  ,fill opacity=1 ] (80.43,47.65) -- (150.43,47.65) -- (150.43,182.65) -- (80.43,182.65) -- cycle ;
%Shape: Boxed Bezier Curve [id:dp026662432462608354] 
%\draw [color={rgb, 255:red, 0; green, 0; blue, 0 }  ,draw opacity=1 ][line width=0.75]    (19.86,116.52) .. controls (54.85,204.5) and (56.87,252.88) .. (19.86,347.41) ;
%Shape: Boxed Bezier Curve [id:dp5896788884007254] 
%\draw [color={rgb, 255:red, 0; green, 0; blue, 0 }  ,draw opacity=1 ][line width=0.75]    (210.36,347.41) .. controls (174.94,260.35) and (172.74,212.4) .. (209.52,118.54) ;
%Shape: Parallelogram [id:dp9785224768978394] 
\draw  [fill={rgb, 255:red, 237; green, 237; blue, 237 }  ,fill opacity=.5 ] (550.6,91) -- (868,91) -- (788.4,175) -- (460,175) -- cycle ;
\draw  [draw opacity=0][fill={rgb, 255:red, 128; green, 128; blue, 128 }  ,fill opacity=1 ] (662.95,91) -- (756.4,91) -- (621.78,174.77) -- (528.32,174.77) -- cycle ;
%Shape: Triangle [id:dp8180074701488836] 
\draw  [draw opacity=0][fill={rgb, 255:red, 128; green, 128; blue, 128 }  ,fill opacity=1 ] (723.63,92.19) -- (723.63,174.77) -- (621.78,174.77) -- cycle ;
%Shape: Triangle [id:dp8431834524697952] 
\draw  [draw opacity=0][fill={rgb, 255:red, 128; green, 128; blue, 128 }  ,fill opacity=1 ] (722.72,173.16) -- (717.5,91) -- (755.63,91) -- cycle ;
%Curve Lines [id:da33696863245108266] 
\draw [color={rgb, 255:red, 208; green, 2; blue, 27 }  ,draw opacity=1 ][fill={rgb, 255:red, 128; green, 128; blue, 128 }  ,fill opacity=1 ][line width=1.5]    (721.47,175.54) .. controls (756.63,132) and (741.63,111) .. (754.95,91) ;
%Curve Lines [id:da3483785067591947] 
\draw [color={rgb, 255:red, 8; green, 2; blue, 200 }  ,draw opacity=1 ][fill={rgb, 255:red, 128; green, 128; blue, 128 }  ,fill opacity=1 ][line width=1.5]    (527,176) .. controls (560.95,111) and (625.55,119.04) .. (662.95,91) ;
%Straight Lines [id:da5222364951819702] 
\draw[->]    (645.34,134.5) -- (722.1,74.73) ;
%Straight Lines [id:da3062179491129744] 
\draw[->]    (645.34,134.5) -- (789.34,134.5) ;
%Straight Lines [id:da8451631871859975] 
\draw[->]    (463,449) -- (463,226) ;

% Text Node
\draw (587,25) node [anchor=north west][inner sep=0.75pt]  [font=\LARGE]  {$\Psi (u)$};
% Text Node
\draw (765,25) node [anchor=north west][inner sep=0.75pt]  [font=\LARGE]  {$\Psi ^{*}( u)$};
% Text Node
\draw (588,240) node [anchor=north west][inner sep=0.75pt]  [font=\huge]  {\textcolor{\darkblue}{$\Sigma$}};
% Text Node
\draw (758,240) node [anchor=north west][inner sep=0.75pt]  [font=\huge]  {\textcolor{\darkred}{$\Sigma^{*}$}};
% Text Node
\draw (716,38.4) node [anchor=north west][inner sep=0.75pt]  [font=\LARGE]  {$x$};
% Text Node
\draw (800,115.4) node [anchor=north west][inner sep=0.75pt]  [font=\LARGE]  {$y$};
% Text Node
\draw (442,201.4) node [anchor=north west][inner sep=0.75pt]  [font=\LARGE]  {$\tau $};

\end{tikzpicture}

%% file: images/dS3_cut.tex
\begin{tikzpicture}[xscale=1.33, yscale=1.17];
%\path[draw= cyan, line width=0.01mm, snake it] (2,1.5) arc (40:-40:2.2cm);
%\path[draw=blue,very thick,<-] (5.32,.06) arc (179:181:1cm);
\draw[thick,dashed,gray] (2.5,-2) -- (4,-0.5);
\draw[thick,dashed,gray] (5.5,-2) -- (4,-0.5);
\draw[thick,dashed,gray] (2.5,2) -- (4,0.5);
\draw[thick,dashed,gray] (5.5,2) -- (4,0.5);
\draw[very thick,purple] (4,-2) -- (4,2);
\draw[lightgray,fill=lightgray,opacity=.75] (2.5,1.65) -- (4,1.93) --(4,2) -- (2.5,2) -- cycle;
\draw[lightgray,fill=lightgray,opacity=.75] (2.5,-1.65)-- (4,-1.93) -- (4,-2) -- (2.5,-2) -- cycle;
\draw[lightgray,fill=lightgray,opacity=.75] (5.5,1.65) -- (4,1.93) -- (4,2) -- (5.5,2) -- cycle;
\draw[lightgray,fill=lightgray,opacity=.75] (5.5,-1.65) -- (4,-1.93) -- (4,-2) -- (5.5,-2) -- cycle;
%\filldraw[\darkblue] (6,-1.55) circle (2pt);
%\draw[thick, \darkblue] (6,1.5) circle (2pt);
\draw (3.6,0) node {\rotatebox{0}{\textcolor{\darkblue}{\Large
$\Sigma$}}};
\draw (4.45,0) node {\rotatebox{0}{\textcolor{\darkred}{\Large
$\Sigma^*$}}};
\draw (4,-2.35) node {\rotatebox{0}{{\large
$\mathscr{I}^-$}}};
\draw (4,2.35) node {\rotatebox{0}{{\large
$\mathscr{I}^+$}}};
\draw[gray, thick] (2.5,-2) -- (5.5,-2);
\draw[gray, thick] (2.5,2) -- (5.5,2);
\draw[thick,gray] (5.5,-2) -- (5.5,2);
\draw[thick,gray] (2.5,-2) -- (2.5,2);
\draw (3,1.86) node {\rotatebox{0}{{\scriptsize
$k=-1$}}};
\draw (3,-1.86) node {\rotatebox{0}{{\scriptsize
$k=-1$}}};
\draw (5,1.86) node {\rotatebox{0}{{\scriptsize
$k=-1$}}};
\draw (5,-1.86) node {\rotatebox{0}{{\scriptsize
$k=-1$}}};
\path[draw=black,thick, fill=lightgray,opacity=.75] (2.5,1.65) arc (270:292:4cm);
\path[draw=black, fill=lightgray,opacity=.75, thick] (5.5,1.65) arc (270:248:4cm);
\path[draw=black, fill=lightgray,opacity=.75, thick] (2.5,-1.65) arc (90:68:4cm);
\path[draw=black, fill=lightgray,opacity=.75, thick] (5.5,-1.65) arc (90:112:4cm);
\end{tikzpicture}

%% file: images/shockwave_with_line.tex
\tikzset{every picture/.style={line width=0.75pt}} %set default line width to 0.75pt        

\begin{tikzpicture}[x=0.75pt,y=0.75pt,yscale=-.9, xscale=1]
%uncomment if require: \path (0,373); %set diagram left start at 0, and has height of 373

%Image [id:dp6968048587767709] 
\draw (329.22,182.69) node  {\includegraphics[width=343.82pt,height=220.53pt]{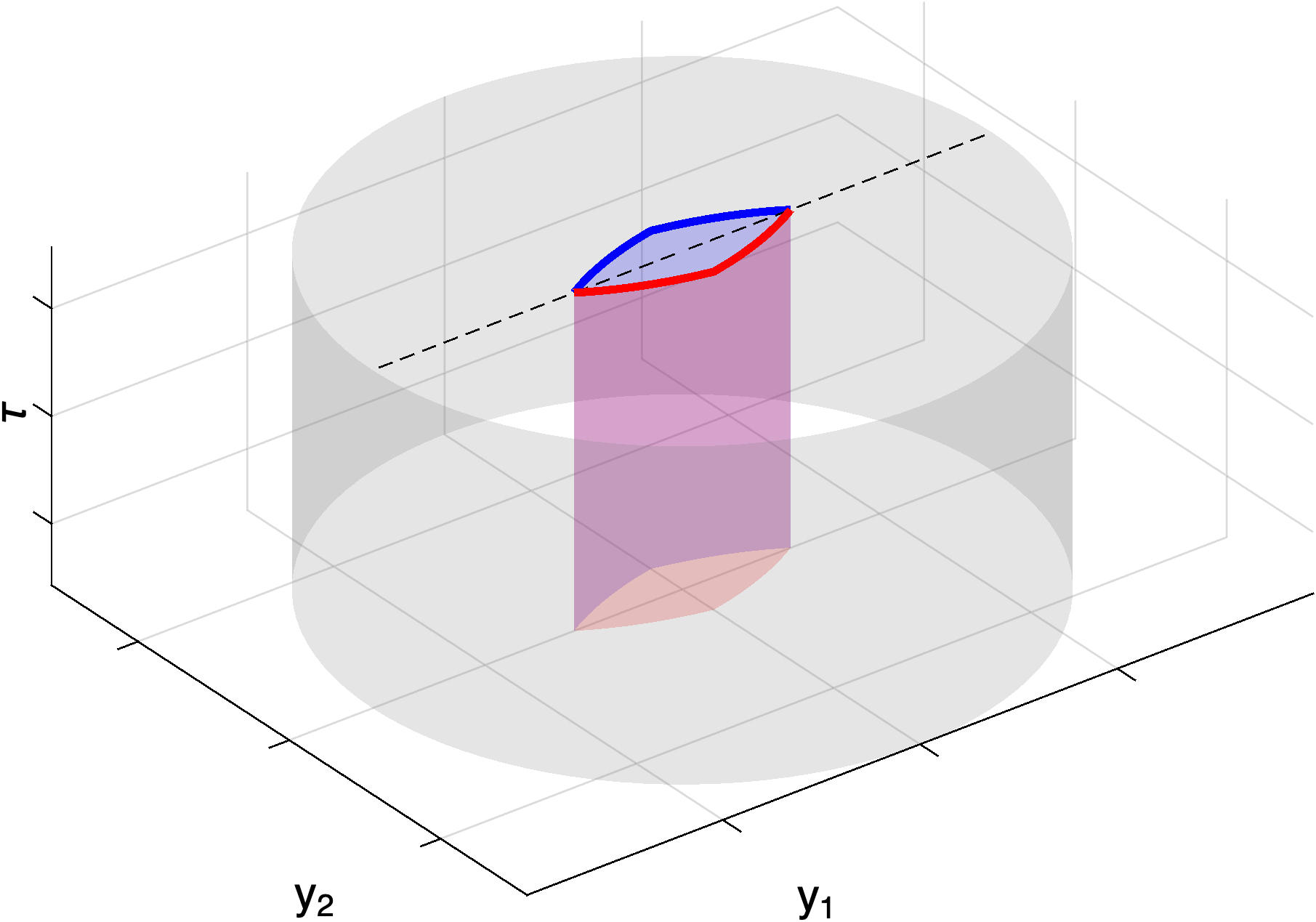}};
%Straight Lines [id:da42116401875726905] 
\draw [color={rgb, 255:red, 126; green, 211; blue, 33 }  ,draw opacity=1 ][fill={rgb, 255:red, 126; green, 211; blue, 33 }  ,fill opacity=1 ][line width=2.25]    (347,118) -- (347,234) ;
%Shape: Circle [id:dp9505617157438597] 
\draw  [draw opacity=0][fill={rgb, 255:red, 126; green, 211; blue, 33 }  ,fill opacity=1 ] (343,118) .. controls (343,115.79) and (344.79,114) .. (347,114) .. controls (349.21,114) and (351,115.79) .. (351,118) .. controls (351,120.21) and (349.21,122) .. (347,122) .. controls (344.79,122) and (343,120.21) .. (343,118) -- cycle ;
%Shape: Circle [id:dp9408549092932629] 
\draw  [draw opacity=0][fill={rgb, 255:red, 126; green, 211; blue, 33 }  ,fill opacity=1 ] (343,234) .. controls (343,231.79) and (344.79,230) .. (347,230) .. controls (349.21,230) and (351,231.79) .. (351,234) .. controls (351,236.21) and (349.21,238) .. (347,238) .. controls (344.79,238) and (343,236.21) .. (343,234) -- cycle ;
%Straight Lines [id:da7442950902677548] 
\draw [color={rgb, 255:red, 126; green, 211; blue, 33 }  ,draw opacity=1 ][fill={rgb, 255:red, 126; green, 211; blue, 33 }  ,fill opacity=1 ][line width=2.25]  [dash pattern={on 6.75pt off 4.5pt}]  (328,103) -- (328,219) ;
%Shape: Circle [id:dp5827812462949915] 
\draw  [draw opacity=0][fill={rgb, 255:red, 126; green, 211; blue, 33 }  ,fill opacity=1 ] (324,103) .. controls (324,100.79) and (325.79,99) .. (328,99) .. controls (330.21,99) and (332,100.79) .. (332,103) .. controls (332,105.21) and (330.21,107) .. (328,107) .. controls (325.79,107) and (324,105.21) .. (324,103) -- cycle ;
%Shape: Circle [id:dp5673652543394301] 
\draw  [draw opacity=0][fill={rgb, 255:red, 126; green, 211; blue, 33 }  ,fill opacity=1 ] (324,219) .. controls (324,216.79) and (325.79,215) .. (328,215) .. controls (330.21,215) and (332,216.79) .. (332,219) .. controls (332,221.21) and (330.21,223) .. (328,223) .. controls (325.79,223) and (324,221.21) .. (324,219) -- cycle ;

\end{tikzpicture}

%% file: images/shockwave_with_line_2.tex
\tikzset{every picture/.style={line width=0.75pt}} %set default line width to 0.75pt        

\begin{tikzpicture}[x=0.75pt,y=0.75pt,yscale=-.9, xscale=1]
%uncomment if require: \path (0,373); %set diagram left start at 0, and has height of 373

%Image [id:dp6968048587767709] 
\draw (329.22,182.69) node  {\includegraphics[width=343.82pt,height=220pt]{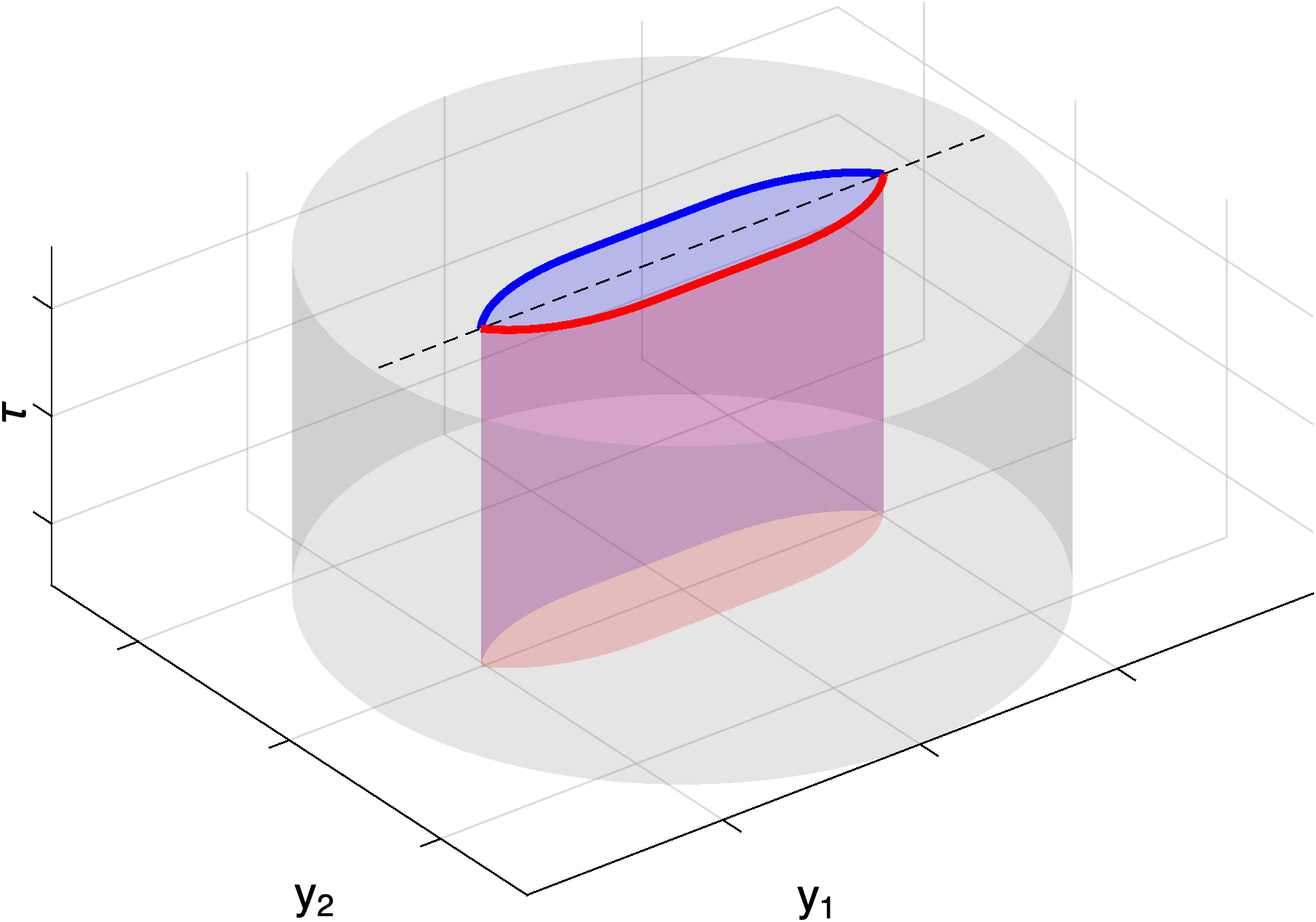}};
%Straight Lines [id:da42116401875726905] 
\draw [color={rgb, 255:red, 126; green, 211; blue, 33 }  ,draw opacity=1 ][fill={rgb, 255:red, 126; green, 211; blue, 33 }  ,fill opacity=1 ][line width=2.25]    (347,118) -- (347,234) ;
%Shape: Circle [id:dp9505617157438597] 
\draw  [draw opacity=0][fill={rgb, 255:red, 126; green, 211; blue, 33 }  ,fill opacity=1 ] (343,118) .. controls (343,115.79) and (344.79,114) .. (347,114) .. controls (349.21,114) and (351,115.79) .. (351,118) .. controls (351,120.21) and (349.21,122) .. (347,122) .. controls (344.79,122) and (343,120.21) .. (343,118) -- cycle ;
%Shape: Circle [id:dp9408549092932629] 
\draw  [draw opacity=0][fill={rgb, 255:red, 126; green, 211; blue, 33 }  ,fill opacity=1 ] (343,234) .. controls (343,231.79) and (344.79,230) .. (347,230) .. controls (349.21,230) and (351,231.79) .. (351,234) .. controls (351,236.21) and (349.21,238) .. (347,238) .. controls (344.79,238) and (343,236.21) .. (343,234) -- cycle ;
%Straight Lines [id:da7442950902677548] 
\draw [color={rgb, 255:red, 126; green, 211; blue, 33 }  ,draw opacity=1 ][fill={rgb, 255:red, 126; green, 211; blue, 33 }  ,fill opacity=1 ][line width=2.25]  [dash pattern={on 6.75pt off 4.5pt}]  (328,103) -- (328,219) ;
%Shape: Circle [id:dp5827812462949915] 
\draw  [draw opacity=0][fill={rgb, 255:red, 126; green, 211; blue, 33 }  ,fill opacity=1 ] (324,103) .. controls (324,100.79) and (325.79,99) .. (328,99) .. controls (330.21,99) and (332,100.79) .. (332,103) .. controls (332,105.21) and (330.21,107) .. (328,107) .. controls (325.79,107) and (324,105.21) .. (324,103) -- cycle ;
%Shape: Circle [id:dp5673652543394301] 
\draw  [draw opacity=0][fill={rgb, 255:red, 126; green, 211; blue, 33 }  ,fill opacity=1 ] (324,219) .. controls (324,216.79) and (325.79,215) .. (328,215) .. controls (330.21,215) and (332,216.79) .. (332,219) .. controls (332,221.21) and (330.21,223) .. (328,223) .. controls (325.79,223) and (324,221.21) .. (324,219) -- cycle ;

\end{tikzpicture}

%% file: images/ds3_with_slice.tex
\begin{tikzpicture}[xscale=.92,yscale=.89,line cap=round,line join=round]
  \definecolor{outline}{RGB}{20,45,70}
  \definecolor{capfill}{RGB}{236,228,219}
  \definecolor{slicefill}{RGB}{150,165,210}
  \definecolor{greenline}{RGB}{60,135,55}
  \def\yt{2.25}
  \def\yb{-2.25}
  \def\rx{2.3}
  \def\ry{0.48}
  \def\hx{1.20}
  \def\hy{2.75}
  \def\ang{55}
  \coordinate (A) at (-1.82,\yt-0.28);   % top-left
  \coordinate (B) at ( 1.82,\yt+0.28);   % top-right
  \coordinate (C) at (-1.62,\yb-0.26);   % bottom-left
  \coordinate (D) at ( 1.98,\yb+0.22);   % bottom-right
  \fill[capfill,opacity=.78] (0,\yt) ellipse[x radius=\rx,y radius=\ry];
  \fill[capfill,opacity=.78] (0,\yb) ellipse[x radius=\rx,y radius=\ry];
  \fill[slicefill,opacity=.35]
    (A)
      .. controls (-1.12,1.25) and (-1.02,-1.05) ..
    (C)
      -- (D)
      .. controls (1.08,-1.00) and (1.02,1.30) ..
    (B)
      -- cycle;
  \draw[outline,thick] (0,\yt) ellipse[x radius=\rx,y radius=\ry];
  \draw[outline,thick] (0,\yb) ellipse[x radius=\rx,y radius=\ry];

  \draw[outline]
    (A) .. controls (-1.12,1.25) and (-1.02,-1.05) .. (C);
  \draw[outline]
    (B) .. controls (1.02,1.30) and (1.08,-1.00) .. (D);
  \draw[red,very thick]
    (A) .. controls (-0.55,\yt-0.08) and (0.85,\yt-0.05) .. (B);
  \draw[red,very thick,opacity=.78]
    (C) .. controls (-0.45,\yb+0.05) and (0.95,\yb+0.12) .. (D);

\path[draw=black,  thick] (-2.27,-2.2) arc (-25:25:5.15cm);
\path[draw=black,  thick] (2.27,2.2) arc (155:205:5.15cm);

  \node[text=blue!70!black] at (0.08,0.22) {\LARGE $\Sigma$};
\end{tikzpicture}

%% file: images/JT_limit_wick.tex
\tikzset{every picture/.style={line width=0.75pt}} %set default line width to 0.75pt        

\begin{tikzpicture}[x=0.75pt,y=0.75pt,yscale=-1,xscale=1]
%uncomment if require: \path (0,380); %set diagram left start at 0, and has height of 380

%Straight Lines [id:da1937308898301574] 
\draw    (476.82,269.6) -- (581.5,317.8) ;
\draw [shift={(583.32,318.64)}, rotate = 204.72] [color={rgb, 255:red, 0; green, 0; blue, 0 }  ][line width=0.75]    (10.93,-3.29) .. controls (6.95,-1.4) and (3.31,-0.3) .. (0,0) .. controls (3.31,0.3) and (6.95,1.4) .. (10.93,3.29)   ;
%Straight Lines [id:da7237986525233786] 
\draw    (676.72,295.23) -- (676.72,215.08) ;
\draw [shift={(676.72,213.08)}, rotate = 90] [color={rgb, 255:red, 0; green, 0; blue, 0 }  ][line width=0.75]    (10.93,-3.29) .. controls (6.95,-1.4) and (3.31,-0.3) .. (0,0) .. controls (3.31,0.3) and (6.95,1.4) .. (10.93,3.29)   ;
%Shape: Parallelogram [id:dp16256257468658275] 
\draw  [draw opacity=0][fill={rgb, 255:red, 150; green, 165; blue, 210 }  ,fill opacity=0.35 ] (38.33,107.83) -- (38.33,317.67) -- (179.32,287.41) -- (179.32,77.58) -- cycle ;
%Curve Lines [id:da8842686695061117] 
\draw [fill={rgb, 255:red, 255; green, 255; blue, 255 }  ,fill opacity=1 ]   (38.33,107.83) .. controls (58.94,163.17) and (62.87,254.94) .. (38.33,317.67) ;
%Curve Lines [id:da0005855360853149749] 
\draw [fill={rgb, 255:red, 255; green, 255; blue, 255 }  ,fill opacity=1 ]   (179.32,77.58) .. controls (162.25,105.52) and (162.25,260.35) .. (179.32,287.41) ;
%Curve Lines [id:da5176484725905033] 
\draw [color={rgb, 255:red, 208; green, 2; blue, 27 }  ,draw opacity=1 ][fill={rgb, 255:red, 255; green, 255; blue, 255 }  ,fill opacity=1 ][line width=1.5]    (38.33,107.83) .. controls (62.55,111.83) and (153.83,95.58) .. (179.32,77.58) ;
%Curve Lines [id:da442836811157855] 
\draw [color={rgb, 255:red, 208; green, 2; blue, 27 }  ,draw opacity=1 ][fill={rgb, 255:red, 255; green, 255; blue, 255 }  ,fill opacity=1 ][line width=1.5]    (38.33,318.67) .. controls (50.71,312.62) and (83.12,303.73) .. (115.39,297.08) .. controls (147.67,290.43) and (163.95,287.33) .. (179.32,287.41) ;
%Shape: Parallelogram [id:dp10059529664518951] 
\draw  [draw opacity=0][fill={rgb, 255:red, 150; green, 165; blue, 210 }  ,fill opacity=0.35 ] (658.67,181.07) -- (658.67,330.93) -- (445.85,234.78) -- (445.85,84.93) -- cycle ;
%Curve Lines [id:da6389322437354911] 
\draw [fill={rgb, 255:red, 255; green, 255; blue, 255 }  ,fill opacity=1 ]   (658.67,181.07) .. controls (579,168) and (495,130) .. (444.83,83) ;
%Curve Lines [id:da012275941726005657] 
\draw [fill={rgb, 255:red, 255; green, 255; blue, 255 }  ,fill opacity=1 ]   (658.82,335.67) .. controls (605,284) and (494,237) .. (445.82,237.6) ;
%Curve Lines [id:da47731511853597597] 
\draw [color={rgb, 255:red, 208; green, 2; blue, 27 }  ,draw opacity=1 ][fill={rgb, 255:red, 255; green, 255; blue, 255 }  ,fill opacity=1 ][line width=1.5]    (658.67,181.07) .. controls (648.82,211.87) and (647.82,309.79) .. (658.82,335.67) ;
%Curve Lines [id:da5269652532725826] 
\draw [color={rgb, 255:red, 208; green, 2; blue, 27 }  ,draw opacity=1 ][fill={rgb, 255:red, 255; green, 255; blue, 255 }  ,fill opacity=1 ][line width=1.5]    (444.83,83) .. controls (458.82,112.43) and (455.64,210.5) .. (445.82,237.6) ;

% Text Node
\draw (510,305) node [anchor=north west][inner sep=0.75pt]  [font=\huge]  {$\sigma $};
% Text Node
\draw (688,250) node [anchor=north west][inner sep=0.75pt]  [font=\huge]  {$\nu $};
% Text Node
\draw (250,160) node [anchor=north west][inner sep=0.75pt]  [font=\huge]  {$ds^{2}\rightarrow -ds^{2}$};
% Text Node
\draw (85,185) node [anchor=north west][inner sep=0.75pt]  [font=\Huge, color=blue!70!black] [align=left] {dS$\displaystyle _{2}$};
% Text Node
\draw (510,195) node [anchor=north west][inner sep=0.75pt]  [font=\Huge, color=blue!70!black] [align=left] {AdS$\displaystyle _{2}$};

\end{tikzpicture}

%% file: sections/boundary_thermodynamics.tex
\vspace{-1mm}
In this section, we  study the thermodynamic properties of $2+1$ near-de Sitter gravity. 
%We have shown how the reparametrization mode $\psi(u)$ describes the effective dynamics of 3D near-de Sitter gravity and how its effective action can be derived from the 3D Einstein-Hilbert action. 
We describe how one can recover the SYK entropy from the 1D effective action and how it relates to the Gibbons-Hawking entropy \cite{Gibbons:1976ue, Gibbons:1977mu}
of the 3D Schwarzschild-de Sitter spacetime with opening angle $2\theta$. 
 The SYK and GH entropies are similar but different
\begin{eqnarray}
\label{eq:syk-gh-entropy}
S_{\rm SYK}(\theta) =  \frac{2\pi\theta -2\theta^2}{\lambda},
\qquad \qquad  S_{GH}(\theta) = \frac{2\theta}{4G_N},
\end{eqnarray}
and it will therefore be instructive to see how both arise from a gravity computation. After a short review of the 3D Schwarzschild-de Sitter geometry in section \ref{subsec:3DSdS_thermodynamics}, we explain in section \ref{subsec:SYK_entropy_dS3} why our choice of $k=-1$ conformal boundary conditions leads to a different notion of gravitational entropy and energy than the one that follows from the standard Gibbons-Hawking setup. 
Finally, in section \ref{subsec:hayward_time}, we give a gravitational interpretation of the fake temperature in SYK by studying the flow generated by the Hayward Hamiltonian we derived in the previous section.

\subsection{3D Schwarzschild-de Sitter thermodynamics}\label{subsec:3DSdS_thermodynamics}
\vspace{-1mm}

As a warm-up, let us briefly recall some basic properties of the Euclidean 3D Schwarzschild-de Sitter (SdS) geometry.
In static coordinates, 3D SdS spacetime is described by the standard Euclidean 3D de Sitter metric %\cite{Spradlin:2001pw}
\begin{eqnarray} \label{lsdsone}
   ds^2\!\is \!  (1 - r^2) d\tau + \frac{dr^2}{1 - r^2} 
 + r^2  d\varphi^2, 
\end{eqnarray}
where the angular coordinate $\varphi$ exhibits the presence of a conical defect, while smoothness of the cosmological horizon implies that the Euclidean time $\tau$ is periodic with $\beta=2\pi$:
\begin{equation}
\label{periods}
\varphi \sim \varphi + 2 \theta ,\qquad \qquad {\tau \sim \tau + 2\pi}\, .
\end{equation}
The conical singularity is sourced by a point mass $M$  related to the opening angle $2\theta$ by \cite{Spradlin:2001pw}
\begin{eqnarray}
\label{epsirel}
%2\pi - \alpha = 2\pi \sqrt{1-8G_NM} \qquad \quad 
2\pi  M 
%= \int_{\Sigma_0} \!\!\sqrt{h} \spc T_{BY}%- M_0)  
% = \frac{1}{16\pi G_N} \bigl(4\pi \alpha  - \alpha^2\bigr) 
=  \frac{\pi}{4 G_N}- \frac{\theta^2}{4\pi G_N}. %\qquad \ \ 2\pi M_0 \equiv %\frac{\pi}{4 G_N}
\end{eqnarray}
One can recover this relationship between the mass $M$ and the opening angle by excising a small tube around the conical defect  and integrating the normal component\footnote{
Here $n^a$ denotes the (normalized) normal vector to the tubular surface $\Sigma_0$.} $T_{BY} = n^a n^b \spc T_{ab}^{BY}$
of the Brown-York stress tensor $
T^{BY}_{ab} = - \frac{1}{8\pi G_N}\bigl(  K_{ab} - K h_{ab}  - {h_{ab}} \bigr)$
around the tubular surface $\Sigma_0$ surrounding the defect.

Following Gibbons and Hawking \cite{Gibbons:1976ue}, we would like to identify the classical Einstein-Hilbert action evaluated on the SdS geometry \eqref{lsdsone} as the semiclassical answer for the free energy of the microscopic quantum theory. Let us assume that this dual quantum system lives on a worldline situated at the defect. This observer uses $\tau$ as proper time. We then read off from the time periodicity in \eqref{periods} that the observer experiences a finite temperature environment with an inverse de Sitter temperature $\beta_{\rm dS} = 2\pi$. Note that this temperature is independent of the opening angle and the total mass $M$. This suggests that the gravitational free energy should include, in addition to the GH entropy, a term $\beta_{\rm dS} E_{\rm BY} = 2\pi M$.

\begin{figure}[t]
  \centering
{\includegraphics[width=0.85\textwidth]{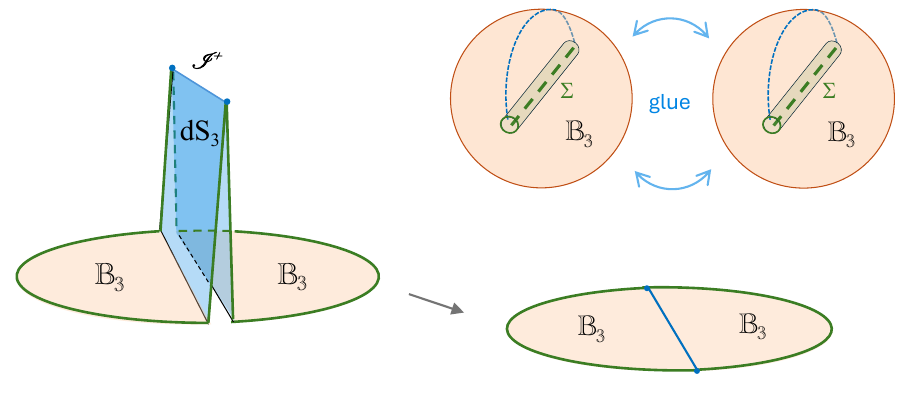}}\
  \caption{Schematic depiction of the standard Euclidean gravity path integral of the Schwarzschild-de Sitter spacetime. The left panel shows the overlap of two Hartle-Hawking wave functions prepared by the tan (Euclidean) and blue (Lorentzian) regions $\mathbb{B}_3$ and dS$_3$. The Lorentzian parts cancel in the overlap, leaving a three sphere obtained by gluing together two three balls $\mathbb{B}_3$ with a small tube around the conical defect removed. The action integral over the 3D volume gives the GH entropy, while the 2D surface integral over the tube gives $\beta_{\rm dS} = 2\pi$ times the Brown-York energy. The thin dashed blue arcs indicate that the opening angle of the SdS spacetime is $2\theta$.}
  \label{fig:ds-glue-one}
 \vspace{-2mm}
 \end{figure}

Following this hint, let us define the total (regulated) gravitational action as follows \cite{Banihashemi_2023}
\begin{eqnarray}
I_{\rm total} = I_{EH} + I_{BY},
\end{eqnarray}
where the Einstein-Hilbert term is evaluated over the three-sphere with a conical defect with opening angle $2\theta$, and with a small tube around the defect removed, and $I_{BY}$ is the Brown-York energy contribution 
\begin{eqnarray}
I_{EH} \is -\frac{1}{16\pi G_N} \int_{S_3} \!\!\sqrt{g}\spc (R\spc -\Lambda) \; = \; \frac{2\theta}{4 G_N} \\[3mm]
I_{BY} \is \int_{\Sigma_0}\! \!\!\sqrt{h} \spc T_{BY} = 2\pi M = \frac{\pi}{4 G_N}- \frac{\theta^2}{4\pi G_N}.
\end{eqnarray}
The first term is essentially the volume of the Euclidean SdS geometry, the three-sphere with the azimuthal angle $\varphi$ restricted to the interval $[0, 2\theta]$. 

Via the thermodynamic interpretation of $I_{\rm total}$ with $\beta_{\rm dS}$ times the free energy, the bulk term $I_{EH}$ is identified with the GH entropy of the SdS spacetime, while the surface term $I_{BY}$ gives the energy contribution\footnote{Note that, in spite of the suggestive notation, the formulas for $M$, $\beta_{\rm dS}$ and $S_{GH}$ do not obey the first law. Instead, imposing the relation
$dM = T_{GH} dS_{GH}$ leads to the standard formula  $\beta_{GH}  = {4\pi^2}/{2\theta}$ for the GH temperature of SdS$_3$.}
\begin{eqnarray}
- I_{\rm total} = S_{GH}  - \beta_{\rm dS}  M, \qquad \ \ S_{GH} = \frac{2\theta}{4G_N}, \qquad \beta_{\rm dS} = 2\pi.
\end{eqnarray}
In section \ref{subsec:SYK_entropy_dS3}, we will repeat this analysis for 3D near-de Sitter gravity. A first suggestive observation \cite{tietto2025microscopicmodelsitterspacetime} is that the total Euclidean gravity action  \begin{eqnarray}
-I_{\rm total} = \frac{1}{4 \pi G_N} \bigl(\pi^2 + 2\pi \theta - \theta^2 \bigr),
\end{eqnarray}
exactly coincides with the semiclassical formula \eqref{eq:syk-gh-entropy}
for the SYK entropy with $\lambda = 4\pi G_N$. Note, however, that in the above discussion $-I_{\rm total}$ is not an entropy but a free energy. Nonetheless, as we will see, this match will play a key role in the following.

\subsection{SYK entropy from dS$_3$ gravity}\label{subsec:SYK_entropy_dS3}
\vspace{-1mm}

We have shown that the three-dimensional Einstein-Hilbert action with near-de Sitter boundary conditions reduces to the 1D effective action given in \eqref{action+hamiltonian} and \eqref{psiparam} for the boundary trajectory $\psi(u)$.
%\begin{eqnarray} \label{eq:onedsefftwo}
%&&\hspace{-3mm} S_{EH}[\psi] \, =\, \frac{1}{2\pi G_N}\int \! du\,  \bigl(p'\phi - H\bigr)\notag \\[-1mm]\\[-1mm]\notag
%H\! \is\! -\sqrt{1-e^{-2\phi}}\cos p, \qquad\quad\ \psi' = \frac{e^{ip}} {\sqrt{e^{2\phi}-1}}.
%\end{eqnarray} Here we turned off the MQ coupling. 
%Hence the above 1D action describe the boundary dynamics of an empty near-de Sitter space time. 
This action is equivalent to the q-Schwarzian action  \cite{Blommaert:2024ydx} derived from 2D sine-dilaton gravity. Our interpretation in pure 3D Einstein gravity adds some important insight into the boundary conditions that instruct the quantization of the 1D model. 
In particular, our dictionary implies that $p$ is naturally periodic with $2\pi$ and that $\phi$ is restricted to take only positive values
\begin{equation}
\label{eq:posphi}
%\boxed{\    
\phi(u)\geq 0. \, %\tiny {}_{\strut}^{\strut}}
\end{equation}
This positivity constraint will play an important role in what follows.\footnote{In the quantum theory, the spectrum of $\phi$ is discrete. In SYK, $\phi$ is related to the chord number $n$ via $\phi = 2\pi G_N n$. }

It is natural to ask if we can use the 1D theory to extract a formula for the entropy of the near-de Sitter geometry. At the technical level, this calculation has already been done for us in \cite{Blommaert:2024ydx} and the final result is known to match the SYK spectral density. It is still instructive to review the analysis of \cite{Blommaert:2024ydx} here and view it through our new holographic lens as a semiclassical computation of an Euclidean path integral in 3D Einstein-de Sitter gravity, so we can compare it with the classic computation \cite{Gibbons:1976ue,Gibbons:1977mu} by Gibbons and Hawking. As we will see, our answer will reproduce the GH entropy minus an important correction term that arises due to the positivity constraint \eqref{eq:posphi}.

We wish to compute the entropy $S(\theta)$ of the 1D effective theory as a function of the energy
via the semiclassical identification
\begin{eqnarray}
S(\theta) = \frac{1}{2\pi G_N} \oint  p \spc d\phi, \qquad  \quad H=-\frac{1}{2\pi G_N}\cos\theta.
\end{eqnarray}
Here the contour integral is presumed to follow a (closed) Euclidean trajectory in the ($p,\phi$) phase space with given energy $E=-\cos \theta/2\pi G_N$. Let us compute this integral, following \cite{Blommaert:2024ydx}.

The classical equations of motion for $\phi$ and $p$ can be conveniently written in the form of a 1D Liouville equation and an energy constraint
\begin{eqnarray}
\phi'' + \frac 1 4 e^{-2\phi} = 0, \qquad\quad
\cos p = \frac{\cos\theta}{\sqrt{1-e^{-2\phi}}}.
\end{eqnarray}
Finding the solution to these equations is straightforward \cite{Blommaert:2024ydx}
\begin{eqnarray}
\label{eq:phisol}
\phi(\eta)\! \is\! - \log\biggl(\frac{\sin \theta}{\cos\eta}\biggr) \qquad \quad \    \eta=i u \sin  \theta -\frac{\pi v}{2}, \\[2mm]
p(\eta) \! \is\! \frac{1}{2}\log\biggl(\frac{z+e^{2i\theta}}{1+ z e^{2i\theta}}\biggr), \qquad\qquad\ \  z = e^{2i \eta}.
\label{eq:psol}
\end{eqnarray}
Here $\eta$ is a rescaled Euclidean 1D time coordinate. Note that the solution is periodic in $\eta$ with period $\pi$. This looks promising, as it appears to support our assumption that our Euclidean trajectory describes a closed contour in phase space. 
However, this conclusion is premature: the positivity constraint \eqref{eq:posphi} implies that $\eta$ is constrained to lie within the range
\begin{equation}
\label{restriction}
-{\pi v} < 2\eta < {\pi v}\quad ( {\rm mod}\ 2\pi), \qquad \quad   \theta = \frac{\pi}{2}(1-v).
\end{equation}
Here $v\in [-1,1]$ and $\theta\in [0,\pi]$. 
The above restriction \eqref{restriction} and positivity constraint \eqref{eq:posphi} were introduced in \cite{Blommaert:2024ydx} as an ad hoc prescription in sine-dilaton gravity needed to reproduce the DSSYK entropy. In our 3D de Sitter setting, both constraints follow naturally from the holographic dictionary. 
\begin{figure}[t]\centering
    \begin{tikzpicture}[scale=2, thick]
	\draw[fill] (0, 0) circle (.7pt);
	\draw [decorate,gray,decoration={zigzag,segment length=4,amplitude=2,post=lineto,post length=0}] (-.87,.5) arc (150:180:1);
	\draw [decorate,gray,decoration={zigzag,segment length=4,amplitude=2,post=lineto,post length=0}] (-1,0) arc (180:210:1);
	\node at (0.8,0.8) {{\textcolor{red}{$C$}}};
	\node at (-0.75,-0.2) {{\textcolor{blue}{$\gamma$}}};
	%\node at (1.5,-0.1) {$\operatorname{Re}\,z$};
	%\node at (-0.3,1.5) {$\operatorname{Im}\,z$};
	\draw [->] (-1.2,0) -- (1.3,0);
	\draw [->] (0,-1.2) -- (0,1.2);
        \draw (1.1,1.1) node {$z$};
	\draw[red] (-.87,-.5) arc (-150:150:1);
	\draw[blue] (-.87,-0.5) -- (-.87,0.5);
	\draw[fill] (-0.87,-0.5) circle (.7pt) node[left] {{$e^{-i\pi v}$}};
	\draw[fill] (-.87,.5) circle (.7pt) node[left] {{$e^{i\pi v}$}};
	\draw[fill=white] (-1,0) circle (1pt);
	\end{tikzpicture}
    \caption{Contours $C$ and $\gamma$ used in evaluating the action contour $\oint p d\phi$, see \cite{Blommaert:2024ydx}}
    \label{fig:contour}
    \vspace{-.5mm}
\end{figure}
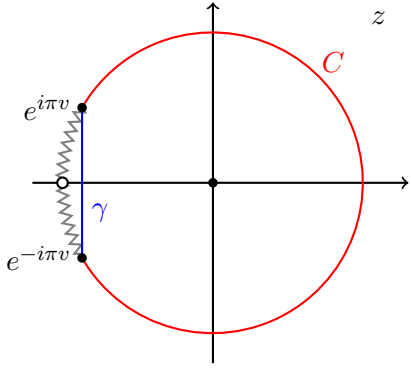

Using equation \eqref{eq:phisol}, we can write
%\begin{eqnarray}
$d\phi = -\tan\!\spc \eta \, d\eta = \left(z^{-1} - {2}({1+z})^{-1}\right)dz$
%\end{eqnarray}
which combined with \eqref{eq:psol} leads to the following integral expression for the semiclassical entropy \cite{Blommaert:2024ydx}
\begin{eqnarray}
\label{actionint}
S(\theta) = \frac{1}{4G_N} \int_C \frac{dz}{2\pi} \biggl(\frac{1}{z} - \frac{2}{1+z}\biggr) \log\biggl(\frac{z+e^{2i\theta}}{1+ z e^{2i\theta}}\biggr) \; \equiv\; \frac{1}{4G_N} \int_C \frac{dz}{2\pi}\, g(z).
\end{eqnarray}
Here $C$ denotes the open contour depicted above. The origin and the point $z=-1$ denote poles in $d\phi$ and the zig-zag line denotes the branch cuts of the logarithm in $p(\eta)$. The red open contour follows an open segment of the unit circle.
To perform the integral, it is convenient to first deform the contour $C$ into the contour $\gamma$, picking up the residue at $z=0$
\begin{eqnarray}
S(\theta) = \frac{2\theta}{4G_N} - \frac{1}{4G_N} \int_\gamma \frac{dz}{2\pi}\, g(z).
\end{eqnarray}
Here in the first term we recognize the standard Gibbons-Hawking entropy %\eqref{eq:gh-entropy} 
of the Schwarzschild-de Sitter spacetime.
The integral over the $\gamma$ contour can be explicitly evaluated by means of dilogarithm functions \cite{Blommaert:2024ydx} but immediately simplifies by means of a standard dilogarithm identity. The final result reproduces the semiclassical formula for the DSSYK spectral density with the identification $\lambda = 4\pi G_N$
\begin{eqnarray}
\label{eq:syk-ent}
S(\theta) = \frac{2\pi\theta -2\theta^2}{4\pi G_N} +{\rm const }.
\end{eqnarray}

Given our 3D derivation of the 1D action, the above can be viewed as a direct calculation of the SYK spectral density from 3D Einstein gravity. 
From the gravity side, it is not surprising that along the way  we made contact with the Gibbons-Hawking formula \eqref{eq:syk-gh-entropy}. The surprise, perhaps, is that the GH and SYK gravitational entropies differ by the extra term given by  the $\gamma$ contour. It is a natural guess that this extra term arises due to our non-standard choice of boundary conditions on $\mathscr{I}^\pm$.
On the SYK side, the extra term from the $\gamma$ contour and the gap in the circular Euclidean trajectory reflect the unusual thermodynamics of the SYK model that lead to the `fake temperature' and `fake disk' \cite{Lin_2023}. It is tempting to think of the contours $C$ (and $\gamma$) as the boundary of the (fake) thermal disk, while on the gravity side we view $C$ as a contour parallel to the great circle of ${\mathscr{I}}^\pm$ that cuts through the angular wedge in figure \ref{fig:membrane1}. The appearance of the branch-cut in figure \ref{fig:contour} then reflects that the interior region inside the wedge has been removed from the spacetime.

\begin{figure}[t]
  \centering
{\includegraphics[width=0.86\textwidth]{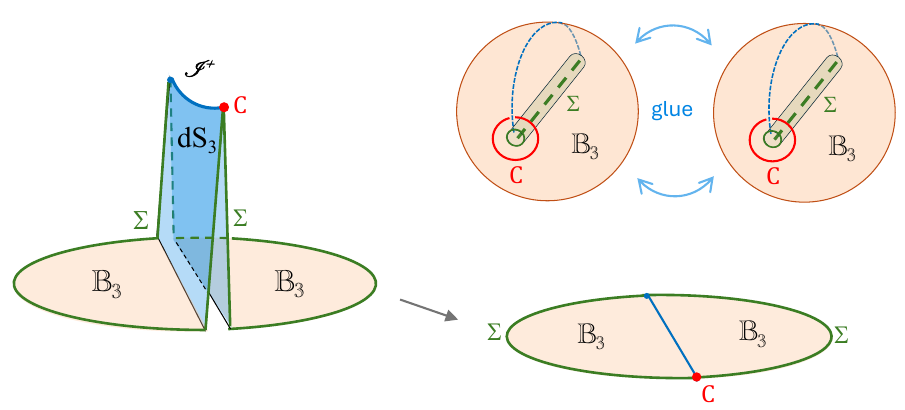}}\
  \caption{The Euclidean path integral of near-de Sitter gravity. The two Hartle-Hawking wave functions are glued together via the $k=-1$ conformal boundary conditions on $\mathscr{I}^\pm$. This requires a Hayward term at the cusp~$\color{red}{C}$. The Lorentzian contributions cancel, but the Euclidean Hayward term survives. The action integrals over the volume and the surface of the green tube combine to contribute the entropy $S_{\rm SYK}$, while the Euclidean part of the Hayward term, shown as the red circle $\color{red}{C}$, contributes the energy term $\beta_{\rm SYK} E_{\rm SYK}$. The red circle corresponds to the thermal solution in figure~\ref{fig:membrane1} continued to Euclidean time, $\psi(u)\to\psi(iu)$.}
  \label{fig:near-ds-glue}
 \end{figure}

It would be instructive to redo the above calculation by starting from an Euclidean 3D geometry and directly performing a Gibbons-Hawking type analysis. We will not attempt to do the full computation here, but only depict an outline of this computation  in figure \ref{fig:near-ds-glue}, which should be compared with figure \ref{fig:ds-glue-one}. It  shows the overlap of two Hartle-Hawking wave functions prepared by the tan (Euclidean) and blue (Lorentzian) regions $\mathbb{B}_3$ and dS$_3$ glued together via the $k=-1$ boundary conditions on $\mathscr{I}^\pm$. This requires the inclusion of a Hayward term. The Lorentzian contributions cancel, but the Euclidean Hayward term survives. The remaining Euclidean action integral receives three contributions: the volume integral and the surface integral over the green tube combine to contribute the SYK entropy $S_{\rm SYK}$. The Euclidean portion of the Hayward term, indicated by the red circle $\color{red}{{C}}$ in figure \ref{fig:near-ds-glue}, contributes the energy term $\beta_{\rm SYK} E_{\rm SYK}$. 
Via the reduction to a one-dimensional effective theory, the red contour $\color{red}{{C}}$ thus plays a double role. The action integral $\oint_{\;{{C}}} pd\phi$ in \eqref{actionint} evaluates the bulk and surface contributions to the 3D gravitational action, while the Hamiltonian term $\oint_{\; {{C}}} H$ evaluates the one-dimensional contribution due to the Hayward-corner term.
%We leave a more detailed study of the difference between the two gravitational entropy formulas \eqref{eq:syk-gh-entropy} for future work. 

\subsection{Hayward time and fake temperature}\label{subsec:hayward_time}

\vspace{-2mm}

According to the dictionary we developed so far, the SYK model lives on the curves $\cal C^\pm$ at the intersection between $\Sigma$ and $\mathscr{I^\pm}$. We will refer to the generator of the `time' flow $\xi_H=\frac{d}{d u}$ along these curves 
 as the Hayward Hamiltonian, since it comes from the Hayward corner term in the gravitational action.
In this subsection, we show that this flow extends to a rescaled time flow for an observer sitting on the poles of the spatial sphere\footnote{Here we already chose the gauge $|\psi'|=\sinh|\text{Im}\psi|$.} 
 \begin{equation}
    H= \sqrt{\gamma}\,\eta_H=  -\sqrt{1-e^{-2\phi}}\cos p,\qquad \xi_H\to\cos\frac{\pi v}{2} \frac{\partial}{\partial \tau}.
\end{equation}
The rescaled time flow provides a natural gravity interpretation of the fake temperature in SYK.

\begin{figure}[t]
\centering
\resizebox{0.35\textwidth}{!}{\input{images/hayward_flow}}
%~~~~~~~~~~~~~~~~  \raisebox{6mm}{\includegraphics[width=0.36\textwidth,height=0.2\textheight]{images/sds-membrane.pdf}}
\caption{The constant temperature solution corresponds to a Schwarzschild-de Sitter spacetime with a conical defect. Translation $\frac{d}{d u}$ along the holographic screen $\mathcal{C}^+$ can be extended to a Killing vector that is space-like in the future patch but time-like in the static patch. On the observer worldline, the Killing vector is proportional to static patch time translation, with $\mathsf{c}=\cos \frac{\pi v}{2}$. }
\end{figure}
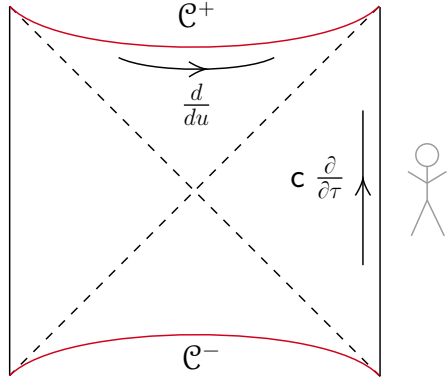

We want to study the flow generated by the Hayward Hamiltonian in the $\mu=0$ case. To this end, it is convenient to consider the dS$_2$ slice \eqref{eq:metric_membrane},
\begin{equation}
    h_{ab}dx^adx^b=-d\tau^2+\cosh^2\tau\tanh^2 y(u) du^2.
\end{equation}
The curves $\cal C^\pm$ are at $\tau = T+\log \coth |y(u)|$; the flow along these curves is given by 
\begin{equation}\label{eq:hayward_flow}
    \xi_H=(\log\coth|y(u)|)'\frac{\partial}{\partial \tau} + \frac{\partial}{\partial u}.
\end{equation}
$\xi_H$ generates an isometry\footnote{If we use $u$ as a coordinate on $\mathcal{C}^\pm$, the induced metric $ds^2=e^{2T}du^2$ is flat in $u$.} of $\cal C^\pm$, and can be extended to a Killing vector on the dS$_2$ slice. Doing this will be cumbersome in general, as $\xi_H$ depends on the solution for $y(u)=\text{Im}\,\psi(u)$. However, we can obtain an explicit expression for the thermal solution near the poles $u\to\infty$.

Let us briefly pause and explain why we are interested in this regime. In figure \ref{fig:membrane1}, we showed that for $\mu=0$, the dual spacetime has a conical deficit located at the poles of the spatial sphere, which correspond to $u\to\pm\infty$. We can introduce an observer sitting at the poles. As we reviewed in section~\ref{subsec:3DSdS_thermodynamics}, this observer is surrounded by a cosmological horizon and experiences a Hawking temperature $\beta_{\text{dS}}=2\pi$. The proper time of this observer is given by the global time $\tau$; we can interpret the inverse temperature as the periodicity in $\tau$ when we continue it to Euclidean time $\tau\to i \tau$.

To obtain the flow \eqref{eq:hayward_flow} close to the poles, we expand the thermal solution \eqref{eq:thermal_solution_psi} near $u\to \infty$,
\begin{equation}
\begin{aligned}
 &    \psi(u)\sim \frac{\pi}{2}-2e^{-\cos(\nspc\frac{\pi v}{2}\nspc) \spc u -i\frac{\pi}{2}(1-v)},\\[2mm]
    & y(u)\sim2e^{-\cos(\nspc \frac{\pi v}{2}\nspc )\spc u }\sin\frac{\pi}{2}(1-v).
\end{aligned}
\end{equation}
From this, we directly deduce that the Hayward Killing flow $\xi_H$ given in \eqref{eq:hayward_flow} reduces at the poles to 
\begin{equation}\label{eq:hayward_static_flow}
    \xi_H = \cos\Bigl(\frac{\pi v}{2}\Bigr)\frac{\partial}{\partial\tau}.
\end{equation}
Here we dropped the $\partial/\partial u$ term since $u=\infty$ is a fixed point of $u$ translations. We derived this result at the pole on $\mathscr{I}^\pm$; we can then extend it to the full observer trajectory. Since $\partial/\partial\tau$ is a Killing vector on the pole, %\footnote{On the pole, $\tau$ coincide with the static patch time, and translation in the latter is a Killing vector.},
we would have obtained the same result by first extending $\xi_H$ to a Killing vector on the dS$_2$ slice, and then restricting it to the pole.

The result \eqref{eq:hayward_static_flow} shows that the Hayward time is rescaled with respect to the time experienced by an observer in de Sitter. Then, a periodicity of $\beta_{\text{dS}}=2\pi$ in (Euclidean) $\tau$, corresponds to a periodicity of
\begin{equation}
    \beta_{ H} = \frac{2\pi}{\cos \frac{\pi v}{2}},
\end{equation}
in Hayward time. This matches the SYK fake temperature \eqref{eq:fake_temperature}. Indeed, our dictionary identifies the Hayward time with the SYK time, and the above result provides a geometric interpretation of the fake temperature phenomenon in SYK.

%% file: images/hayward_flow.tex
\tikzset{every picture/.style={line width=0.75pt}} %set default line width to 0.75pt        

\begin{tikzpicture}[x=0.75pt,y=0.75pt,yscale=-1.05,xscale=1.05]

%Straight Lines [id:da32095143341626586] 
\draw  [dash pattern={on 4.5pt off 4.5pt}]  (106,31) -- (344,269) ;
%Straight Lines [id:da9290295604905993] 
\draw  [dash pattern={on 4.5pt off 4.5pt}]  (343.5,31) -- (106,268.5) ;
%Straight Lines [id:da36526225540387414] 
\draw [color={rgb, 255:red, 0; green, 0; blue, 0 }  ,draw opacity=1 ]   (106,31) -- (106,268.5) ;
%Straight Lines [id:da1292435705585987] 
\draw [color={rgb, 255:red, 0; green, 0; blue, 0 }  ,draw opacity=1 ]   (344,31.5) -- (344,269) ;
%Curve Lines [id:da6135996877952525] 
\draw [color={rgb, 255:red, 208; green, 2; blue, 27 }  ,draw opacity=1 ]   (106,31) .. controls (140.16,66.55) and (309.16,65.55) .. (343.5,31) ;
%Shape: Boxed Bezier Curve [id:dp8230189003801441] 
\draw [color={rgb, 255:red, 208; green, 2; blue, 27 }  ,draw opacity=1 ]   (343.5,268.5) .. controls (309.34,232.95) and (140.34,233.95) .. (106,268.5) ;
%Shape: Circle [id:dp5967269083004614] 
\draw  [color={rgb, 255:red, 155; green, 155; blue, 155 }  ,draw opacity=1 ] (367,126.73) .. controls (367,122.72) and (370.25,119.46) .. (374.27,119.46) .. controls (378.28,119.46) and (381.54,122.72) .. (381.54,126.73) .. controls (381.54,130.75) and (378.28,134) .. (374.27,134) .. controls (370.25,134) and (367,130.75) .. (367,126.73) -- cycle ;
%Straight Lines [id:da5548996130359731] 
\draw [color={rgb, 255:red, 155; green, 155; blue, 155 }  ,draw opacity=1 ]   (374.27,134) -- (374.27,155.67) ;
%Shape: Boxed Line [id:dp8223670254131826] 
\draw [color={rgb, 255:red, 155; green, 155; blue, 155 }  ,draw opacity=1 ]   (374.27,155.67) -- (385.29,178.67) ;
%Shape: Boxed Line [id:dp8827768367309806] 
\draw [color={rgb, 255:red, 155; green, 155; blue, 155 }  ,draw opacity=1 ]   (374.27,155.67) -- (364.1,178.67) ;
%Straight Lines [id:da9388274813684849] 
\draw [color={rgb, 255:red, 155; green, 155; blue, 155 }  ,draw opacity=1 ]   (361.94,137.33) -- (374.27,144.83) ;
%Straight Lines [id:da5620526403924213] 
\draw [color={rgb, 255:red, 155; green, 155; blue, 155 }  ,draw opacity=1 ]   (374.27,144.83) -- (386.6,137.33) ;
%Curve Lines [id:da4174700902189118] 
\draw    (176,64) .. controls (196.17,74) and (256.17,74) .. (276,64) ;
\draw [shift={(232.11,71.43)}, rotate = 179.28] [color={rgb, 255:red, 0; green, 0; blue, 0 }  ][line width=0.75]    (10.93,-4.9) .. controls (6.95,-2.3) and (3.31,-0.67) .. (0,0) .. controls (3.31,0.67) and (6.95,2.3) .. (10.93,4.9)   ;
%Shape: Boxed Line [id:dp278783030720125] 
\draw    (333,197.5) -- (333,98) ;
\draw [shift={(333,141.75)}, rotate = 90] [color={rgb, 255:red, 0; green, 0; blue, 0 }  ][line width=0.75]    (10.93,-4.9) .. controls (6.95,-2.3) and (3.31,-0.67) .. (0,0) .. controls (3.31,0.67) and (6.95,2.3) .. (10.93,4.9)   ;

% Text Node
\draw (212,26.4) node [anchor=north west][inner sep=0.75pt]  [font=\LARGE]  {$\mathcal{C}^{+}$};
% Text Node
\draw (215,79.4) node [anchor=north west][inner sep=0.75pt]  [font=\LARGE]  {$\frac{d}{du}$};
% Text Node
\draw (285,126.4) node [anchor=north west][inner sep=0.75pt]  [font=\LARGE]  {$\mathsf{c} \ \frac{\partial }{\partial \tau }$};
% Text Node
\draw (215,246.4) node [anchor=north west][inner sep=0.75pt]  [font=\LARGE]  {$\mathcal{C}^{-}$};

\end{tikzpicture}

%% file: sections/towards_an_holographic_dictionary.tex
%In this section we will add some more entries to the holographic dictionary.  In particular, we will establish a correspondence between the two-point functions of SYK scaling operators and the boundary-to-boundary scalar Green's function in the near-de Sitter spacetime.

%\subsection{Two-point function as geodesic length}\label{subsec:geodesic_dual}

In this section, we relate the semiclassical DSSYK two-point functions to the boundary-to-boundary scalar Green functions, expressed as exponentials of the geodesic length, in the dual de Sitter spacetime.  We will find that our proposed holographic dictionary involves taking the product of two SYK two-point functions. This identification is a required first step in building a more complete holographic dictionary between SYK correlation functions and cosmological correlators of local operators on our 1D holographic screen.

In our setup, we allowed the MQ coupling $\mu$ to be of order $1/\lambda$, while the probe two-point functions involve scaling operators with dimension $\Delta$ of order 1. To leading order in $\lambda = 4\pi G_N$, we can thus ignore the backreaction due to the propagating mode created by ${\cal O}_\Delta$. Since our dual spacetime implements the Israel junction conditions, it incorporates the full semiclassical backreaction due to the energy sheet $\mu(u)$. It would then seem reasonable to conclude that the semiclassical SYK two-point functions are identical to the Green function of a free scalar field evaluated in this backreacted spacetime. 

Consider a scalar field  of mass $m$ satisfying the scalar wave equation in 2+1-D de Sitter spacetime
\begin{eqnarray}
\qquad \bigl(\Box - m^2\bigr) \Phi_\Delta(X) = 0, \qquad \  \ m^2 \!= \Delta(1-\Delta).
\end{eqnarray}
More generally, we can consider this scalar field equation in the Schwarzschild-de Sitter spacetime with a conical defect, or in case $\mu(u)$ is non-zero, in the backreacted spacetime with an energy sheet.
Let $X_1$ and $X_2$ denote two points on our 1D holographic screen $\cal{C}^+$ at future infinity and let
\begin{eqnarray}
\label{eq:gds}
G_{\rm dS}(X_1,X_2) \, \is \, \bigl\langle \Phi_\Delta(X_1) \Phi_\Delta(X_2)\bigr\rangle_{{\rm dS}}
\end{eqnarray}
denote the corresponding boundary-to-boundary Green function of the free scalar field $\Phi_\Delta$. For a boundary-to-boundary Green function, we have 
\begin{eqnarray}
\label{eq:gds1}
G_{\rm dS}(X_1,X_2) \, \is e^{-\Delta\, \ell(X_1,X_2)}
\end{eqnarray}
where $\ell(X_1,X_2)$ denotes the length of the dominant geodesic between the points $X_1$ and $X_2$ on $\mathscr{I}^+$.

Following the standard GKPW holographic dictionary, modified to our setting and taking into account that the geodesic dictionary involves taking the absolute value squared of the SYK two-point function, we will be led to identify the boundary-to-boundary Green function \eqref{eq:gds} of the free scalar field in the backreacted de Sitter-Schwarzschild geometry with the (SYK)$^2$ two-point function\footnote{Here we identify $(X_1,X_2) = (X[\psi(u_1)],X[\psi(u_2)])$, see below.}
\begin{eqnarray}
\label{eq:gsyksq}
G_{\rm SYK}(X_1,X_2) = \bigl\langle \mathbb{O}_\Delta(u_1) \mathbb{O}(u_2)\bigr\rangle \qquad \qquad
\mathbb{O}_\Delta(u) \equiv \mathcal{O}_\Delta(u) \overline{\mathcal{O}}_\Delta(u).
\end{eqnarray}
Here $\mathcal{O}_\Delta(u)$ and $\overline{\mathcal{O}}_\Delta(u)$ denote two (complex conjugate) scaling operators in two identical but independent SYK models. We will now verify this dictionary.

\subsection{Two-point functions as geodesic lengths}
\label{subsec:geodesic_dual}
As outlined in Appendix \ref{sec:geodlength}, the embedding coordinates $X$ of global 3D de Sitter spacetime can be conveniently represented as 2$\times$2 matrices ${\bf X}$ with determinant equal to  $-1$. Points
${X}^+$ at future infinity $\mathscr{I}^+$ then map to matrices of the general form
\begin{align}
\label{eq:xfactorpsi}
    \mathbf{X}^+&\rightarrow e^T\, \mathbf{z}[\psi]\, \mathbf{z}[\psi]^\dag,\qquad \quad
\mathsf{z}[\psi]=\frac1{\sqrt{2\sinh{\rm Im}\psi}}\left(\!\begin{array}{c}e^{i\psi/2}\!\\ e^{-i\psi/2}\end{array}\!\right).
\end{align}
This is the familiar statement that points at infinity factorize into a product of spinors.
A similar decomposition holds for points $X^-$ on $\mathscr{I}^-$.  We denote the point corresponding to the spinor $\mathsf{z}[\psi]$ by $X^\pm[\psi]$.
In global de Sitter space, the geodesic length between two points $X_1^+$ and $X_2^\pm$ at asymptotic infinity can be expressed as the log of the inner product $X_1^+\!\cdot X_2^\pm$ of the embedding coordinates. 

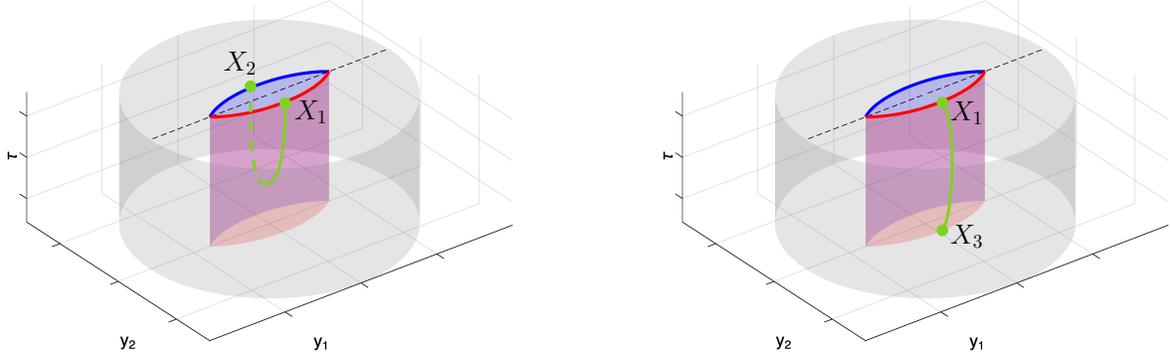
\begin{figure}[t]
  \begin{center}
    \resizebox{0.92\textwidth}{!}{\input{images/geodesic_membrane}}
  \caption{The two types of geodesics that connect points on the holographic screen correspond to the two types of SYK two-point functions ${g}_{RR}$ and ${g}_{RL}$. The lengths of the geodesics are computed in the standard de Sitter geometry. The geodesic that reconnects with $\mathscr{I}^+$ is complex.}
    \label{fig:geodesics}
  \end{center}
  \vspace{-2mm}
\end{figure}

In the appendix, we show that the factorization \eqref{eq:xfactorpsi} thus allows us to decompose this inner product into two factors
\begin{eqnarray}
\label{eq:xdot}
X_1^+\!\cdot X_2^+ \is  e^{\ell(X_1^+,X_2^{+})} \; = \; e^{-\mbox{\footnotesize $\frac{g(\psi_1,\psi_2)}{2}$}}e^{-\mbox{\footnotesize $\frac{g(\psi_2,\psi_1)}{2}$}}
\end{eqnarray}
where each factor on the right-hand side is given by the inner product of two spinors $\mathsf{z}[\psi_1]$ and $\mathsf{z}[\psi_2]$.
As we work out below for both types of geodesic lengths shown in figure \ref{fig:geodesics}, the two factors in \eqref{eq:xdot} assemble into the product of two SYK two-point functions.

We want to study the two types of two-point functions drawn in figure \ref{fig:geodesics}. Let $X^+_1 = X^+[\psi(u_1)]$ denote the point on $\cal{C}$ on $\mathscr{I}^+$, $X^{+*}_2 = X^+[\psi^*(u_2)]$  a point on $\mathcal{C}^*$ on $\mathscr{I}^+$, and $X^-_3 = X^-[\psi(u_3)]$ a point on $\cal{C}$ on $\mathscr{I}^-$, as shown in the figure.
%We will often use the short-hand notation  $X_1^+ = X^+[\psi(u_1)],$ and $ X_2^\pm = X^\pm[\psi(u_2)].$

Motivated by the spinor factorization 
\eqref{eq:xfactorpsi} of the asymptotic coordinates $X^\pm$, we propose the following semiclassical holographic identification between the product of two SYK two-point functions and the Green function of a free scalar field at the corresponding locations on the intersection curves~${\cal C}^\pm$
\begin{eqnarray}
\label{eq:gdictionary}
\boxed{\ \ \begin{array}{c}{G_{\rm dS}\bigl(X^+_1, X^{+*}_2\bigr) \, =\, e^{-\Delta\, \ell(X^+_1, X^{+*}_2) } =\;
    \bigl\langle \mathbb{O}^R_\Delta(u_1) \mathbb{O}^R_\Delta(u_2)\bigr\rangle \,\;  %\overline{\!\bigl\langle {\cal O}^R_\Delta(u_1) {\cal O}^R_\Delta(u_2)\bigr\rangle\!}
    \large {}^{\strut}}, \\[5mm]
{G_{\rm dS}\bigl(X^+_1, X^{-}_3\bigr)\; =\;  e^{-\Delta\, \ell(X^+_1, X^{-}_2) }\; =
\,   \;  \bigl\langle \mathbb{O}^R_\Delta(u_1) \mathbb{O}^L_\Delta(u_3)\bigr\rangle \,  %{\bigl\langle {\cal O}^L_\Delta(u_1) {\cal O}^R_\Delta(u_3)\bigr\rangle}
\large {}_{\strut}}.\end{array}\ \ } 
\end{eqnarray}
 %In equation \eqref{eq:gdictionary}, 
Here we used the fact that in the semiclassical limit, the boundary-to-boundary two-point function is given by the exponential of the geodesic lengths $\ell(X^+_1, X^{+*}_2)$ and  $\ell(X^+_1, X^{-}_3)$ between the relevant points on $\mathscr{I}^\pm$. The two types of geodesics are schematically drawn in figure~\ref{fig:geodesics}.

Our semiclassical holographic dictionary \eqref{eq:gdictionary} identifies the geodesic lengths with the sum of the corresponding SYK two-point functions\footnote{Here an overline on $g$ means that we exchange $\psi$ and $\psi^*$ in the solution \eqref{eq:grrl} , i.e. $\overline{g}_{RL} = ({g}_{RL})^*$, $\overline{g}_{RR} = ({g}_{RR})^*-2i\pi$.}
\begin{equation}
\begin{split}
\label{eq:geodict}
    \ell(X^+_1, X^{+*}_2) =& \ 2T \, - \,\frac{{g}_{RR}(u_1, u_2)+\overline{g}_{RR}(u_1, u_2)}{2},\\[3mm]
    \ell(X^+_1, X_3^-) =&\  2T\, -\, \frac{{g}_{RL}(u_1, u_3)+\overline{g}_{RL}(u_1, u_3)}{2}.
\end{split}
\end{equation}
where $T\to \infty$ parametrizes the approach of the cut-off surface to infinity. We see that the same-side and opposite-side Green functions label the two types of paths between $\mathscr{I}^\pm$ indicated in figure~\ref{fig:geodesics}.
In Appendix~\ref{sec:geodlength} we obtain the explicit expression of the geodesic length \eqref{eq:geodiscs_length} in 3D de Sitter spacetime between points at future and past infinity $\mathscr{I}^\pm$, in the $(\tau, x, y)$ coordinate system. 
Let us verify that both sides are identical in the limit of small $\lambda = 4\pi G_N$. 

Defining the complex coordinate $\psi=x+iy$, we find that the length of boundary-to-boundary geodesics between points on $\mathscr{I}^\pm$ takes the asymptotic form\footnote{In equation \eqref{eq:geodiscs_length}, we specify to points on the cutoff surface $\tau = \pm(T-\log \tanh y)$. } 
\begin{eqnarray}
    {\ell(X^+_1, X^{+*}_2)} & \simeq &\,  
     \; \log \biggl(\frac{\sin\bigl(\frac 1 2 (\psi_1\!-\psi^*_2)\bigr)\sin\bigl(\frac 1 2 (\psi_1^*\!-\psi_2)\bigr)}{\cosh y_1\cosh y_2}\biggr) + \tau_1+\tau_2 +i\pi ,\\[3.5mm]
       {\ell(X^+_1, X^-_2)}  & \simeq& \, \;  \ %\left(
       \log\biggl( \frac{\cos\bigl(\frac 1 2 (\psi_1\!-\psi^*_2)\bigr)\, \cos\bigl(\frac 1 2 (\psi_1^*\!-\psi_2)\bigr)}{\cosh y_1\cosh y_2}%\right) 
       \biggr)
       + \tau_1-\tau_2.
    \label{eq:g_psi_dual}
\end{eqnarray}
These geodesic lengths are computed in the undeformed de Sitter spacetime and hold for any pair of points on $\mathscr{I}^\pm$. In a moment, we will specialize the endpoints to lie on the 1D holographic screen ${\cal C} = \{x+iy = \psi(u)\}$, so we can compare with the SYK two-point function. 

Note that the geodesic that reconnects with $\mathscr{I}^+$ is in fact complex: %\footnote{%As we review in Appendix \ref{sec:3ddS}, 
there is no real geodesics connecting different points on $\mathscr{I}^+$. Instead, the geodesic that computes the propagator follows a complex trajectory that starts out as a real timelike trajectory from point $1$ at $\mathscr{I}^+$ towards the point antipodal to $2$ at $\mathscr{I}^-$. When the geodesic encounters the $\tau=0$ slice, it enters an Euclidean (see Appendix \ref{app:spinor_coordinates}) $S^3$ part of the geometry, traversing it along a great circle, until it re-emerges on the other side as a real time geodesic directed towards point $2$ at $\mathscr{I}^+$.

In Sec.~\ref{sec:MQ_coupling} we reviewed the semiclassical two-point functions \eqref{eq:g_psi} of the DSSYK fermions and found that they are expressed in terms of the complex reparametrization mode via
\begin{equation}
\begin{split}
\label{eq:grrl}
    {g}_{RR}(u_1, u_2) =& \log \biggl(\frac{\psi'(u_1)\psi'^{*}(u_2)} {\sin^2\bigl(\frac 1 2 (\psi(u_1)-\psi^*(u_2))\bigr)}\biggr) - i\pi,\\[2mm]
   {g}_{RL}(u_1, u_2) = & \; \log\biggl( \frac{\psi'(u_1)\psi'^{*}(u_2)}{\cos^2\bigl(\frac 1 2 (\psi(u_1)-\psi^*(u_2))\bigr)}\biggr).
\end{split}
\end{equation}
 Comparing the above two equations \eqref{eq:grrl} and \eqref{eq:g_psi_dual}, while  using the relation \eqref{eq:gauge_choice} and setting $\tau_1$ and $\tau_2$ equal to the respective cutoff values \eqref{eq:scridef} at $\mathscr{I}^\pm$, we confirm our holographic dictionary \eqref{eq:gdictionary}-\eqref{eq:geodict}.

\def\XX{\mbox{\small $X$}}

%\begin{eqnarray}
%\label{eq:greenidone}
%G_{\rm SYK}(X_1,X_2) \, \stackrel{\mbox{\small ?}}{=} \, G_{\rm dS}(X_1,X_2) 
%\end{eqnarray}
To conclude, we have shown that  
\begin{eqnarray}
\label{eq:greenidtwo}
G_{\rm SYK}(X_1,X_2) = e^{-\Delta \spc \ell(X_1,X_2)}
\end{eqnarray}
where $\ell(X_1,X_2)$ denotes the length of the dominant\footnote{Let us make this statement more precise.
Consider two points at $\mathscr{I}^+$ labeled by $u_1$ and $u_2$. Each point has different coordinates as seen from the left or right of the gluing surface. Let point $1$ be on $\Sigma$, i.e. $x_1+iy_1=\psi(u_1)$. Then we have two geodesics: the one connecting point $1$ to point $2$ on $\Sigma$ with coordinates $x_2+iy_2 =\psi(u_2)$ and the one connecting to  point $2$ on $\Sigma^*$ with coordinates $x_2+iy_2=\psi^*(u_2)$. Which one dominates? The relevant geodesics in Lorentzian signature are timelike. Hence, the one that maximizes proper time dominates. We can verify this geodesic is the one that connects points on $\Sigma$ to points on $\Sigma^*$. This is the length that SYK computes.} geodesic between the points $X[\psi(u_1)]$ and $X[\psi(u_2)]$ on $\mathscr{I}^+$ in the undeformed 3D de Sitter spacetime. We then argued that the gravitational de Sitter backreaction is encoded via the SYK backreaction of the MQ coupling on the soft mode $\psi(u)$, displacing the location of the holographic screen ${\cal C}$. This is a true statement when applied to the length $\ell(X_1,X_2)$ of the dominant geodesic 
\begin{eqnarray}
\label{eq:greenidthree}
G_{\rm SYK}(X_1,X_2) =  \Bigl\langle \exp\biggl(i m \int_{\gamma{\nspc}_{X_1\to X_2}^{}}\hspace{-7mm} \sqrt{g_{\mu\nu} \dot{\XX}^\mu\dot{\XX}^\nu}\, \biggr) \Bigr\rangle, %= e^{-\Delta \spc \ell(X_1,X_2)}
\end{eqnarray}
where the expectation value on the right-hand side indicates a path integral over all trajectories with the same topology as the dominant timelike geodesic with length $\ell(X_1,X_2)$. In pure 3D de Sitter spacetime, the right-hand sides of equations \eqref{eq:greenidtwo} and \eqref{eq:greenidthree} indeed coincide. The SYK two-point function \eqref{eq:gsyksq} should thus be interpreted as a gravitationally dressed de Sitter two-point function with an explicit gravitational Wilson line connecting the beginning and end point.

%Another hint that the SYK two point function captures the effect of backreaction is the following. 

\tdplotsetmaincoords{250}{100}

\subsection{Two-time ordering and boundary locality}
\label{subsec:twotime}

Our holographic dictionary allows for explicit study of the gravitational interaction between, say, a heavy particle and a light particle propagating in its background spacetime. The heavy particle is represented by a delta-function source $\mu(u)=\hat{\mu}\, \delta(u-u_0)$, thus creating a shockwave geometry in the form of a kink in the boundary trajectory $\psi(u)$, as shown in figure \ref{fig:membrane2} and \ref{fig:membrane2b}. As seen from the figures, the backreaction of the heavy particle moves the two poles of the Schwarzschild-de Sitter geometry (the two points where the wedge ends) closer to each other. This effect makes signal propagation between the poles possible. This is a characteristic property of de Sitter backreaction. We study this scattering scenario in a separate publication. 

 Here we comment on two interpretational questions. The first is the apparent puzzle that SYK operators at different time points do not commute while in the dual de Sitter spacetime they are identified with spacelike separated points on past or future infinity. How can this be consistent with locality? A logically related question is: Should we in fact adopt the alternative holographic dictionary based on two-time physics, in which the SYK time evolution charts a one-dimensional path on the boundary of AdS$_{1+2}$ spacetime with two time directions? We start with the first question. 

Operator ordering matters in SYK, but why would it matter in the gravity dual given that we chose the holographic screen ${\mathcal{C}}$ to be spacelike? A partial explanation comes from the above described backreaction effect. Since local SYK operators are confined to lie on the gluing curve ${\cal C}$, their two-point functions exhibit a discontinuity at the location of the kink created by a heavy particle, as shown in figures \ref{fig:membrane2} and \ref{fig:membrane2b}. This discontinuity makes an OTOC configuration different from the TOC configuration. The exponential de Sitter expansion furthermore leads to a  Lyapunov behavior, analogous to that in near-AdS${}_2$ gravity,  except that the exponentially growing displacement takes place  along the space direction. By itself, this does not violate locality or causality: detecting a particle at a location $X_1$ on $\mathscr{I}^+$ implies that it was present in the past. Its backreaction affects the space-like trajectory of the holographic screen $\mathcal{C}^+$ along $\mathscr{I}^+$. This explains why TOCs and OTOCs look different. 

To gain more insight, let us temporarily switch our perspective and adopt the alternative dual gravity interpretation in terms of an AdS$_{1+2}$ spacetime with two time directions, by performing a triple Wick rotation %between dS$_{2+1}$ and AdS$_{1+2}$ 
that multiplies the 3D metric with an overall minus sign. This aligns the time flow of the SYK model with that in the dual spacetime. In the bulk, the lightcone is now rotated by 90 degrees as indicated in figure \ref{fig:ads2plus1}. However, all non-trivial dynamics is confined to the 2D gluing surface $\Sigma$, which is now a Lorentzian AdS$_2$ membrane with a 1D timelike boundary. The fact that $\Sigma$ is embedded inside an ambient AdS spacetime with two time directions has little bearing on the causality of the 2D physics on $\Sigma$ itself. So the two-time perspective looks perfectly tenable.

Let us look more closely at the interpretation of the two-time theory from the DSSYK point of view. If we fix the choice of time-dependent coupling $\mu(t)$ and consider the Heisenberg operators ${\cal O}_\Delta(t)$, in the $\lambda\rightarrow 0$ limit they have the two-point function given by equation \eqref{greenone} with a fixed $\psi(t)$ trajectory in the complex plane. 
We can also consider the larger algebra of operators by including all possible choices of $\mu(t)$. Let us denote ${\cal O}_\Delta(t)[\mu(t')]$ as the Heisenberg operator at time $t$ for trajectory $\mu(t')$. Written explicitly, if we denote the initial time as $t=0$, we have
\begin{align}
    {\cal O}_\Delta(t)[\mu(t')]&=U_\mu(0,t){\cal O}_\Delta^SU_\mu(t,0), %\text{~with~} 
    \qquad \ U_\mu(t,0)=Te^{-i\int_0^tH[\mu(t')]dt'}.
\end{align}
For a generic quantum mechanical model (such as an SYK model with finite $q$), 
this operator depends on the entire trajectory $\mu(t')$ for $t'\leq t$. However, the DSSYK two-point function (\ref{greenone}) for $\lambda\to 0$ has a more local time dependence than the most generic situation: the two-point function $G_\Delta(t_1,t_2)$ only depends on $\psi(t)$ and its derivative $\psi'(t)$ at $t=t_1,t_2$. 

\begin{figure}[t]
\centering
    \includegraphics[width=3.2in]{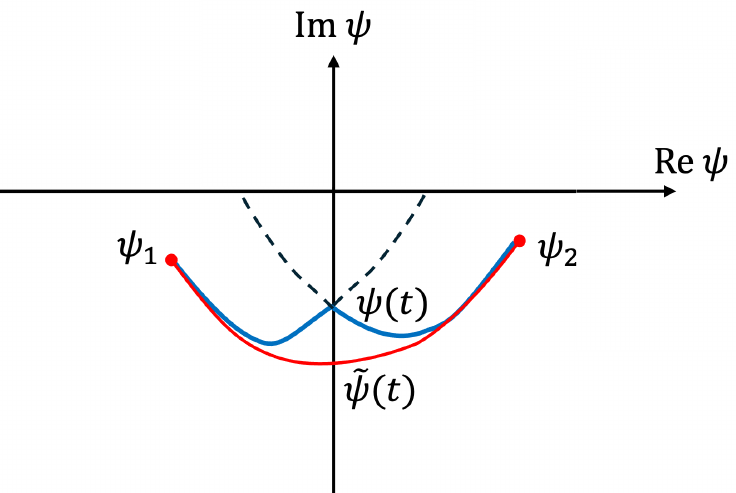}
    \caption{Illustration of two paths in $\psi$ plane $\psi(t)$ and $\tilde{\psi}(t)$ which have the same end points and are tangential at the end points, so that $\psi$ and $p'$ are the same for the two paths at the end points. }\label{fig: equivalence_of_operators}
\end{figure}

Consider two trajectories $\psi(t)$ and $\tilde{\psi}(t)$ corresponding to different time-dependent couplings $\mu(t)$ and $\tilde{\mu}(t)$, as is illustrated in figure \ref{fig: equivalence_of_operators}. If there exists a pair of time points $t_1,t_2$ and $\tilde{t}_1,\tilde{t}_2$ such that
\begin{align}
    \psi(t_{1,2})=\tilde{\psi}(\tilde{t}_{1,2}),~\psi'(t_{1,2})=\tilde{\psi}'(\tilde{t}_{1,2})
\end{align}
(i.e. the two paths intersect tangentially at these two points), then the two-point function evaluated at time points $t_1,t_2$ and $\tilde{t}_1,\tilde{t}_2$ in both situations are equal
\begin{align}
    \bigl\langle {\cal O}_\Delta(t_1)[\mu]{\cal O}_\Delta(t_2)[\mu]\bigr\rangle=\left\langle {\cal O}_\Delta(\tilde{t}_1)[\tilde{\mu}]{\cal O}_\Delta(\tilde{t}_2)[\tilde{\mu}]\right\rangle.
\end{align}
This is illustrated in figure \ref{fig: equivalence_of_operators}. Considering that $|\psi'(t)|$ is determined by $\psi(t)$, only the phase factor $p$ is an independent degree of freedom. Therefore, the two-point function suggests that we can denote ${\cal O}_\Delta (t)[\mu]={\cal O}_\Delta(\psi,p)$ which only depends on the value of $\psi,p$ and does not explicitly depend on $t$. Furthermore, in this notation the two-point function satisfies 
\begin{align}
    \left\langle{\cal O}_\Delta(\psi_1,p_1){\cal O}_\Delta(\psi_2,p_2)\right\rangle=\left\langle{\cal O}_\Delta(\psi_1,0){\cal O}_\Delta(\psi_2,0)\right\rangle e^{i\Delta(p_1-p_2)}.
\end{align}
Therefore this operator algebra can be generated by ${\cal O}_\Delta(\psi,0)e^{\pm i\Delta p}$. These operators satisfy a non-trivial braiding property similar to the holomorphic and anti-holomorphic fields in a 2D Euclidean CFT. This is what one would expect of chiral edge modes of 3D gravity. 

From the point of view of von Neumann algebra, the bulk semiclassical limit ($G_N\rightarrow 0$) corresponds to a large $N$ limit of the boundary theory which leads to type III$_1$ algebra for a time interval \cite{leutheusser2025subregion}. Physically, this means that simple operators at different times are viewed as independent from each other (not (anti-)commuting but linearly independent), because the Heisenberg equation that relates operators at different times becomes untractable. In the same limit, since operators at different times can be viewed as independent, one can also introduce more parameters ($\mu(t)$ in our case) and define a larger von Neumann algebra. In DSSYK with $\lambda\rightarrow 0$, the correlation function suggests that this larger von Neumann algebra is equivalent to that of chiral fields in Euclidean 2d CFT. %In Lagrangian 2d QFT, local operators have integer or half-integer spin (bosons or fermions). 
Chiral CFT fields with a generic $\Delta$ are nonlocal. This enables the possibility of nontrivial commutation relations even if the Euclidean geometry looks like a spatial direction\footnote{As a comparison, we can consider a 2d compact boson CFT with Lagrangian density $\mathcal{L}=\frac{K}{4\pi}\partial_\mu\varphi\partial^\mu\varphi$ with $\varphi$ having periodicity of $2\pi$. In this theory, operators $e^{in\varphi(x)}$ with integer $n$ are local operators. Operators $\mathcal{O}_\Delta =e^{i\Delta\varphi(x)}$ with a non-integer $\Delta$, that is, with has a generic scaling dimension, are not single-valued. To remove the ambiguity introduced by the non-single-valuedness of $\varphi(x)$, when we define the two-point function $\left\langle\mathcal{O}_\Delta(x)^\dagger\mathcal{O}_\Delta(y)\right\rangle$, we need to explicitly define the path between $x$ and $y$. For a given path $P$ from $x$ to $y$ on the two-dimensional plane, we can define
\begin{align}
\bigl\langle\mathcal{O}_\Delta(x)^\dagger\mathcal{O}_\Delta(y)\bigr\rangle=e^{i\Delta\int_Pdz^\mu\partial_\mu \varphi(z)}
\end{align}
This path dependence should be compared to how the DSSYK$_{\lambda\rightarrow 0}$ two-point function depends on the path in the $\psi$ plane.}.

We end this discussion with a comment on how this structure may need to be modified at finite $\lambda$. It was shown \cite{Berkooz:2018jqr} that the DSSYK model has the structure of a chiral CFT with an R-matrix given by the 6j-symbol of the quantum group U$_{\mathsf{q}}$(SU(1,1)) with $\mathsf{q}=e^{-\lambda}$. The same quantum group  also describes the braiding properties of partial waves in 2+1-D de Sitter gravity and conformal blocks in complex Liouville CFT \cite{Mengyang_Verlinde:2024zrh,Collier:2025lux}, in which case it accounts for the exact quantum gravitational interactions at finite $G_N$.  Matching the two structures leads to the same identification \eqref{lambdagn} between the SYK coupling and the 3D Newton constant.
This correspondence, combined with
our result that the de Sitter Green function factorizes into a square of DSSYK two-point functions, suggests that the full de Sitter gravity is dual to a pair of SYK models that each play the role of a chiral sector of a complex Liouville CFT dual to dS$_3$ gravity. In this combined theory, it is possible to formulate a holographic dictionary in which space-like separated operators on $\mathscr{I}^\pm$ commute. The formal argument goes as follows \cite{Jackson_2015}. 

Locality of a non-chiral CFT is ensured by the fact that it is defined to solve the conformal bootstrap. 
The matrix elements of non-chiral operators factorize into matrix elements of chiral vertex operators
\begin{eqnarray} 
\langle\spc j\spc | \mathbb{O}(\psi,\bar{\psi})|\spc i\spc \rangle \; = \; \overline{\!\langle \spc j\spc | \mathcal{O}(\psi)|\spc i\spc \rangle\!}\;\, \langle\spc i\spc | \mathcal{O}(\psi)|\spc j\spc \rangle.
\end{eqnarray}
This leads to the notion of chiral conformal blocks with non-trivial braiding properties. In this sense, a chiral 2D CFT looks like 1D quantum mechanics. The matrix elements of a product of two non-chiral operators $\mathbb{O}_A$ and $\mathbb{O}_B$ factorizes into chiral conformal blocks
\begin{eqnarray}
\label{nca}
\langle\spc j\spc | \spc \mathbb{O}_A \spc {\mathbb{O}}_B \spc | \spc i\spc \rangle\is \sum_{\alpha}\, \bar{\Psi}_{\nspc\alpha} \, {\Psi}_{\nspc\alpha},\qquad \ \ \Psi_\alpha \,\equiv\,
\langle \spc j \spc  | \spc \mathcal{O}_A\, \Pi_\alpha \spc {\mathcal{O}}_B \spc | i\rangle,
\notag \\[-2.5mm]\\[-2.5mm]\notag
\label{ncb}
\langle\spc j\spc | \spc {\mathbb{O}}_B \spc \mathbb{O}_A \spc | \spc i\spc \rangle\is \sum_{\beta}\, \bar{\Phi}_{\nspc\beta}\, {\Phi}_{\nspc\beta},\qquad \quad 
\Phi_\beta\, \equiv \, 
\langle \spc j\spc | \spc {\mathcal{O}}_B \, \Pi_\beta \mathcal{O}_A|\spc i\spc \rangle\,,
\end{eqnarray}
with $\Pi_\alpha$ the projection on the intermediate channel with given energy and spin.\footnote{For a general CFT, the chiral sectors may be labeled by other quantum numbers besides energy. For an irrational CFT like Liouville theory with only Virasoro symmetry, the only conserved quantum numbers are energy and spin. } 
The blocks $\Psi_\alpha$ and $\Phi_\beta$ are related via a braiding operation ${\cal R}$ that interchanges $\mathcal{O}_A$ and $\mathcal{O}_B$. The action of ${\cal R}$ on the blocks takes the form of a unitary basis transformation implemented by the chiral R-matrices
\begin{eqnarray}
\label{euclidr}
\Psi_\alpha\, = \, \sum_\beta \, {\cal R}_{\alpha\beta}\, \Phi_\beta, \quad & & \quad
\bar\Psi_\alpha\, =\, \sum_\beta\, \bar{\cal R}_{\alpha\beta} \,\bar\Phi_\beta.
\end{eqnarray}
Using the reality condition $\bar{\cal R}^T = {\cal R}^\dag$, we are ensured that (\ref{nca}) and (\ref{ncb}) are indeed identical provided we impose the unitarity condition ${\cal R}^\dag {\cal R} =1$. Implementing this construction requires that the quantum numbers of the two chiral sectors are paired up via an equal energy constraint \cite{narovlansky2023doublescaledsyksitterholography}.

%\XLQc{End of my edits. I am not sure how to make the discussion above for DSSYK. Is it only applicable to $\lambda\rightarrow 0$ limit?}
%We will return to this point in the Conclusion.
%when we return to the question whether our holographic dual should have two time directions or not.

%% file: images/geodesic_membrane.tex
\tikzset{every picture/.style={line width=0.75pt}} %set default line width to 0.75pt        

\begin{tikzpicture}[x=0.75pt,y=0.75pt,yscale=-1,xscale=1]
%uncomment if require: \path (0,251); %set diagram left start at 0, and has height of 251

%Image [id:dp9216899928412403] 
\draw (176.67,122.69) node  {\includegraphics[width=262pt,height=182.53pt]{images/TFD-2.png}};
%Shape: Circle [id:dp9505617157438597] 
\draw  [draw opacity=0][fill={rgb, 255:red, 126; green, 211; blue, 33 }  ,fill opacity=1 ] (190,73) .. controls (190,70.79) and (191.79,69) .. (194,69) .. controls (196.21,69) and (198,70.79) .. (198,73) .. controls (198,75.21) and (196.21,77) .. (194,77) .. controls (191.79,77) and (190,75.21) .. (190,73) -- cycle ;
%Shape: Circle [id:dp9408549092932629] 
\draw  [draw opacity=0][fill={rgb, 255:red, 126; green, 211; blue, 33 }  ,fill opacity=1 ] (166,61.33) .. controls (166,59.12) and (167.79,57.33) .. (170,57.33) .. controls (172.21,57.33) and (174,59.12) .. (174,61.33) .. controls (174,63.54) and (172.21,65.33) .. (170,65.33) .. controls (167.79,65.33) and (166,63.54) .. (166,61.33) -- cycle ;
%Image [id:dp199368378704342] 
\draw (629.67,122.31) node  {\includegraphics[width=262pt,height=182.53pt]{images/TFD-2.png}};
%Shape: Circle [id:dp6542585337167109] 
\draw  [draw opacity=0][fill={rgb, 255:red, 126; green, 211; blue, 33 }  ,fill opacity=1 ] (644,72.63) .. controls (644,70.42) and (645.79,68.63) .. (648,68.63) .. controls (650.21,68.63) and (652,70.42) .. (652,72.63) .. controls (652,74.84) and (650.21,76.63) .. (648,76.63) .. controls (645.79,76.63) and (644,74.84) .. (644,72.63) -- cycle ;
%Shape: Circle [id:dp5077580098768175] 
\draw  [draw opacity=0][fill={rgb, 255:red, 126; green, 211; blue, 33 }  ,fill opacity=1 ] (644,160.96) .. controls (644,158.75) and (645.79,156.96) .. (648,156.96) .. controls (650.21,156.96) and (652,158.75) .. (652,160.96) .. controls (652,163.17) and (650.21,164.96) .. (648,164.96) .. controls (645.79,164.96) and (644,163.17) .. (644,160.96) -- cycle ;
%Curve Lines [id:da2730345983996231] 
\draw [color={rgb, 255:red, 126; green, 211; blue, 33 }  ,draw opacity=1 ][line width=1.5]    (648,72.63) .. controls (657,80.96) and (657,146.96) .. (648,160.96) ;
%Curve Lines [id:da42287543306106956] 
\draw [color={rgb, 255:red, 126; green, 211; blue, 33 }  ,draw opacity=1 ][line width=1.5]    (181.33,129) .. controls (190.33,129) and (194.33,102) .. (194,73) ;
%Curve Lines [id:da16680956969214256] 
\draw [color={rgb, 255:red, 126; green, 211; blue, 33 }  ,draw opacity=1 ][line width=1.5]  [dash pattern={on 5.63pt off 4.5pt}]  (181.33,129) .. controls (169.33,131) and (170.33,91) .. (170,61.33) ;

% Text Node
\draw (199,69.4) node [anchor=north west][inner sep=0.75pt] [font=\Large]    {$X_{1}$};
% Text Node
\draw (150,36) node [anchor=north west][inner sep=0.75pt]  [font=\Large]   {$X_{2}$};
% Text Node
\draw (652,155.4) node [anchor=north west][inner sep=0.75pt][font=\Large]    {$X_{3}$};
% Text Node
\draw (652,70) node [anchor=north west][inner sep=0.75pt] [font=\Large]    {$X_{1}$};

\end{tikzpicture}

%% file: sections/discussion.tex
In this paper, we have established a detailed semiclassical correspondence between the non-linear dynamics of the soft mode of DSSYK and that of 2+1-D near-de Sitter gravity with $k=-1$ conformal boundary conditions at future and past infinity. The boundary conditions reduce the 2+1-D gravitational dynamics to that of a single complex reparametrization mode $\psi(u)$. The effective action and the equations of motion of this gravitational mode $\psi(u)$ match the soft mode dynamics of DSSYK \cite{Lensky_2021}, provided we identify the dimensionless DSSYK coupling $\lambda $ with the 3D Newton constant via
\begin{eqnarray}
\label{lambdagn}
\lambda = {2q^2}/N = {4 \pi G_N}.
\end{eqnarray}
Our results hold to leading order in $1/\lambda$ while allowing the energy and temperature to scale as $1/\lambda$. On the dual side, our computations are all done at the level of classical general relativity. However, since both DSSYK and pure de Sitter gravity are exactly solvable, it should be possible to extend our results to all orders in $\lambda$. Here we collect some comments and known results that help solidify our dictionary.

\def\syk{{\mbox{\tiny SYK}}}
\subsubsection*{Relation with 2D sine-dilaton gravity}
\vspace{-1mm}

Our holographic dictionary has several unusual properties. First, it relates a 1D quantum system to a 3D gravity theory and thus adds a total of two extra holographic dimensions. It is logical to look for an intermediate 2D theory that connects with both sides of our duality. The natural candidate for this 2D theory is sine-dilaton gravity \cite{Blommaert:2024ydx} or equivalently, complex Liouville gravity \cite{Mengyang_Verlinde:2024zrh,Collier:2024kwt_worldsheet} placed on a  M\"obius strip \cite{Blommaert:2025eps}. As shown in \cite{Blommaert:2024ydx}, sine-dilaton gravity on a strip can be reduced to a 1D effective theory identical to equation \eqref{action+hamiltonian} by employing a covariant small phase space approach \cite{harlow2019factorizationproblemjackiwteitelboimgravity}. The variable $\phi$ then describes the width of the strip. 
An alternative and equally direct way to see the link between the two theories is to note that the 1D effective action \eqref{action+hamiltonian} solves the functional differential equation
\begin{eqnarray}
\Bigl(\pi_\phi - \frac{2}{\lambda} \spc p'\Bigr)\Bigl(\pi_p + \frac{2}{\lambda}\spc  \dot\phi\Bigr)\! \is \! \frac{2}{\lambda^2} \, e^{-2\phi}\sin 2p, %\notag\\[-2mm]\\[-2mm]
\qquad \ \ 
\pi_\phi\! = \frac{\delta{S}_\syk\!\!}{\delta \phi}, \ \ \ \pi_p\! = \frac{\delta{S}_\syk\!\!}{\delta p}.
\end{eqnarray}
which can be recognized as the Hamilton-Jacobi-WDW equation of 2D sine-dilaton gravity in the conformal gauge, where $p$ plays the role of the dilaton and $\phi$ of the conformal factor of the 2D metric. Hence $\Psi_\syk(p,\phi) \equiv e^{-\frac{2}{\lambda} S_{\rm SYK}(p,\phi)}$ naturally plays the role of a sine-dilaton gravity wavefunction. 

Sine-dilaton gravity can be lifted \cite{Mengyang_Verlinde:2024zrh,Collier:2025lux} to 3D by considering the 3D near-de Sitter gravity path integral discussed in section~\ref{subsec:gravact}. %As mentioned there, the path integral with Neumann boundary conditions \eqref{eq:neumannbc} can be obtained by starting with conformal boundary conditions and then integrating over all conformal equivalence classes of boundary metrics. 
As shown in \cite{Mengyang_Verlinde:2024zrh,Collier:2025lux,Cotler_2020} 
the conformal bra and ket wavefunctions of 3D de Sitter gravity can be identified with the partition function of two conjugate complex Liouville CFTs with complex central charge $c_\pm = 13 \pm  \frac{3i }{2G_N}$. Then, the 3D near-de Sitter partition function can be represented as the 2D path integral of complex Liouville gravity theory with $c_+ + c_- = 26$ or, equivalently,  sine-dilaton gravity.

\def\darkblue{blue!85!black}
\def\darkred{red!80!black}
\def\darkgreen{green!70!black}

\subsubsection*{Bulk operators and worldline holography}
\label{subsec:bulk}\vspace{-1mm}
%At zero  (or infinite) temperature, the (opposite sided) two point function of two DSSYK scaling operators simplifies to
%\begin{eqnarray}\label{twopt}G_\Delta(u) = \bigl\langle {\cal O}_\Delta(u) {\cal O }_\Delta(0) \bigr\rangle  = \biggl(\frac 1{\sinh^2\bigl(\frac 1 2 (u-i\epsilon)\bigr)} \biggr)^{\!\Delta}\end{eqnarray}
As argued above, local operators in a single DSSYK model have properties that look like those of chiral operators in a 2D Euclidean CFT. Moreover, we have shown that exponentials $e^{-\Delta \, \ell(X_1,X_2)}$ of the length $\ell$ of boundary-to-boundary geodesics map to the absolute value squared of SYK two-point functions. All this suggests that a complete SYK/de Sitter dictionary should involve combining two identical DSSYK models, in the same way that a 2D Euclidean CFT consists of combining its holomorphic and antiholomorphic sectors, and that we can define bulk operators by adapting the HKLL construction \cite{Hamilton:2006az} to de Sitter spacetime \cite{Xiao:2014uea,Doi:2024nty, GMVX}
\begin{eqnarray}
\mathbb{O}_\Delta(X) \is \int_{\mathscr{I}^\pm} \!\!\! \dd^2u\, K_\Delta(X,u,\bar{u})\, \overline{\mathcal{O}}_\Delta(\bar u)\, \mathcal{O}_\Delta(u),
\label{eq:HKLLbilocal}
%\\[1mm]\label{eq:K_explicit}& &\ \ K_\Delta(X,u,\bar{u}) = \biggl(\frac{\eta}{\eta^2 - (x -u)(\bar{x}-\bar{u})}\biggr)^{2\Delta}.
\end{eqnarray}
where $K_\Delta(X,u,\bar{u})$ denotes the de Sitter version of the HKLL kernel. Here $\mathcal{O}_\Delta(u)$ and $\overline{\mathcal{O}}_\Delta(\bar{u})$ denote local scaling operators in the two DSSYK models. As shown in \cite{Xiao:2014uea,Doi:2024nty,GMVX}, the above class of operators defines a generalized free field in de Sitter spacetime. The derivation follows the same logic as for the usual HKLL construction, except for the fact that the integral over $u$ and $\bar{u}$ now runs over the full real axis. In de Sitter space, this means that the integral runs over all of $\mathscr{I}^+$.

A closely related dictionary for worldline de Sitter holography was proposed in \cite{narovlansky2023doublescaledsyksitterholography}. In this setup, the physical bulk observables acting along the geodesic worldline of an idealized de Sitter observer are defined by the convolution integral
\begin{equation}
\label{eq:solips}
    \mathbb{O}_{\Delta}(\tau) \, = \, \int du\,\widetilde{\mathcal{O}}_{1-\Delta}(u)\,\mathcal{O}_{\Delta}(\tau-u)
\end{equation}
of two scaling operators ${\cal O}_\Delta$ and $\tilde{O}_{1-\Delta}$ in two identical but independent SYK models. 
%In \cite{narovlansky2023doublescaledsyksitterholography} this class of operators were identified as `physical operators' that commute with a zero total energy constraint.
 The relative time variable $\tau$ in \eqref{eq:solips} measures the Lorentzian time experienced by a static de Sitter observer. Hence, simultaneous forward time evolution in the doubled SYK system corresponds to physical time for the observer \cite{narovlansky2023doublescaledsyksitterholography}.
The above observables \eqref{eq:solips} have been studied in the context of static patch holography in \cite{Anninos_2012, narovlansky2023doublescaledsyksitterholography} and shown to reproduce the (antipodal) de Sitter two-point function evaluated along the observer worldline.\footnote{ As shown in \cite{GMVX}, the formulas \eqref{eq:solips} and \eqref{eq:HKLLbilocal} are related by placing the point $X$ along the observer trajectory and writing the operator $\widetilde{O}_{1-\Delta}({u})$ as a 1D shadow transform of the operator $\overline{O}_{\Delta}(\bar{u})$.} The result of Sec.~\ref{subsec:hayward_time}, where we showed that the u-flow along the curve $\cal{C}^+$ at future infinity $\mathscr{I}^+$ can be extended to a time flow along the trajectory of a static observer, is consistent with this interpretation.  

As explained in \cite{Anninos_2012,GMVX}, the above construction of a generalized free field in de Sitter spacetime relies only on 1D conformal symmetry, large $N$ factorization, and the split representation of de Sitter Green functions \cite{Sleight:2019mgd}. All these properties receive corrections as soon as one turns on the coupling $\lambda$ of the DSSYK boundary theory or gravitational interactions in the bulk. The results in this paper indicate that the two types of corrections are in fact each other's holographic duals via the identification $\lambda = 4\pi G_N$. If this proposal is correct, then the $n$-point functions of the physical operators \eqref{eq:solips}, computed using the dynamical rules given in this paper, should match those of a generalized free field on a worldline in the backreacted spacetime \cite{kolchmeyer2024chaosemergencecosmologicalhorizon}.

\section*{Acknowledgments}

\vspace{-1mm}

We thank Shoaib Akhtar, Ahmed Almheiri, Andreas Blommaert, Aidan Herderschee, David Kolchmeyer, Henry Lin, Juan Maldacena, Vladimir Narovlansky, Beatrix M\"uhlmann, Andrew Sontag, Haifeng Tang, Damiano Tietto, and Zhenbin Yang for useful discussions. Part of this work was carried out while XLQ was on sabbatical at the Institute for Advanced Study (IAS). XLQ thanks IAS, and in particular his host Juan Maldacena, for their hospitality. XLQ is supported by the Simons Foundation.

%% file: sections/spinor_coordinates.tex
In this appendix, we review the spinor decomposition of coordinates  and compute the length of boundary-to-boundary geodesics on the asymptotic boundary of 3D de Sitter spacetime. 

\subsection{3D de Sitter in embedding space}
\vspace{-1mm}
De Sitter spacetime is the maximally symmetric solution to the Einstein equations with  positive cosmological constant.  Three-dimensional de Sitter dS$_3$ can be embedded in $\mathbb{R}^{3,1}$ as the hyperboloid
\begin{equation}
    -X_0^2+X_1^2+X_2^2+X_3^2=R^2_{\rm dS},
\end{equation}
where $R_{\rm dS}$ is the de Sitter radius; here and in the main text, we choose units such that $R_{\rm dS}=1$. 
The geodesic distance $\ell(X, Y)$ between two points with embedding coordinates $X$ and  $Y$ is
\begin{equation}
\begin{split}
    \ell(X, Y) = \text{cos}^{-1}(X \cdot Y) \qquad &\text{if $X,Y$ are spacelike;}\\
   \ell(X, Y) = \text{cosh}^{-1}(X\cdot Y) \qquad &\text{if $X,Y$ are timelike.}
\end{split}
\label{eq:geodesic_embedding}
\end{equation}
Here $X\cdot Y$ is the inner product with signature $(-, +, +, +)$. Note that if $X\cdot Y<-1$ there are no real geodesics connecting $X$ and $Y$.

The embedding coordinates $X_i$ can be organized into a  matrix 
\begin{equation}
    \mathbf{X}=\begin{pmatrix}
        X_0-X_1 &X_2+iX_3\\
        X_2-i X_3 & X_0+X_1
    \end{pmatrix}\qquad \text{det}\mathbf{X}=-1.
\end{equation}
In matrix notation, the inner product of two vectors $X, Y$  is
\begin{equation}
    X\cdot Y  = \frac{1}{2}\text{tr}(\mathbf{X}\epsilon \mathbf{Y}^t \epsilon ),
\end{equation}
with $\epsilon$ the 2d $\epsilon$-symbol. This matrix notation will be useful in the study of geodesics that connect points near the asymptotic past or future boundary $\mathscr{I}^\pm$ of  3D de Sitter space.

The topology of dS$_3$ is $S^2\times \mathbb{R}$, with $S^2$ the spatial sphere. 
We can cover the hyperboloid using different coordinate systems. A global coordinate system is
\begin{flalign}
\quad
\begin{cases}
X_0 &= \sinh{\tau} \\
X_1 &= \cosh\tau\cos\theta \\
X_2 &= \cosh\tau\sin\theta\cos\varphi \\
X_3 &= \cosh\tau\sin\theta\sin\varphi
\end{cases}
&&
ds^2 &= -d\tau^2 + \cosh^2{\tau}\; (d\theta^2+\sin^2\theta d\varphi^2). &&
\label{eq:global_coordinates}
\end{flalign}
 We also used the following coordinates on the spatial sphere,
\begin{equation}
    x=\varphi\qquad y = \log\cot\theta/2, \qquad ds^2 = -d\tau^2 +  \frac{\cosh^2{\tau}}{\cosh^2y}(dx^2+dy^2).
    \label{eq:complex_coordinates}
\end{equation}

\medskip
\subsection{Spinor description of boundary-to-boundary geodesic length}\label{sec:geodlength}

The form of the coordinate matrix $\mathbf{X}$ simplifies when we approach $\mathscr{I}^\pm$, where we can factor out a term that diverges as $X_0\to \pm \infty$. In terms of the coordinates we introduced above, we have
%\begin{eqnarray}
%    \mathbf{X}^+& \to & \frac{e^\tau}{2}\frac{1}{\cos\frac{\psi-\psi^*}2 }\left(\begin{array}{c}e^{i\psi/2}\\e^{-i\psi/2}\end{array}\right)\left(e^{-i\psi^*/2}~e^{i\psi^*/2}\right)\\[-6mm]&& \qquad\qquad\qquad\qquad\qquad\qquad\qquad\qquad\qquad\qquad\qquad \begin{array}{c}{\psi = x+iy}\\[2mm]{\psi^*=x-iy}\end{array} \notag\\[-6mm]
%    \mathbf{X}^-& \to & \frac{e^{-\tau}}{2}\frac{1}{\cos\frac{\psi-\psi^*}2 }\left(\begin{array}{c}e^{i\psi^*/2}\\-e^{-i\psi^*/2}\end{array}\right)\left(-e^{-i\psi/2}~e^{i\psi/2}\right)
%\end{eqnarray}
\begin{eqnarray}
    \mathbf{X}^+ &\to& \; \frac{e^T}{2}\frac1{\sinh{|\rm Im}\psi|}\left(\begin{array}{c}e^{i\psi/2}\\e^{-i\psi/2}\end{array}\right)\left(e^{-i\psi^*/2}~e^{i\psi^*/2}\right) \qquad\ \ \, {\rm on}\, \ \mathscr{I}^+\\[3mm]
    \mathbf{X}^-&\to& \frac{e^{T}}{2}\frac{1}{\sinh{|\rm Im}\psi|}\left(\begin{array}{c}e^{i\psi^*/2}\\-e^{-i\psi^*/2}\end{array}\right)\left(-e^{-i\psi/2}~e^{i\psi/2}\right) \qquad {\rm on}\, \ \mathscr{I}^-
\end{eqnarray}
Here we introduced the complex variable $\psi=x+i y$,
which plays an important role in the main text, and used the definition \ref{eq:scridef}, $\tau= \pm(T-\log\tanh |\rm{Im}\psi|)$, of $\mathscr{I}^\pm$\footnote{If we consider the hyperbolic slicing defined by constant $X_1$,
\begin{align}
    X_1=\cosh\tau\cos\theta=\cosh T,
\end{align}
then approaching $\mathscr{I}^\pm$ amounts to sending $T\rightarrow +\infty,\tau\rightarrow \pm \infty$ with finite $\cos\theta$. We then have
\begin{align}
    e^{\pm \tau}=\frac{\cosh T}{\cos\theta}+\sqrt{\frac{\cosh^2 T}{\cos^2\theta}-1}\simeq \frac{e^T}{\cos\theta}=\frac{e^T}{\tanh y}.
\end{align}}.
%In the $T\rightarrow +\infty$ limit we have
%\begin{eqnarray}
%    \mathbf{X}^+ &\to& \; \frac{e^T}{2}\frac1{\sinh{\rm Im}\psi}\left(\begin{array}{c}e^{i\psi/2}\\e^{-i\psi/2}\end{array}\right)\left(e^{-i\psi^*/2}~e^{i\psi^*/2}\right) \qquad\ \ \, {\rm on}\, \ \mathscr{I}^+\\[3mm]
%    \mathbf{X}^-&\to& \frac{e^{T}}{2}\frac{1}{\sinh{\rm Im}\psi}\left(\begin{array}{c}e^{i\psi^*/2}\\-e^{-i\psi^*/2}\end{array}\right)\left(-e^{-i\psi/2}~e^{i\psi/2}\right) \qquad {\rm on}\, \ \mathscr{I}^-
%\end{eqnarray}
The above decomposition motivates introducing the spinor
\begin{align}
\mathsf{z}[\psi]&=\frac1{\sqrt{2\sinh{|\rm Im}\psi|}}\left(\begin{array}{c}e^{i\psi/2}\\e^{-i\psi/2}\end{array}\right),
\end{align}
in terms of which, we have
\begin{eqnarray}
\label{eq:matrices_at_scrai}
    \mathbf{X}^+[\psi]\rightarrow \ e^T\mathsf{z}[\psi] \,\mathsf{z}[\psi]^\dag,\qquad\quad  &&\quad
    \mathbf{X}^-[\psi]\!\rightarrow -e^T\mathsf{z}[\psi^*+\pi] \mathsf{z}[\psi^*+\pi]^\dagger.
\end{eqnarray}

If we consider the length between points at $\mathscr{I}^\pm$, the expressions again simplify. First, using equation \eqref{eq:matrices_at_scrai}, we have
\begin{equation}
\begin{split}
    \text{tr}(\mathbf{X}^+_1\epsilon \mathbf{X}^{+t}_2 \epsilon )&= -\frac{e^{2T}}{\sinh {|\rm Im}\psi_1|\sinh {|\rm Im}\psi_2|} \sin\left(\frac{\psi_1-\psi_2}{2}\right)\sin\left(\frac{\psi_1^*-\psi_2^*}{2}\right);\\[3mm]
    \text{tr}(\mathbf{X}^+_1\epsilon \mathbf{X}^{-t}_2 \epsilon )&= \frac{e^{2T}}{\sinh {|\rm Im}\psi_1|\sinh {|\rm Im}\psi_2|} \cos\left(\frac{\psi_1-\psi^*_2}{2}\right)\cos\left(\frac{\psi_1^*-\psi_2}{2}\right).
\end{split}
\end{equation}
As $T\to \infty$, the geodesic length diverges and we can approximate $\cosh^{-1}(x)\simeq \ln 2 x$. Then, the geodesic length between points at $\mathscr{I}^\pm$ is
\begin{equation}
\begin{split}
    \ell(X^+_1, X^+_2)\simeq&  \log\left(\frac{\sin\left((\psi_1-\psi_2)/2\right)\sin\left((\psi_1^*-\psi^*_2)/2\right)}{\sinh {|\rm Im}\psi_1|\sinh {|\rm Im}\psi_2|}\right) +2T +i\pi,\\[3mm]
        \ell(X^+_1, X^-_2)\simeq&  \log\left(\frac{\cos\left((\psi_1-\psi_2^*)/2\right)\cos\left((\psi_1^*-\psi_2)/2\right)}{\sinh {|\rm Im}\psi_1|\sinh{|\rm Im}\psi_2|}\right) +2T.
    \label{eq:geodiscs_length}
    \end{split}
\end{equation}

Let us briefly comment on the meaning of the first expression in equation \eqref{eq:geodiscs_length} above, which presents two subtleties. Firstly, it is complex; secondly, we used  $  \ell(X_1, X_2) = \text{cosh}^{-1}(X_1\cdot X_2) $, which  holds for timelike separated points, but  points at $\mathscr{I}^+$ are spacelike separated. Indeed, from equation ~\eqref{eq:geodesic_embedding}, we see that there is no real geodesic connecting two points $X_1$, $X_2$, when $X_1\cdot X_2$ is less than $-1$, which is  the case for different points at $\mathscr{I}^+$\footnote{In global coordinates we have $
    X_1\cdot X_2 = -\sinh \tau_1\sinh \tau_2 + \vec{n}_1\cdot\vec{n}_2\cosh \tau_1\cosh\tau_2,$
where $\vec{n}_1$ and $\vec{n}_2$ denote the two points on the spatial sphere. When both points are at $\mathscr{I}^+$, $\tau_1, \tau_2\to\infty$, we have $ X_1\cdot X_2 \to e^{\tau_1+\tau2}(-1 + \vec{n}_1\cdot\vec{n}_2)<-1.$}.  
However, the expression above is still useful, as it is the  geodesic distance that computes two-point functions of operators inserted at $\mathscr{I}^+$, in the Bunch Davies state.  We can interpret it as the length of a geodesic in a complex spacetime. This spacetime is obtained by continuing to Euclidean time $\tau\to i\tau$ on the $\tau=0$ slice, and describes a half $S^3$ glued to the $\tau=0$ slice of de Sitter. The Euclidean part of the geometry  prepares the Bunch-Davies state. The geodesic connecting two points $X_1^+$, $X_2^+$ at $\mathscr{I}^+$ starts from $X_1^+$ and follows the trajectory that would end at the point antipodal to $X_2^+$ at $\mathscr{I}^-$; when the geodesic encounter the $\tau=0$ slice it enters the Euclidean part of the geometry, and follows a great circle on the sphere; finally it emerges on the Lorentzian part and reaches $X_2^+$ at $\mathscr{I}^+$. The $i\pi$ is the length on the Euclidean sphere.

We conclude by noting that, in spinor notation, the SYK two-point functions \eqref{eq:g_psi} are given by
\begin{eqnarray}
    e^{g_{RR}(u_1,u_2)}\is {\left(\mathsf{z}[\psi(u_2)]^\dagger \sigma_z\mathsf{z}[\psi(u_1)]\right)^{-2}}\notag\\[-2mm]\\[-2mm]\notag
    \ e^{g_{RL}(u_1,u_2)}\is {\left(\mathsf{z}[\psi(u_2)]^\dagger \mathsf{z}[\psi(u_1)]\right)^{-2}}.
\end{eqnarray}
The dictionary \eqref{eq:geodict} follows from these expressions.

%% file: sections/more_on_the_action.tex
\vspace{-2mm}

In this appendix, we explicitly evaluate each term of the gravitational action \eqref{eq:action}, 
\begin{equation}
    S_{EH} =\frac{1}{16\pi G_N} \left[\int_{\mathcal{V}}\!\! \sqrt{-g}\spc(R-2)+2\int_{\Sigma}\!\! \sqrt{-h}\spc K_h - \int_{\mathscr{I}^\pm}\!\!\sqrt{\sigma}\spc K_\sigma+2\int_{\cal C^\pm}\!\!\!\sqrt{\gamma}\spc \eta_H\right],
\end{equation}
and show that the boundary terms enforce the boundary conditions we outlined in Sec.~\ref{subsec:effective_action}. Here, all the integrals are performed over both sides, i.e. on $\cal V$, $\cal V^*$, $\Sigma$, $\Sigma^*$, $\dots$. 

\subsection{Evaluation of the Einstein-Hilbert action}
\vspace{-2mm}

To evaluate the bulk contribution, we use that $R=6$ away from the $dS_2$ slice $\Sigma$, and that the time integral runs from $-(T+\phi)$ to $(T+\phi)$ in the presence of the cutoff \eqref{eq:scridef},
\begin{equation}
        \int_{\mathcal{V}}\!\!\sqrt{-g}\,(R-2) =4\int \frac{ dx  dy}{\cosh^2y}\int_{-T-\phi(y) }^{T+\phi(y)}d\tau \cosh ^2\tau = 
e^{2T}\int \frac{dxdy}{\sinh^2y}+4\int \frac{dxdy}{\cosh^2y}(T+\phi). 
\end{equation}
To evaluate the contribution from the $dS_2$ slice, we use (minus\footnote{In Sec.~\ref{subsec:cutting_ds} we computed the inward pointing normal, whereas in the action we use the outward pointing normal. }) \eqref{eq:extrinsic_curvature} for the extrinsic curvature, and \eqref{eq:metric_membrane} for the induced metric on the slice,
\begin{equation}
\begin{split}
    \int_{\Sigma}\!\!\sqrt{-h}K_h &=-2 \int \frac{du|\psi'|\cosh\tau }{\cosh y}\frac{\cosh y}{\cosh \tau |\psi'|}(p'\!+|\psi'|\tanh \text{Im}\psi\cos p) \int_{-T-\phi}^{T+\phi} d\tau\\
    &=-4\int du (T+\phi) (p'+|\psi'|\tanh \text{Im}\psi\cos p).
\end{split}
\end{equation}
To evaluate the boundary terms at $\mathscr{I}^\pm$, we need to compute the extrinsic curvature. Let $m^\mu$ be the outward pointing normal to the cutoff surface; we have 
\begin{equation}
    m_\mu= -\frac{\cosh \tau \sinh |y|}{\sqrt{\cosh\tau^2\sinh y^2-1}}\left(1, 0, \frac{1}{\sinh y\cosh y} \right),\qquad K_\sigma=\nabla_\mu m^\mu=2+4 e^{-2T}+\dots
\end{equation}
The induced metric at $\mathscr{I}^\pm$ is
\begin{equation}
    \sigma_{ij}dx^idx^j = \frac{e^{2T}}{4}\frac{dx^2+dy^2}{\sinh^2y} +\frac{dx^2}{2\cosh^2y}+ \frac{dy^2}{\cosh^2 y}\left(\frac{1}{2}-\frac{1}{\sinh^2y}\right)+\dots,
\end{equation}
and the contribution to the action evaluates to
\begin{equation}
\begin{split}
      \int_{\mathscr{I}^\pm}\!\!\sqrt{\sigma} K_\sigma&=2\int dxdy\left(\frac{e^{2T}}{4\sinh^2y}+\frac{1}{2\cosh^2y}\left(1-\frac{1}{\sinh^2y}\right)\right)(2+4e^{-2T})\\[1mm]
      &=e^{2T}\int\frac{dxdy}{\sinh^2 y} +4\int\frac{dxdy}{\cosh^2y}.
\end{split}
\end{equation}
Finally, to evaluate the corner terms, we need the boost parameter $\eta$ and the induced metric $\gamma$ at the intersection between the slice and $\mathscr{I}^\pm$,
\begin{equation}
    \eta_H=\sinh^{-1}m_\mu n^\mu= 2e^{-T} \frac{\cos p}{\cosh \text{Im}\psi}, \qquad     \gamma_{ij}dx^idx^j = \frac{e^{2T}}{4}\frac{|\psi'|^2du^2}{\sinh^2 \text{Im}\psi}.
\end{equation}
Here we used the normal \eqref{eq:tangent_normal} to the $dS_2$ slice, and substituted  $y=\text{Im}\psi$ on the intersection. Then, the contribution of the corner term is
\begin{equation}
\int_{\cal C^\pm}\!\!\!\sqrt{\gamma}\spc \eta_H=%\int_{\Sigma^\pm\cap\mathscr{I}^\pm}\sqrt{\sigma}\sinh^{-1}m_\mu n^\mu=%
    4\int\! du \, \frac{|\psi'|}{\sinh |\text{Im}\psi|}\frac{\cos p}{\cosh \text{Im}\psi}=\color{white}\boxed{\color{black}4\int\! du \, \frac{|\psi'|}{\sinh |\text{Im}\psi|} \sqrt{1-e^{-2\phi}}\cos p.}
\end{equation}

\vspace{-2mm}

\subsection{Boundary action for Neumann boundary conditions}
In Sec.~\ref{subsec:effective_action} we defined the gravitational dynamics by choosing Dirichlet boundary conditions on $\Sigma$, and conformal boundary conditions on $\mathscr{I}^\pm$. We want to find an action whose variation vanishes on shell, once we impose these boundary conditions. We first find the appropriate action for Neumann boundary conditions at $\mathscr{I}^\pm$, and then show that the same action is appropriate for conformal boundary conditions.

Consider the gravitational action appropriate for Dirichlet boundary conditions on $\Sigma$~and~$\mathscr{I}^\pm$,
\begin{eqnarray}
    S' =\frac{1}{16\pi G_N} \left[\int_{\mathcal{V}^\pm}\!\! \sqrt{-g}\spc(R-2)+2\int_{\Sigma^{\pm}}\!\! \sqrt{-h}\spc K_h - 2\int_{\mathscr{I}^\pm}\!\!\sqrt{\sigma}\spc K_\sigma+2\int_{\Sigma^\pm\cap\mathscr{I}^\pm}\!\!\!\sqrt{\gamma}\spc \eta\right].
\end{eqnarray}
The first term is the Einstein-Hilbert action. The second and third terms are the Gibbons-Hawking-York terms \cite{Gibbons:1976ue, PhysRevLett.28.1082}, and enforce Dirichlet boundary conditions both on $\Sigma$ and $\mathscr{I}^\pm$. The last term is a Hayward term \cite{Hayward:1993my}, and is necessary to eliminate the variation of the normal at the intersections between $\Sigma$ and $\mathscr{I}^\pm$. Indeed, when the equations of motion are satisfied \cite{ Hayward:1993my, BlauGRNotes}, 
\begin{equation}
    \delta S'= -\int_{\Sigma_\pm} \Pi^{ab}_{h} \delta h_{ab}+\int_{\mathscr{I}^\pm} \Pi_\sigma^{ij}\delta \sigma_{ij}+2\int_{\Sigma^\pm\cap\mathscr{I}^\pm}\eta\spc\delta \sqrt{\gamma}.
\end{equation}
Here the $\Pi^{ij}$s are the canonical momenta conjugate to the induced metrics, e.g.
\begin{equation}
    16\pi G_N\Pi^{ij}_{\sigma}\equiv\sqrt{\sigma}(K^{ij}_{\sigma}-K_\sigma\sigma^{ij}).
\end{equation}
Since we are not fixing the metric at $\mathscr{I}^\pm$, $\delta\sigma_{ij}\neq 0$, and $\delta S'$ does not vanish on shell. Instead, we impose $\delta \Pi^{ij}_{\sigma}=0$. Then, we need to Legendre transform  with respect to $\Pi^{ij}$, $\sigma_{ij}$ \cite{BlauGRNotes,  Krishnan_2017}  
\begin{equation}
\label{eq:action_variation}
    S_{EH}=S'-\int_{\mathscr{I}^\pm}\Pi^{ij}_{\sigma}\sigma_{ij}, \qquad \delta S_{EH} = -\int_{\Sigma_\pm} \Pi^{ab}_{h} \delta h_{ab}-\int_{\mathscr{I}^\pm}  \sigma_{ij}\delta \Pi_\sigma^{ij}+2\int\eta\spc\delta \sqrt{\gamma}.
\end{equation}
The variation of $S_{EH}$ vanishes once we impose $\delta h_{ab}=\delta \gamma=\delta \Pi^{ij}_\sigma=0$. Finally, using $\Pi^{ij}_{\sigma}\sigma_{ij}=-K_{\sigma}$, we verify that $S_{EH}$ is indeed the gravitational action \eqref{eq:action}.

\subsection{Conformal boundary conditions}
In the main text, we claimed that the gravitational action \eqref{eq:action} is the appropriate action for conformal boundary conditions on $\mathscr{I}^\pm$. Let us show that the variation \eqref{eq:action_variation}, and in particular the term $\sigma_{ij}\delta \Pi_{\sigma}^{ij}$, vanishes if we fix the conformal structure and the trace of the extrinsic curvature at $\mathscr{I}^\pm$,
\begin{equation}
    \sigma_{ij} = e^{\omega(x^i)}\hat{\sigma}_{ij}, \qquad \delta \hat \sigma_{ij} = 0,\quad \delta K_{\sigma}=0.
\end{equation}
We have
\begin{equation}
        \delta \Pi_{\sigma}^{ij} \,\propto \,\sqrt{\hat\sigma}\delta(e^{\omega}K_{\sigma}^{ij}-K_{\sigma}\hat\sigma_{ij}) = \sqrt{\sigma} (K_{\sigma}^{ij}\delta \omega+\delta K^{ij}_{\sigma}),
\end{equation}
and contracting with $\sigma_{ij}$, we obtain
\begin{equation}
    \sigma_{ij}\delta\Pi_{\sigma}^{ij}\, \propto\,\sqrt{\sigma}( K_{\sigma}\delta \omega +\sigma_{ij}\delta K^{ij}_{\sigma})=\sqrt{\sigma}\delta K_{\sigma}=0.
\end{equation}
Here we used
\begin{equation}
    \delta K_{\sigma} = \delta \sigma_{ij}K^{ij}_{\sigma}+\sigma_{ij}\delta K^{ij}_\sigma =K_{\sigma}\delta \omega + \sigma_{ij}\delta K_{\sigma}^{ij}.
\end{equation}